\documentclass[longauth]{aaEC}

\usepackage{lastpage}
\usepackage{graphicx}
\usepackage{natbib}
\usepackage{scalerel}

\usepackage[table]{xcolor}

\usepackage{txfonts}
\usepackage[pdfencoding=auto,psdextra]{hyperref}
\hypersetup{
    colorlinks=true,
    linkcolor=blue,
    filecolor=magenta,
    urlcolor=blue,
    citecolor=blue
}
\makeatletter
\renewcommand*\aa@pageof{, page \thepage{} of \pageref*{LastPage}}
\makeatother

\newcommand{\mat}[1]{\tens{#1}}

\usepackage[utf8]{inputenc}

\usepackage[switch, modulo]{lineno}

\usepackage{euclid}
\usepackage{ulem}
\defcitealias{hernandez20}{H20}
\defcitealias{congedo24}{C24}

\begin{document}
\title{\Euclid: Early Release Observations. Weak gravitational lensing analysis of Abell 2390\thanks{This paper is published on behalf of the Euclid Consortium.}}

\newcommand{\orcid}[1]{}
\author{T.~Schrabback\orcid{0000-0002-6987-7834}\thanks{\email{tim.schrabback@uibk.ac.at}}\inst{\ref{aff1},\ref{aff2}}
\and G.~Congedo\orcid{0000-0003-2508-0046}\inst{\ref{aff3}}
\and R.~Gavazzi\orcid{0000-0002-5540-6935}\inst{\ref{aff4},\ref{aff5}}
\and W.~G.~Hartley\inst{\ref{aff6}}
\and H.~Jansen\orcid{0009-0002-1332-7742}\inst{\ref{aff1}}
\and Y.~Kang\orcid{0009-0000-8588-7250}\inst{\ref{aff6}}
\and F.~Kleinebreil\inst{\ref{aff1}}
\and H.~Atek\orcid{0000-0002-7570-0824}\inst{\ref{aff5}}
\and E.~Bertin\inst{\ref{aff7}}
\and J.-C.~Cuillandre\orcid{0000-0002-3263-8645}\inst{\ref{aff7}}
\and J.~M.~Diego\orcid{0000-0001-9065-3926}\inst{\ref{aff8}}
\and S.~Grandis\orcid{0000-0002-4577-8217}\inst{\ref{aff1}}
\and H.~Hoekstra\orcid{0000-0002-0641-3231}\inst{\ref{aff9}}
\and M.~K\"ummel\orcid{0000-0003-2791-2117}\inst{\ref{aff10}}
\and L.~Linke\orcid{0000-0002-2622-8113}\inst{\ref{aff1}}
\and H.~Miyatake\orcid{0000-0001-7964-9766}\inst{\ref{aff11},\ref{aff12},\ref{aff13}}
\and N.~Okabe\orcid{0000-0003-2898-0728}\inst{\ref{aff14},\ref{aff15},\ref{aff16}}
\and S.~Paltani\orcid{0000-0002-8108-9179}\inst{\ref{aff6}}
\and M.~Schefer\inst{\ref{aff6}}
\and P.~Simon\inst{\ref{aff2}}
\and F.~Tarsitano\orcid{0000-0002-5919-0238}\inst{\ref{aff6}}
\and A.~N.~Taylor\inst{\ref{aff3}}
\and J.~R.~Weaver\orcid{0000-0003-1614-196X}\inst{\ref{aff17}}
\and R.~Bhatawdekar\orcid{0000-0003-0883-2226}\inst{\ref{aff18}}
\and M.~Montes\orcid{0000-0001-7847-0393}\inst{\ref{aff19}}
\and P.~Rosati\orcid{0000-0002-6813-0632}\inst{\ref{aff20},\ref{aff21}}
\and S.~Toft\orcid{0000-0003-3631-7176}\inst{\ref{aff22},\ref{aff23}}
\and B.~Altieri\orcid{0000-0003-3936-0284}\inst{\ref{aff18}}
\and A.~Amara\inst{\ref{aff24}}
\and L.~Amendola\orcid{0000-0002-0835-233X}\inst{\ref{aff25}}
\and S.~Andreon\orcid{0000-0002-2041-8784}\inst{\ref{aff26}}
\and N.~Auricchio\orcid{0000-0003-4444-8651}\inst{\ref{aff21}}
\and C.~Baccigalupi\orcid{0000-0002-8211-1630}\inst{\ref{aff27},\ref{aff28},\ref{aff29},\ref{aff30}}
\and M.~Baldi\orcid{0000-0003-4145-1943}\inst{\ref{aff31},\ref{aff21},\ref{aff32}}
\and A.~Balestra\orcid{0000-0002-6967-261X}\inst{\ref{aff33}}
\and S.~Bardelli\orcid{0000-0002-8900-0298}\inst{\ref{aff21}}
\and P.~Battaglia\orcid{0000-0002-7337-5909}\inst{\ref{aff21}}
\and R.~Bender\orcid{0000-0001-7179-0626}\inst{\ref{aff34},\ref{aff10}}
\and A.~Biviano\orcid{0000-0002-0857-0732}\inst{\ref{aff28},\ref{aff27}}
\and E.~Branchini\orcid{0000-0002-0808-6908}\inst{\ref{aff35},\ref{aff36},\ref{aff26}}
\and M.~Brescia\orcid{0000-0001-9506-5680}\inst{\ref{aff37},\ref{aff38}}
\and J.~Brinchmann\orcid{0000-0003-4359-8797}\inst{\ref{aff39},\ref{aff40},\ref{aff41}}
\and S.~Camera\orcid{0000-0003-3399-3574}\inst{\ref{aff42},\ref{aff43},\ref{aff44}}
\and G.~Ca\~nas-Herrera\orcid{0000-0003-2796-2149}\inst{\ref{aff45},\ref{aff46},\ref{aff9}}
\and G.~P.~Candini\orcid{0000-0001-9481-8206}\inst{\ref{aff47}}
\and V.~Capobianco\orcid{0000-0002-3309-7692}\inst{\ref{aff44}}
\and C.~Carbone\orcid{0000-0003-0125-3563}\inst{\ref{aff48}}
\and V.~F.~Cardone\inst{\ref{aff49},\ref{aff50}}
\and J.~Carretero\orcid{0000-0002-3130-0204}\inst{\ref{aff51},\ref{aff52}}
\and S.~Casas\orcid{0000-0002-4751-5138}\inst{\ref{aff53}}
\and F.~J.~Castander\orcid{0000-0001-7316-4573}\inst{\ref{aff19},\ref{aff54}}
\and M.~Castellano\orcid{0000-0001-9875-8263}\inst{\ref{aff49}}
\and G.~Castignani\orcid{0000-0001-6831-0687}\inst{\ref{aff21}}
\and S.~Cavuoti\orcid{0000-0002-3787-4196}\inst{\ref{aff38},\ref{aff55}}
\and K.~C.~Chambers\orcid{0000-0001-6965-7789}\inst{\ref{aff56}}
\and A.~Cimatti\inst{\ref{aff57}}
\and C.~Colodro-Conde\inst{\ref{aff58}}
\and C.~J.~Conselice\orcid{0000-0003-1949-7638}\inst{\ref{aff59}}
\and L.~Conversi\orcid{0000-0002-6710-8476}\inst{\ref{aff60},\ref{aff18}}
\and Y.~Copin\orcid{0000-0002-5317-7518}\inst{\ref{aff61}}
\and A.~Costille\inst{\ref{aff4}}
\and F.~Courbin\orcid{0000-0003-0758-6510}\inst{\ref{aff62},\ref{aff63}}
\and H.~M.~Courtois\orcid{0000-0003-0509-1776}\inst{\ref{aff64}}
\and M.~Cropper\orcid{0000-0003-4571-9468}\inst{\ref{aff47}}
\and A.~Da~Silva\orcid{0000-0002-6385-1609}\inst{\ref{aff65},\ref{aff66}}
\and H.~Degaudenzi\orcid{0000-0002-5887-6799}\inst{\ref{aff6}}
\and G.~De~Lucia\orcid{0000-0002-6220-9104}\inst{\ref{aff28}}
\and H.~Dole\orcid{0000-0002-9767-3839}\inst{\ref{aff67}}
\and M.~Douspis\orcid{0000-0003-4203-3954}\inst{\ref{aff67}}
\and F.~Dubath\orcid{0000-0002-6533-2810}\inst{\ref{aff6}}
\and X.~Dupac\inst{\ref{aff18}}
\and S.~Dusini\orcid{0000-0002-1128-0664}\inst{\ref{aff68}}
\and S.~Escoffier\orcid{0000-0002-2847-7498}\inst{\ref{aff69}}
\and M.~Farina\orcid{0000-0002-3089-7846}\inst{\ref{aff70}}
\and R.~Farinelli\inst{\ref{aff21}}
\and S.~Farrens\orcid{0000-0002-9594-9387}\inst{\ref{aff7}}
\and F.~Faustini\orcid{0000-0001-6274-5145}\inst{\ref{aff49},\ref{aff71}}
\and S.~Ferriol\inst{\ref{aff61}}
\and F.~Finelli\orcid{0000-0002-6694-3269}\inst{\ref{aff21},\ref{aff72}}
\and P.~Fosalba\orcid{0000-0002-1510-5214}\inst{\ref{aff54},\ref{aff19}}
\and M.~Frailis\orcid{0000-0002-7400-2135}\inst{\ref{aff28}}
\and E.~Franceschi\orcid{0000-0002-0585-6591}\inst{\ref{aff21}}
\and M.~Fumana\orcid{0000-0001-6787-5950}\inst{\ref{aff48}}
\and S.~Galeotta\orcid{0000-0002-3748-5115}\inst{\ref{aff28}}
\and K.~George\orcid{0000-0002-1734-8455}\inst{\ref{aff10}}
\and W.~Gillard\orcid{0000-0003-4744-9748}\inst{\ref{aff69}}
\and B.~Gillis\orcid{0000-0002-4478-1270}\inst{\ref{aff3}}
\and C.~Giocoli\orcid{0000-0002-9590-7961}\inst{\ref{aff21},\ref{aff32}}
\and J.~Gracia-Carpio\inst{\ref{aff34}}
\and A.~Grazian\orcid{0000-0002-5688-0663}\inst{\ref{aff33}}
\and F.~Grupp\inst{\ref{aff34},\ref{aff10}}
\and S.~V.~H.~Haugan\orcid{0000-0001-9648-7260}\inst{\ref{aff73}}
\and J.~Hoar\inst{\ref{aff18}}
\and W.~Holmes\inst{\ref{aff74}}
\and I.~M.~Hook\orcid{0000-0002-2960-978X}\inst{\ref{aff75}}
\and F.~Hormuth\inst{\ref{aff76}}
\and A.~Hornstrup\orcid{0000-0002-3363-0936}\inst{\ref{aff77},\ref{aff78}}
\and P.~Hudelot\inst{\ref{aff5}}
\and K.~Jahnke\orcid{0000-0003-3804-2137}\inst{\ref{aff79}}
\and M.~Jhabvala\inst{\ref{aff80}}
\and B.~Joachimi\orcid{0000-0001-7494-1303}\inst{\ref{aff81}}
\and E.~Keih\"anen\orcid{0000-0003-1804-7715}\inst{\ref{aff82}}
\and S.~Kermiche\orcid{0000-0002-0302-5735}\inst{\ref{aff69}}
\and M.~Kilbinger\orcid{0000-0001-9513-7138}\inst{\ref{aff7}}
\and B.~Kubik\orcid{0009-0006-5823-4880}\inst{\ref{aff61}}
\and K.~Kuijken\orcid{0000-0002-3827-0175}\inst{\ref{aff9}}
\and M.~Kunz\orcid{0000-0002-3052-7394}\inst{\ref{aff83}}
\and H.~Kurki-Suonio\orcid{0000-0002-4618-3063}\inst{\ref{aff84},\ref{aff85}}
\and R.~Laureijs\inst{\ref{aff86}}
\and A.~M.~C.~Le~Brun\orcid{0000-0002-0936-4594}\inst{\ref{aff87}}
\and D.~Le~Mignant\orcid{0000-0002-5339-5515}\inst{\ref{aff4}}
\and S.~Ligori\orcid{0000-0003-4172-4606}\inst{\ref{aff44}}
\and P.~B.~Lilje\orcid{0000-0003-4324-7794}\inst{\ref{aff73}}
\and V.~Lindholm\orcid{0000-0003-2317-5471}\inst{\ref{aff84},\ref{aff85}}
\and I.~Lloro\orcid{0000-0001-5966-1434}\inst{\ref{aff88}}
\and G.~Mainetti\orcid{0000-0003-2384-2377}\inst{\ref{aff89}}
\and D.~Maino\inst{\ref{aff90},\ref{aff48},\ref{aff91}}
\and E.~Maiorano\orcid{0000-0003-2593-4355}\inst{\ref{aff21}}
\and O.~Mansutti\orcid{0000-0001-5758-4658}\inst{\ref{aff28}}
\and S.~Marcin\inst{\ref{aff92}}
\and O.~Marggraf\orcid{0000-0001-7242-3852}\inst{\ref{aff2}}
\and M.~Martinelli\orcid{0000-0002-6943-7732}\inst{\ref{aff49},\ref{aff50}}
\and N.~Martinet\orcid{0000-0003-2786-7790}\inst{\ref{aff4}}
\and F.~Marulli\orcid{0000-0002-8850-0303}\inst{\ref{aff93},\ref{aff21},\ref{aff32}}
\and R.~J.~Massey\orcid{0000-0002-6085-3780}\inst{\ref{aff94}}
\and S.~Maurogordato\inst{\ref{aff95}}
\and E.~Medinaceli\orcid{0000-0002-4040-7783}\inst{\ref{aff21}}
\and S.~Mei\orcid{0000-0002-2849-559X}\inst{\ref{aff96},\ref{aff97}}
\and Y.~Mellier\inst{\ref{aff98},\ref{aff5}}
\and M.~Meneghetti\orcid{0000-0003-1225-7084}\inst{\ref{aff21},\ref{aff32}}
\and E.~Merlin\orcid{0000-0001-6870-8900}\inst{\ref{aff49}}
\and G.~Meylan\inst{\ref{aff99}}
\and J.~J.~Mohr\orcid{0000-0002-6875-2087}\inst{\ref{aff100}}
\and A.~Mora\orcid{0000-0002-1922-8529}\inst{\ref{aff101}}
\and M.~Moresco\orcid{0000-0002-7616-7136}\inst{\ref{aff93},\ref{aff21}}
\and L.~Moscardini\orcid{0000-0002-3473-6716}\inst{\ref{aff93},\ref{aff21},\ref{aff32}}
\and R.~Nakajima\orcid{0009-0009-1213-7040}\inst{\ref{aff2}}
\and C.~Neissner\orcid{0000-0001-8524-4968}\inst{\ref{aff102},\ref{aff52}}
\and R.~C.~Nichol\orcid{0000-0003-0939-6518}\inst{\ref{aff24}}
\and S.-M.~Niemi\orcid{0009-0005-0247-0086}\inst{\ref{aff45}}
\and C.~Padilla\orcid{0000-0001-7951-0166}\inst{\ref{aff102}}
\and F.~Pasian\orcid{0000-0002-4869-3227}\inst{\ref{aff28}}
\and K.~Pedersen\inst{\ref{aff103}}
\and W.~J.~Percival\orcid{0000-0002-0644-5727}\inst{\ref{aff104},\ref{aff105},\ref{aff106}}
\and V.~Pettorino\inst{\ref{aff45}}
\and S.~Pires\orcid{0000-0002-0249-2104}\inst{\ref{aff7}}
\and G.~Polenta\orcid{0000-0003-4067-9196}\inst{\ref{aff71}}
\and M.~Poncet\inst{\ref{aff107}}
\and L.~A.~Popa\inst{\ref{aff108}}
\and L.~Pozzetti\orcid{0000-0001-7085-0412}\inst{\ref{aff21}}
\and F.~Raison\orcid{0000-0002-7819-6918}\inst{\ref{aff34}}
\and A.~Renzi\orcid{0000-0001-9856-1970}\inst{\ref{aff109},\ref{aff68}}
\and J.~Rhodes\orcid{0000-0002-4485-8549}\inst{\ref{aff74}}
\and G.~Riccio\inst{\ref{aff38}}
\and E.~Romelli\orcid{0000-0003-3069-9222}\inst{\ref{aff28}}
\and M.~Roncarelli\orcid{0000-0001-9587-7822}\inst{\ref{aff21}}
\and C.~Rosset\orcid{0000-0003-0286-2192}\inst{\ref{aff96}}
\and R.~Saglia\orcid{0000-0003-0378-7032}\inst{\ref{aff10},\ref{aff34}}
\and Z.~Sakr\orcid{0000-0002-4823-3757}\inst{\ref{aff25},\ref{aff110},\ref{aff111}}
\and D.~Sapone\orcid{0000-0001-7089-4503}\inst{\ref{aff112}}
\and B.~Sartoris\orcid{0000-0003-1337-5269}\inst{\ref{aff10},\ref{aff28}}
\and M.~Schirmer\orcid{0000-0003-2568-9994}\inst{\ref{aff79}}
\and P.~Schneider\orcid{0000-0001-8561-2679}\inst{\ref{aff2}}
\and A.~Secroun\orcid{0000-0003-0505-3710}\inst{\ref{aff69}}
\and G.~Seidel\orcid{0000-0003-2907-353X}\inst{\ref{aff79}}
\and M.~Seiffert\orcid{0000-0002-7536-9393}\inst{\ref{aff74}}
\and S.~Serrano\orcid{0000-0002-0211-2861}\inst{\ref{aff54},\ref{aff113},\ref{aff19}}
\and C.~Sirignano\orcid{0000-0002-0995-7146}\inst{\ref{aff109},\ref{aff68}}
\and G.~Sirri\orcid{0000-0003-2626-2853}\inst{\ref{aff32}}
\and A.~Spurio~Mancini\orcid{0000-0001-5698-0990}\inst{\ref{aff114}}
\and L.~Stanco\orcid{0000-0002-9706-5104}\inst{\ref{aff68}}
\and J.~Steinwagner\orcid{0000-0001-7443-1047}\inst{\ref{aff34}}
\and P.~Tallada-Cresp\'{i}\orcid{0000-0002-1336-8328}\inst{\ref{aff51},\ref{aff52}}
\and I.~Tereno\orcid{0000-0002-4537-6218}\inst{\ref{aff65},\ref{aff115}}
\and N.~Tessore\orcid{0000-0002-9696-7931}\inst{\ref{aff81}}
\and R.~Toledo-Moreo\orcid{0000-0002-2997-4859}\inst{\ref{aff116}}
\and F.~Torradeflot\orcid{0000-0003-1160-1517}\inst{\ref{aff52},\ref{aff51}}
\and I.~Tutusaus\orcid{0000-0002-3199-0399}\inst{\ref{aff110}}
\and E.~A.~Valentijn\inst{\ref{aff86}}
\and L.~Valenziano\orcid{0000-0002-1170-0104}\inst{\ref{aff21},\ref{aff72}}
\and J.~Valiviita\orcid{0000-0001-6225-3693}\inst{\ref{aff84},\ref{aff85}}
\and T.~Vassallo\orcid{0000-0001-6512-6358}\inst{\ref{aff10},\ref{aff28}}
\and G.~Verdoes~Kleijn\orcid{0000-0001-5803-2580}\inst{\ref{aff86}}
\and A.~Veropalumbo\orcid{0000-0003-2387-1194}\inst{\ref{aff26},\ref{aff36},\ref{aff35}}
\and Y.~Wang\orcid{0000-0002-4749-2984}\inst{\ref{aff117}}
\and J.~Weller\orcid{0000-0002-8282-2010}\inst{\ref{aff10},\ref{aff34}}
\and G.~Zamorani\orcid{0000-0002-2318-301X}\inst{\ref{aff21}}
\and F.~M.~Zerbi\inst{\ref{aff26}}
\and E.~Zucca\orcid{0000-0002-5845-8132}\inst{\ref{aff21}}
\and M.~Bolzonella\orcid{0000-0003-3278-4607}\inst{\ref{aff21}}
\and C.~Burigana\orcid{0000-0002-3005-5796}\inst{\ref{aff118},\ref{aff72}}
\and L.~Gabarra\orcid{0000-0002-8486-8856}\inst{\ref{aff119}}
\and J.~Mart\'{i}n-Fleitas\orcid{0000-0002-8594-569X}\inst{\ref{aff120}}
\and S.~Matthew\orcid{0000-0001-8448-1697}\inst{\ref{aff3}}
\and A.~Pezzotta\orcid{0000-0003-0726-2268}\inst{\ref{aff121},\ref{aff34}}
\and V.~Scottez\orcid{0009-0008-3864-940X}\inst{\ref{aff98},\ref{aff122}}
\and M.~Sereno\orcid{0000-0003-0302-0325}\inst{\ref{aff21},\ref{aff32}}
\and M.~Viel\orcid{0000-0002-2642-5707}\inst{\ref{aff27},\ref{aff28},\ref{aff30},\ref{aff29},\ref{aff123}}
\and D.~Scott\orcid{0000-0002-6878-9840}\inst{\ref{aff124}}}

\institute{Universit\"at Innsbruck, Institut f\"ur Astro- und Teilchenphysik, Technikerstr. 25/8, 6020 Innsbruck, Austria\label{aff1}
\and
Universit\"at Bonn, Argelander-Institut f\"ur Astronomie, Auf dem H\"ugel 71, 53121 Bonn, Germany\label{aff2}
\and
Institute for Astronomy, University of Edinburgh, Royal Observatory, Blackford Hill, Edinburgh EH9 3HJ, UK\label{aff3}
\and
Aix-Marseille Universit\'e, CNRS, CNES, LAM, Marseille, France\label{aff4}
\and
Institut d'Astrophysique de Paris, UMR 7095, CNRS, and Sorbonne Universit\'e, 98 bis boulevard Arago, 75014 Paris, France\label{aff5}
\and
Department of Astronomy, University of Geneva, ch. d'Ecogia 16, 1290 Versoix, Switzerland\label{aff6}
\and
Universit\'e Paris-Saclay, Universit\'e Paris Cit\'e, CEA, CNRS, AIM, 91191, Gif-sur-Yvette, France\label{aff7}
\and
Instituto de F\'isica de Cantabria, Edificio Juan Jord\'a, Avenida de los Castros, 39005 Santander, Spain\label{aff8}
\and
Leiden Observatory, Leiden University, Einsteinweg 55, 2333 CC Leiden, The Netherlands\label{aff9}
\and
Universit\"ats-Sternwarte M\"unchen, Fakult\"at f\"ur Physik, Ludwig-Maximilians-Universit\"at M\"unchen, Scheinerstrasse 1, 81679 M\"unchen, Germany\label{aff10}
\and
Kobayashi-Maskawa Institute for the Origin of Particles and the Universe, Nagoya University, Chikusa-ku, Nagoya, 464-8602, Japan\label{aff11}
\and
Institute for Advanced Research, Nagoya University, Chikusa-ku, Nagoya, 464-8601, Japan\label{aff12}
\and
Kavli Institute for the Physics and Mathematics of the Universe (WPI), University of Tokyo, Kashiwa, Chiba 277-8583, Japan\label{aff13}
\and
Physics Program, Graduate School of Advanced Science and Engineering, Hiroshima University, 1-3-1 Kagamiyama, Higashi-Hiroshima, Hiroshima 739-8526, Japan\label{aff14}
\and
Hiroshima Astrophysical Science Center, Hiroshima University, 1-3-1 Kagamiyama, Higashi-Hiroshima, Hiroshima 739-8526, Japan\label{aff15}
\and
Core Research for Energetic Universe, Hiroshima University, 1-3-1, Kagamiyama, Higashi-Hiroshima, Hiroshima 739-8526, Japan\label{aff16}
\and
Department of Astronomy, University of Massachusetts, Amherst, MA 01003, USA\label{aff17}
\and
ESAC/ESA, Camino Bajo del Castillo, s/n., Urb. Villafranca del Castillo, 28692 Villanueva de la Ca\~nada, Madrid, Spain\label{aff18}
\and
Institute of Space Sciences (ICE, CSIC), Campus UAB, Carrer de Can Magrans, s/n, 08193 Barcelona, Spain\label{aff19}
\and
Dipartimento di Fisica e Scienze della Terra, Universit\`a degli Studi di Ferrara, Via Giuseppe Saragat 1, 44122 Ferrara, Italy\label{aff20}
\and
INAF-Osservatorio di Astrofisica e Scienza dello Spazio di Bologna, Via Piero Gobetti 93/3, 40129 Bologna, Italy\label{aff21}
\and
Cosmic Dawn Center (DAWN)\label{aff22}
\and
Niels Bohr Institute, University of Copenhagen, Jagtvej 128, 2200 Copenhagen, Denmark\label{aff23}
\and
School of Mathematics and Physics, University of Surrey, Guildford, Surrey, GU2 7XH, UK\label{aff24}
\and
Institut f\"ur Theoretische Physik, University of Heidelberg, Philosophenweg 16, 69120 Heidelberg, Germany\label{aff25}
\and
INAF-Osservatorio Astronomico di Brera, Via Brera 28, 20122 Milano, Italy\label{aff26}
\and
IFPU, Institute for Fundamental Physics of the Universe, via Beirut 2, 34151 Trieste, Italy\label{aff27}
\and
INAF-Osservatorio Astronomico di Trieste, Via G. B. Tiepolo 11, 34143 Trieste, Italy\label{aff28}
\and
INFN, Sezione di Trieste, Via Valerio 2, 34127 Trieste TS, Italy\label{aff29}
\and
SISSA, International School for Advanced Studies, Via Bonomea 265, 34136 Trieste TS, Italy\label{aff30}
\and
Dipartimento di Fisica e Astronomia, Universit\`a di Bologna, Via Gobetti 93/2, 40129 Bologna, Italy\label{aff31}
\and
INFN-Sezione di Bologna, Viale Berti Pichat 6/2, 40127 Bologna, Italy\label{aff32}
\and
INAF-Osservatorio Astronomico di Padova, Via dell'Osservatorio 5, 35122 Padova, Italy\label{aff33}
\and
Max Planck Institute for Extraterrestrial Physics, Giessenbachstr. 1, 85748 Garching, Germany\label{aff34}
\and
Dipartimento di Fisica, Universit\`a di Genova, Via Dodecaneso 33, 16146, Genova, Italy\label{aff35}
\and
INFN-Sezione di Genova, Via Dodecaneso 33, 16146, Genova, Italy\label{aff36}
\and
Department of Physics "E. Pancini", University Federico II, Via Cinthia 6, 80126, Napoli, Italy\label{aff37}
\and
INAF-Osservatorio Astronomico di Capodimonte, Via Moiariello 16, 80131 Napoli, Italy\label{aff38}
\and
Instituto de Astrof\'isica e Ci\^encias do Espa\c{c}o, Universidade do Porto, CAUP, Rua das Estrelas, PT4150-762 Porto, Portugal\label{aff39}
\and
Faculdade de Ci\^encias da Universidade do Porto, Rua do Campo de Alegre, 4150-007 Porto, Portugal\label{aff40}
\and
European Southern Observatory, Karl-Schwarzschild-Str.~2, 85748 Garching, Germany\label{aff41}
\and
Dipartimento di Fisica, Universit\`a degli Studi di Torino, Via P. Giuria 1, 10125 Torino, Italy\label{aff42}
\and
INFN-Sezione di Torino, Via P. Giuria 1, 10125 Torino, Italy\label{aff43}
\and
INAF-Osservatorio Astrofisico di Torino, Via Osservatorio 20, 10025 Pino Torinese (TO), Italy\label{aff44}
\and
European Space Agency/ESTEC, Keplerlaan 1, 2201 AZ Noordwijk, The Netherlands\label{aff45}
\and
Institute Lorentz, Leiden University, Niels Bohrweg 2, 2333 CA Leiden, The Netherlands\label{aff46}
\and
Mullard Space Science Laboratory, University College London, Holmbury St Mary, Dorking, Surrey RH5 6NT, UK\label{aff47}
\and
INAF-IASF Milano, Via Alfonso Corti 12, 20133 Milano, Italy\label{aff48}
\and
INAF-Osservatorio Astronomico di Roma, Via Frascati 33, 00078 Monteporzio Catone, Italy\label{aff49}
\and
INFN-Sezione di Roma, Piazzale Aldo Moro, 2 - c/o Dipartimento di Fisica, Edificio G. Marconi, 00185 Roma, Italy\label{aff50}
\and
Centro de Investigaciones Energ\'eticas, Medioambientales y Tecnol\'ogicas (CIEMAT), Avenida Complutense 40, 28040 Madrid, Spain\label{aff51}
\and
Port d'Informaci\'{o} Cient\'{i}fica, Campus UAB, C. Albareda s/n, 08193 Bellaterra (Barcelona), Spain\label{aff52}
\and
Institute for Theoretical Particle Physics and Cosmology (TTK), RWTH Aachen University, 52056 Aachen, Germany\label{aff53}
\and
Institut d'Estudis Espacials de Catalunya (IEEC),  Edifici RDIT, Campus UPC, 08860 Castelldefels, Barcelona, Spain\label{aff54}
\and
INFN section of Naples, Via Cinthia 6, 80126, Napoli, Italy\label{aff55}
\and
Institute for Astronomy, University of Hawaii, 2680 Woodlawn Drive, Honolulu, HI 96822, USA\label{aff56}
\and
Dipartimento di Fisica e Astronomia "Augusto Righi" - Alma Mater Studiorum Universit\`a di Bologna, Viale Berti Pichat 6/2, 40127 Bologna, Italy\label{aff57}
\and
Instituto de Astrof\'{\i}sica de Canarias, V\'{\i}a L\'actea, 38205 La Laguna, Tenerife, Spain\label{aff58}
\and
Jodrell Bank Centre for Astrophysics, Department of Physics and Astronomy, University of Manchester, Oxford Road, Manchester M13 9PL, UK\label{aff59}
\and
European Space Agency/ESRIN, Largo Galileo Galilei 1, 00044 Frascati, Roma, Italy\label{aff60}
\and
Universit\'e Claude Bernard Lyon 1, CNRS/IN2P3, IP2I Lyon, UMR 5822, Villeurbanne, F-69100, France\label{aff61}
\and
Institut de Ci\`{e}ncies del Cosmos (ICCUB), Universitat de Barcelona (IEEC-UB), Mart\'{i} i Franqu\`{e}s 1, 08028 Barcelona, Spain\label{aff62}
\and
Instituci\'o Catalana de Recerca i Estudis Avan\c{c}ats (ICREA), Passeig de Llu\'{\i}s Companys 23, 08010 Barcelona, Spain\label{aff63}
\and
UCB Lyon 1, CNRS/IN2P3, IUF, IP2I Lyon, 4 rue Enrico Fermi, 69622 Villeurbanne, France\label{aff64}
\and
Departamento de F\'isica, Faculdade de Ci\^encias, Universidade de Lisboa, Edif\'icio C8, Campo Grande, PT1749-016 Lisboa, Portugal\label{aff65}
\and
Instituto de Astrof\'isica e Ci\^encias do Espa\c{c}o, Faculdade de Ci\^encias, Universidade de Lisboa, Campo Grande, 1749-016 Lisboa, Portugal\label{aff66}
\and
Universit\'e Paris-Saclay, CNRS, Institut d'astrophysique spatiale, 91405, Orsay, France\label{aff67}
\and
INFN-Padova, Via Marzolo 8, 35131 Padova, Italy\label{aff68}
\and
Aix-Marseille Universit\'e, CNRS/IN2P3, CPPM, Marseille, France\label{aff69}
\and
INAF-Istituto di Astrofisica e Planetologia Spaziali, via del Fosso del Cavaliere, 100, 00100 Roma, Italy\label{aff70}
\and
Space Science Data Center, Italian Space Agency, via del Politecnico snc, 00133 Roma, Italy\label{aff71}
\and
INFN-Bologna, Via Irnerio 46, 40126 Bologna, Italy\label{aff72}
\and
Institute of Theoretical Astrophysics, University of Oslo, P.O. Box 1029 Blindern, 0315 Oslo, Norway\label{aff73}
\and
Jet Propulsion Laboratory, California Institute of Technology, 4800 Oak Grove Drive, Pasadena, CA, 91109, USA\label{aff74}
\and
Department of Physics, Lancaster University, Lancaster, LA1 4YB, UK\label{aff75}
\and
Felix Hormuth Engineering, Goethestr. 17, 69181 Leimen, Germany\label{aff76}
\and
Technical University of Denmark, Elektrovej 327, 2800 Kgs. Lyngby, Denmark\label{aff77}
\and
Cosmic Dawn Center (DAWN), Denmark\label{aff78}
\and
Max-Planck-Institut f\"ur Astronomie, K\"onigstuhl 17, 69117 Heidelberg, Germany\label{aff79}
\and
NASA Goddard Space Flight Center, Greenbelt, MD 20771, USA\label{aff80}
\and
Department of Physics and Astronomy, University College London, Gower Street, London WC1E 6BT, UK\label{aff81}
\and
Department of Physics and Helsinki Institute of Physics, Gustaf H\"allstr\"omin katu 2, 00014 University of Helsinki, Finland\label{aff82}
\and
Universit\'e de Gen\`eve, D\'epartement de Physique Th\'eorique and Centre for Astroparticle Physics, 24 quai Ernest-Ansermet, CH-1211 Gen\`eve 4, Switzerland\label{aff83}
\and
Department of Physics, P.O. Box 64, 00014 University of Helsinki, Finland\label{aff84}
\and
Helsinki Institute of Physics, Gustaf H{\"a}llstr{\"o}min katu 2, University of Helsinki, Helsinki, Finland\label{aff85}
\and
Kapteyn Astronomical Institute, University of Groningen, PO Box 800, 9700 AV Groningen, The Netherlands\label{aff86}
\and
Laboratoire d'etude de l'Univers et des phenomenes eXtremes, Observatoire de Paris, Universit\'e PSL, Sorbonne Universit\'e, CNRS, 92190 Meudon, France\label{aff87}
\and
SKA Observatory, Jodrell Bank, Lower Withington, Macclesfield, Cheshire SK11 9FT, UK\label{aff88}
\and
Centre de Calcul de l'IN2P3/CNRS, 21 avenue Pierre de Coubertin 69627 Villeurbanne Cedex, France\label{aff89}
\and
Dipartimento di Fisica "Aldo Pontremoli", Universit\`a degli Studi di Milano, Via Celoria 16, 20133 Milano, Italy\label{aff90}
\and
INFN-Sezione di Milano, Via Celoria 16, 20133 Milano, Italy\label{aff91}
\and
University of Applied Sciences and Arts of Northwestern Switzerland, School of Computer Science, 5210 Windisch, Switzerland\label{aff92}
\and
Dipartimento di Fisica e Astronomia "Augusto Righi" - Alma Mater Studiorum Universit\`a di Bologna, via Piero Gobetti 93/2, 40129 Bologna, Italy\label{aff93}
\and
Department of Physics, Institute for Computational Cosmology, Durham University, South Road, Durham, DH1 3LE, UK\label{aff94}
\and
Universit\'e C\^{o}te d'Azur, Observatoire de la C\^{o}te d'Azur, CNRS, Laboratoire Lagrange, Bd de l'Observatoire, CS 34229, 06304 Nice cedex 4, France\label{aff95}
\and
Universit\'e Paris Cit\'e, CNRS, Astroparticule et Cosmologie, 75013 Paris, France\label{aff96}
\and
CNRS-UCB International Research Laboratory, Centre Pierre Bin\'etruy, IRL2007, CPB-IN2P3, Berkeley, USA\label{aff97}
\and
Institut d'Astrophysique de Paris, 98bis Boulevard Arago, 75014, Paris, France\label{aff98}
\and
Institute of Physics, Laboratory of Astrophysics, Ecole Polytechnique F\'ed\'erale de Lausanne (EPFL), Observatoire de Sauverny, 1290 Versoix, Switzerland\label{aff99}
\and
University Observatory, LMU Faculty of Physics, Scheinerstrasse 1, 81679 Munich, Germany\label{aff100}
\and
Telespazio UK S.L. for European Space Agency (ESA), Camino bajo del Castillo, s/n, Urbanizacion Villafranca del Castillo, Villanueva de la Ca\~nada, 28692 Madrid, Spain\label{aff101}
\and
Institut de F\'{i}sica d'Altes Energies (IFAE), The Barcelona Institute of Science and Technology, Campus UAB, 08193 Bellaterra (Barcelona), Spain\label{aff102}
\and
DARK, Niels Bohr Institute, University of Copenhagen, Jagtvej 155, 2200 Copenhagen, Denmark\label{aff103}
\and
Waterloo Centre for Astrophysics, University of Waterloo, Waterloo, Ontario N2L 3G1, Canada\label{aff104}
\and
Department of Physics and Astronomy, University of Waterloo, Waterloo, Ontario N2L 3G1, Canada\label{aff105}
\and
Perimeter Institute for Theoretical Physics, Waterloo, Ontario N2L 2Y5, Canada\label{aff106}
\and
Centre National d'Etudes Spatiales -- Centre spatial de Toulouse, 18 avenue Edouard Belin, 31401 Toulouse Cedex 9, France\label{aff107}
\and
Institute of Space Science, Str. Atomistilor, nr. 409 M\u{a}gurele, Ilfov, 077125, Romania\label{aff108}
\and
Dipartimento di Fisica e Astronomia "G. Galilei", Universit\`a di Padova, Via Marzolo 8, 35131 Padova, Italy\label{aff109}
\and
Institut de Recherche en Astrophysique et Plan\'etologie (IRAP), Universit\'e de Toulouse, CNRS, UPS, CNES, 14 Av. Edouard Belin, 31400 Toulouse, France\label{aff110}
\and
Universit\'e St Joseph; Faculty of Sciences, Beirut, Lebanon\label{aff111}
\and
Departamento de F\'isica, FCFM, Universidad de Chile, Blanco Encalada 2008, Santiago, Chile\label{aff112}
\and
Satlantis, University Science Park, Sede Bld 48940, Leioa-Bilbao, Spain\label{aff113}
\and
Department of Physics, Royal Holloway, University of London, TW20 0EX, UK\label{aff114}
\and
Instituto de Astrof\'isica e Ci\^encias do Espa\c{c}o, Faculdade de Ci\^encias, Universidade de Lisboa, Tapada da Ajuda, 1349-018 Lisboa, Portugal\label{aff115}
\and
Universidad Polit\'ecnica de Cartagena, Departamento de Electr\'onica y Tecnolog\'ia de Computadoras,  Plaza del Hospital 1, 30202 Cartagena, Spain\label{aff116}
\and
Infrared Processing and Analysis Center, California Institute of Technology, Pasadena, CA 91125, USA\label{aff117}
\and
INAF, Istituto di Radioastronomia, Via Piero Gobetti 101, 40129 Bologna, Italy\label{aff118}
\and
Department of Physics, Oxford University, Keble Road, Oxford OX1 3RH, UK\label{aff119}
\and
Aurora Technology for European Space Agency (ESA), Camino bajo del Castillo, s/n, Urbanizacion Villafranca del Castillo, Villanueva de la Ca\~nada, 28692 Madrid, Spain\label{aff120}
\and
INAF - Osservatorio Astronomico di Brera, via Emilio Bianchi 46, 23807 Merate, Italy\label{aff121}
\and
ICL, Junia, Universit\'e Catholique de Lille, LITL, 59000 Lille, France\label{aff122}
\and
ICSC - Centro Nazionale di Ricerca in High Performance Computing, Big Data e Quantum Computing, Via Magnanelli 2, Bologna, Italy\label{aff123}
\and
Department of Physics and Astronomy, University of British Columbia, Vancouver, BC V6T 1Z1, Canada\label{aff124}}

\date{Received 08 Jul 2025 / Accepted ...}

   \abstract{
   The \Euclid space telescope of the  European Space Agency (ESA) is designed to provide sensitive and accurate measurements of weak gravitational lensing distortions over wide areas on the sky.
   Here we present a weak gravitational lensing analysis of early \Euclid observations obtained for the field around the massive galaxy cluster Abell 2390 as part of the \Euclid Early Release Observations programme.
   We conduct
   galaxy shape measurements
   using three independent
   algorithms (\texttt{LensMC}, \texttt{KSB+}, and \texttt{SourceXtractor++}).
  Incorporating multi-band photometry from \Euclid and Subaru/Suprime-Cam,
  we estimate photometric  redshifts to preferentially select background sources from tomographic redshift bins,
  for which we calibrate the redshift distributions using the self-organising map approach and
  data from the Cosmic Evolution Survey (COSMOS).
  We quantify the residual cluster member contamination and correct for it in bins of photometric redshift and magnitude using their source density profiles, including corrections for source obscuration and magnification.
 We reconstruct the cluster mass distribution and jointly fit the
 tangential reduced shear profiles of the different
 tomographic bins
with spherical Navarro--Frenk--White profile
 predictions to constrain the cluster mass, finding consistent results for the three shape catalogues and good agreement with earlier measurements.
        As an important
 validation test       we compare
 these joint constraints to mass measurements obtained individually for the different tomographic bins,
finding good consistency.
    More detailed
    constraints on the cluster properties are presented in a companion paper
    that additionally incorporates strong lensing measurements.
    Our analysis provides a first demonstration of the outstanding capabilities of \Euclid for tomographic  weak lensing measurements.
   }

   \titlerunning{\Euclid\/: Weak lensing analysis of Abell 2390}

   \authorrunning{Schrabback, Congedo, Gavazzi, Hartley, Jansen, Kang, Kleinebreil et al.}

   \maketitle

\section{\label{sc:Intro}Introduction}
The primary objective of the European Space Agency's new space telescope {\Euclid} is to test cosmological models using  measurements of galaxy clustering and weak gravitational lensing \citep{EuclidSkyOverview}.
For this, {\Euclid} will observe approximately 14\,000 deg$^2$ of the extragalactic sky in the Euclid Wide Survey \citep[EWS,][]{scaramella22} using its visual Charge-Coupled Device (CCD) imager VIS \citep{EuclidSkyVIS} and
its near-infrared
instrument NISP \citep{EuclidSkyNISP}.
With their  fine pixel sampling (0\farcs1 pixel scale) and space-based resolution,
the VIS images will be used to measure the shapes of approximately 1.5 billion galaxies in order to
constrain weak lensing (WL) distortions caused
by the gravitational potential of foreground structures
\citep[for an introduction to WL see, e.g.,][]{bartelmann01}.
Typically these distortions are weak and change the axis ratios of galaxy images at the per-cent level only.
In this regime, which is typically referred to as `cosmic shear', cosmological parameters are  inferred
by measuring
correlations in galaxy ellipticities as a function of their separation, averaged over
large sky areas
\citep[e.g.,][]{hamana20,amon22,asgari21}.
However, WL data can also result in
competitive cosmological constraints \citep[see e.g.,][]{mantz15,bocquet19,bocquet24b,ghirardini24}
when they are used to calibrate the mass scale
\citep[e.g.,][]{schrabback21b,zohren22,chiu22,grandis24,kleinebreil25}
of galaxy cluster samples that have a cosmologically well modelled selection function \citep[e.g.,][]{bleem15,bleem20,bleem24,hilton21,bulbul24,aymerich24}.
Here, WL data break  degeneracies that exist between  parameters describing the cosmological model on the one hand and cluster mass-observable scaling relations on the other \citep[e.g.,][]{grandis19,bocquet24}.

Massive galaxy clusters create  WL distortions
that are
strong enough to
be
detected
 for a single target
if deep high-resolution imaging is employed, providing a high density of background  galaxies with WL shape measurements \citep[e.g.,][]{vonderlinden14,vonderlinden14b,hoekstra15,sereno17,herbonnet20,kim21}.
Such observations were taken by {\Euclid} for the extremely massive galaxy cluster Abell 2390 \citep[A2390 hereafter, see][]{abell89},
located at redshift \mbox{$z=0.228$} \citep{sohn20},
as part of the
\citet[][ERO]{EROcite}
programme
`Magnifying Lens'
\citep{EROLensData}.
These observations provide an excellent opportunity to showcase {\Euclid}'s outstanding capability to measure the WL signature of a massive galaxy cluster, which is the main goal of this paper.
Simultaneously, this paper
demonstrates some analysis approaches for tomographic \Euclid cluster WL studies that can be employed in future investigations of larger samples.

Initial WL constraints based the   \Euclid observations of A2390 were  reported in the wider overview paper by \citet{EROLensData}.
We significantly improve upon this analysis by incorporating two additional shape measurement methods, a source selection via tomographic redshift bins, improved calibrations, and a correction for cluster member contamination.
Other earlier WL studies of this cluster were limited to ground-based observations, including early work by \citet{squires96},
the `Weighing the Giants' project \citep[WtG,][]{vonderlinden14,applegate14}, the Local Cluster Substructure Survey \citep[LoCuSS,][]{okabe16},
the Canadian Cluster Comparison Project \citep[CCCP,][]{hoekstra15,herbonnet20}, and the recent analysis of WIYN-ODI data by \citet{dutta24}.
Of these, WtG and LoCuSS employed ground-based observations from  Subaru/Suprime-Cam, which we incorporate into our analysis
for the photometric source selection (see Sect.\thinspace\ref{sc:Data:ground_based_data}).

This paper is organised as follows. In Sect.\thinspace\ref{sc:Data} we describe the data used in our study, including the {\Euclid} observations and complementary archival ground-based data.
This is followed by Sect.\thinspace\ref{sc:photoz}, which details the computation of photometric redshifts and their calibration.
Section \ref{sc:Shapes} summarises our measurements of WL galaxy shapes, where we employ and compare three different shape measurement algorithms.
We quantify and account for cluster member contamination in Sect.\thinspace\ref{sc:cluster_member_contamination},
followed by the presentation of the WL mass constraints and reconstruction in Sect.\thinspace\ref{sc:wl_results}.
We discuss our results and compare them to previous WL measurements of the cluster in Sect.\thinspace\ref{sc:discussion}, followed by conclusions in Sect.\thinspace\ref{sc:conclusions}.

 Throughout our analysis we assume a standard flat $\Lambda$CDM cosmology characterised through parameters $\Omega_\Lambda=0.7$, $\Omega_\mathrm{m}=0.3$ and
  $H_0 = 70\,\mathrm{km} \,\mathrm{s}^{-1} \,\mathrm{Mpc}^{-1}$.
 For the computation of WL noise caused by large-scale structure projections (see  Sect.\thinspace\ref{sec:WLconstraintsanduncertainties}), we additionally assume  \mbox{$\sigma_8=0.8$}, \mbox{$\Omega_\mathrm{b}=0.046$}, and \mbox{$n_\mathrm{s}=0.96$}.
 All magnitudes given in this paper are in the AB system.

\section{\label{sc:Data}Data}

\subsection{\label{sc:Data:Euclid}{\Euclid} observations}
{\Euclid}'s A2390 observations were obtained on 28 November
2023
during  {\Euclid}'s performance verification (PV) phase
as part of the
ERO
programme \citep{EROData}.
They consist of three dithered  {\Euclid} Reference Observing Sequences \citep[ROS; see][]{scaramella22,EuclidSkyOverview} of 70.2\thinspace min each.
 To fill detector gaps each ROS contains four dither positions.
At each dither position   a 566$\thinspace$s
exposure was taken with VIS in its
broad optical band-pass
\citep[approximately $540$--$920$ nm, referred to as $\IE$, see][]{EuclidSkyVIS}, while simultaneously a 574$\thinspace$s
spectroscopic exposure was  obtained with NISP.
These were followed by NISP images in the
\YE, \JE, and \HE
filters \citep{EuclidSkyNISP}, each with an exposure time of 112$\thinspace$s.
For the ERO observations, an additional short (95$\thinspace$s)
VIS exposure was taken during each
\YE
exposure, leading to a total integration time of
7932$\thinspace$s
for A2390 with VIS.

As detailed in  \citet{EROData},
stacks were created for each filter using the ERO reduction pipeline and the \texttt{AstrOmatic SWarp} software \citep{bertin02} at the native pixel scales of the corresponding instruments (0\farcs1 for VIS and 0\farcs3 for NISP).
In particular, there are two  flavours of stacks, where the `Flattened' version employs a 64 pixel mesh and $3\times$ smoothing factor to model and subtract the background \citep{EROData}.
This approach is optimised for the
photometry
of faint and compact objects
and therefore employed for the computation of multi-band photometry
(see Sect.\thinspace\ref{sc:Data:photometric_catalog}).
An alternative set of stacks is optimised for the analysis of low surface brightness (LSB) sources (`LSB' version) and therefore does not apply a background subtraction.
For initial tests we conducted  WL shape measurements (see
Sect.\thinspace\ref{sc:Shapes}) on both versions of the stacks.
Given that we found only minimal differences, we conduced the main analysis using the `Flattened' version, to be consistent with the photometric analysis.
Further details on the A2390 ERO data are provided in \citet{EROLensData}, including estimates
of the $5\sigma$ limiting magnitudes of the stacked images, which amount to
\mbox{$\IE=27.01$} for the VIS stack  (assuming apertures with diameter 0\farcs3) and
\mbox{$\YE=25.18$},  \mbox{$\JE=25.22$}, and \mbox{$\HE=25.12$}
for the NISP stacks  (all assuming apertures with diameter 0\farcs6).

\subsection{\label{sc:Data:ground_based_data}Earlier  ground-based observations}

Archival ground-based multi-band imaging data are available for the A2390 field thanks to earlier programmes studying the WL signature of the cluster  (see Sect.\thinspace\ref{sc:Intro}).
While we expect the {\Euclid} VIS images to be superior for the measurement of galaxy shapes, given their outstanding resolution, the inclusion of
multi-band ground-based data
is still important
for the photometric selection of background sources.
From this we incorporate existing
$(B,V,R_\mathrm{c},i,I_\mathrm{c},z^\prime)$ imaging obtained with the Suprime-Cam instrument on the 8.2$\thinspace$m Subaru telescope \citep{miyazaki02}
and also consider Canada-France-Hawaii Telescope (CFHT) Megacam $u$-band imaging taken with the first generation MP9301 $u$-filter.
These data were previously employed by the WtG project \citep{vonderlinden14} and in part by the LoCuSS project \citep{okabe16}.

In our analysis we make use of a custom reduction of the Suprime-Cam data using the \texttt{SDFRED} pipeline \citep{yagi02,ouchi04},
 processed at the Suprime-Cam Legacy Archive at the Canadian Astronomy Data Centre \citep{gwyn20}. Likewise, CFHT Megacam $u$-band images were reduced with the {\tt Elixir} pipeline\footnote{\url{www.cfht.hawaii.edu/Instruments/Elixir}} \citep{magnier2002,magnier2004}.
All these frames were jointly astrometrically registered using {\tt scamp} \citep{bertin02} and \textit{Gaia}-DR3 as a
 reference sample,
yielding a typical 10--20 mas absolute astrometric accuracy in each of the RA and Dec directions.
We created models of the spatially varying point-spread function (PSF) for all of the ground-based data bands  using {\tt PSFEx} \citep{bertin11},  facilitating
PSF
photometry on all stars
that properly accounts for seeing variations  and eases photometric calibration.
The latter was conducted on individual frames using
photometric reference catalogues from the Sloan Digital Sky Survey \citep{ahumada2020} for the $u$ band
and from the Pan-STARRS $3\pi$ survey \citep{chambers2016} for all  Suprime-Cam bands.
The colour terms are more uncertain for the Johnson $B,V$ and Cousins $R_\mathrm{c},I_\mathrm{c}$ filters compared to the Sloan $i$ and $z$ bands,
but overall a very uniform photometric calibration is reached  \citep{gwyn20}. Before exposure stacking, we used the {\tt MaxiMask} \citep{paillassa20} tool to flag cosmic rays, hot pixels, satellite trails, bad columns, saturation bleeds near bright stars, and other defects. Taking those flags into account in the weighting scheme, exposures were then stacked using {\tt swarp} \citep{bertin02} on a $0\farcs19$ pixel scale, common to all ground-based filters. At this stage, no attempt to correct for Galactic extinctions was made.

\begin{figure}[htbp!]
\centering
\includegraphics[angle=0,width=1.0\hsize]{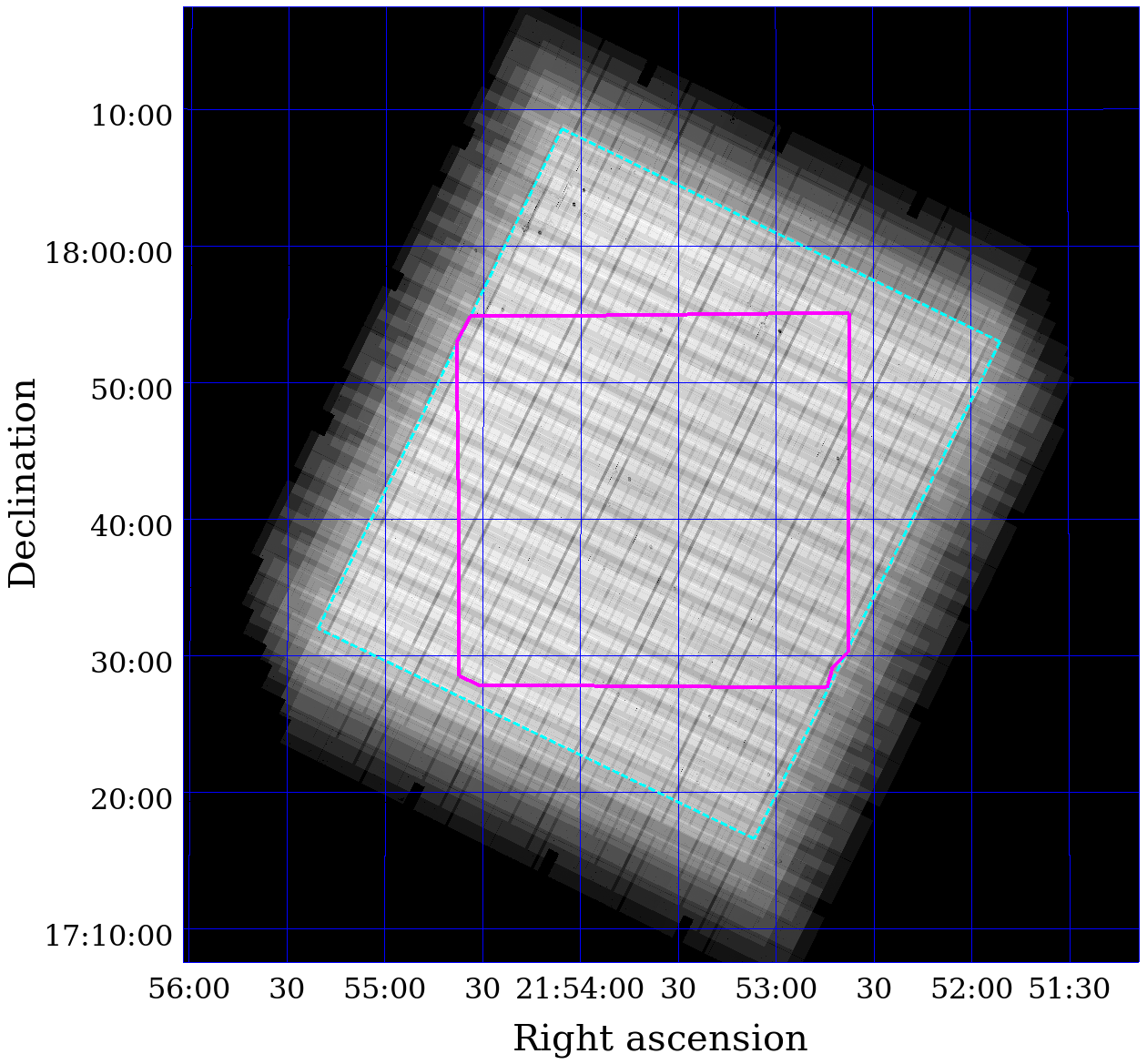}
\caption{Field coverage. The grey-scale shows the  weight image of the VIS image stack on a linear scale. The magenta solid polygon indicates the main region of interest for this WL analysis, where both the {\Euclid} image stacks have their greatest depth and multi-band ground-based observations are available. The cyan dashed polygon indicates the full-depth VIS area employed in the source injection analysis (see Sect.\thinspace\ref{sc:obscuration_correction}).}
\label{fig:weight_layout}
\end{figure}

Most of the $28^\prime\times 34^\prime$ area covered by the Suprime-Cam images
overlaps with the full-depth area of the  {\Euclid} VIS stack
(see Fig.\thinspace\ref{fig:weight_layout}). We regard this overlap area, which is also fully encompassed by the Megacam $u$-band image, as the primary region of interest for our WL study.
This is both due to the full multi-wavelength coverage and the fact that this area already provides complete azimuthal coverage out to a projected radius of 3.07$\thinspace$Mpc
from the cluster centre (at the reference cosmology),
well beyond the expected virial radius of the cluster.
In contrast, measurements at significantly larger radii would probe the WL signal  in the regime of the two-halo term, which is more difficult to model accurately and therefore typically excluded in cluster scaling relation and cosmology analyses \citep[e.g.,][]{dietrich19,grandis21}.

\subsection{\label{sc:Data:zirrus}Removal of foreground Galactic cirrus}
A2390 resides at relatively low Galactic latitude ($b=-27.813^\circ$) leading to prominent foreground emission due to dust cirrus.
Ellien et al. (in prep.), who  use the same {\Euclid} ERO data to investigate the low-surface-brightness intra-cluster light distribution of A2390,
describe the properties of the
cirrus emission in more detail.
Here we wish to ensure minimal impact of the cirrus emission on the WL analysis and therefore
subtract it off via an  advanced background model.

This is achieved using the
\texttt{DeNeb}
tool, a new deep-learning software package designed to perform single-channel source separation on astronomical images (Bertin et al., in prep.).
\texttt{DeNeb}
was trained on a large gallery of labelled images from various origins to perform subtraction of extended features comprised of reflection haloes from bright stars, residual flat-fielding and fringing patterns, and diffuse emission and reflection from Galactic dust,
while preserving stellar and galaxy images.

All the stacks, either {\Euclid} or ground-based, were independently processed with the default
\texttt{DeNeb}
tool.
Due to the lack of network training with $u$-band data, the method resulted in a slightly poorer removal of diffuse extended components in that filter, with an occasional removal of parts of very extended foreground galaxies. These are, however, irrelevant for our scientific goals.
On the contrary, for all other filters the subtraction was very effective.
For the remainder of the analysis we will work exclusively with `denebulised' images.
These provide the major advantage of a much flatter background, which leads to a more robust object detection and deblending.

Notably,
the patchy foreground emission
caused by the scattering
of starlight off Galactic dust also comes with equally complex
extinction variations.
These are not corrected via the procedure applied here and remain a possible concern for photometric redshift estimation (see Sect.\thinspace\ref{sc:photoz}) .

\subsection{\label{sc:Data:photometric_catalog}Object detection and \texttt{SourceXtractor++} measurements}

Here we employ the photometric catalogue first presented in \citet{EROLensData}. This catalogue was generated using \texttt{SourceXtractor++} \citep[][henceforth \texttt{SE++}]{bertin22,kuemmel22}, a  recent re-implementation of \texttt{SExtractor} \citep{bertin96}. We ran \texttt{SE++} in two settings, but source detection was always performed in the VIS $\IE$ band. In the first run, only the VIS image was used to constrain a single S\'ersic profile. This provides shapes that can readily be used for WL,
and best-fit sizes provide a complementary star/galaxy discriminator, since the best effective radius for unsaturated stars is consistently small. The WL
exploitation of the single S\'ersic models is further described in Sect.\thinspace\ref{sc:Shapes:SourceExtractor}.

For the photometric catalogue, objects are detected on the VIS $\IE$ stack, followed by a joint fit
of the {\Euclid} and ground-based images using a
 two-component galaxy model with a de Vaucouleurs-profile bulge and an
exponential disc.
This fit assumed identical bulge and disc orientations, but allowed for free axis ratios. Furthermore, the half-light radii of the bulge and disc were modelled to be wavelength independent, with varying bulge-to-total flux ratios between bands.

{With version {\tt 0.19},
our analysis employs  the latest version of \texttt{SE++} available at the time of the data processing.
It groups neighbouring sources in order to jointly fit their surface brightness profiles, which reduces the impact of blending compared to traditional approaches such as PSF-homogenised aperture photometry.
Since the processing of these ERO data, the development of  \texttt{SE++} has continued, leading to further improvements that will be implemented for the analysis of future \Euclid data sets. In particular, a full
operational decoupling of the detection and model-fitting steps is expected to improve the performance in the case of objects that are fully separated in VIS, but still partially blended in the ground-based data.
However, for our analysis we expect this to be a minor issue, given the superb image quality of the ground-based data \citep{EROLensData}.
Furthermore, in the computation of photometric redshifts optical colours are only incorporated based on ground-based instruments (see Sect.\thinspace\ref{sc:photoz}), which further reduces the potential impact of any mismatches between the ground- and space-based data for the photometry.

More details about the \texttt{SE++} runs, including overall photometric accuracy and star/galaxy separation with single S\'ersic fits can be found in Appendix~\ref{appendix:SE++-details}. All such runs rely on a common model of the \Euclid PSF, which is described in the next subsection.

\subsection{\label{sc:PSFEXmodel}Point-spread function modelling}

In the context of these ERO observations,  PSF models of the different image stacks were obtained using \texttt{PSFEx} \citep{bertin11}, while a more advanced model of the {\Euclid} VIS PSF is being developed
for future WL analyses of larger samples \citep[see section\thinspace 7.6.4. in][]{EuclidSkyOverview}.
In order to limit the potential impact of  brighter-fatter  effects \citep[see e.g.,][]{guyonnet15},
we halved the \texttt{PSFEx} input values
for
detector saturation compared to their actual values. With this precaution, only stars with a photometric signal-to-noise ratio\footnote{Here, ${\rm S/N}_\mathrm{flux}$ is defined via \texttt{SExtractor} parameters as ${\rm S/N}_\mathrm{flux}={\tt FLUX\_AUTO}/{\tt FLUXERR\_AUTO}$.} $70\le {\rm S/N}_\mathrm{flux}\lesssim1500$ are  retained to build the model.
This is most relevant for VIS, which provides most of the morphological information. An image of the rendered VIS PSF is shown in Fig.~\ref{fig:vispsf}. The mean PSF full-width at half maximum (FWHM) is $0\farcs156$, while average PSF
ellipticities
($\epsilon^\mathrm{PSFEx}_1$, $\epsilon^\mathrm{PSFEx}_2$)
amount to
($-0.0152$, $0.0017$) in the stacked image pixel frame.
\texttt{PSFEx} models were rendered with a finer pixel scale of $0\farcs05$, $0\farcs15$, and $0\farcs095$ for VIS, NISP, and ground-based images, respectively, assuming a third-order polynomial to capture spatial variations across the whole focal plane.
We do not attempt to model the wavelength dependence of the VIS PSF for this single-target study, given its moderate accuracy requirements.
In the case of the VIS PSF, our model extends to $2\farcs3$ and only misses 3\% of the encircled energy when compared to Table 3 of \citet{EROData}.
Given that \texttt{SE++} normalises the PSF convolution kernel to unity, this should be taken into account when employing computed model magnitudes of point sources. In the case of NISP bands, the PSF model extends to $6\farcs9$ and encloses between
98\% and 98.5\%
of the total energy.
Modest additional corrections, taking into account the energy enclosed in very extended diffraction spikes, should be considered for stellar photometry \citep{EROData}.
In the case of ground-based data, our PSF models extend to a $5\farcs3$ radius
for all filters.
Given the exquisite seeing conditions ($\sim 0\farcs6$ FWHM) of the employed ground-based data
\citep{EROLensData},
corrections are  negligible for stellar photometry
in those ground-based filters.

\begin{figure}[htbp!]
\centering
\includegraphics[angle=0,width=\hsize]{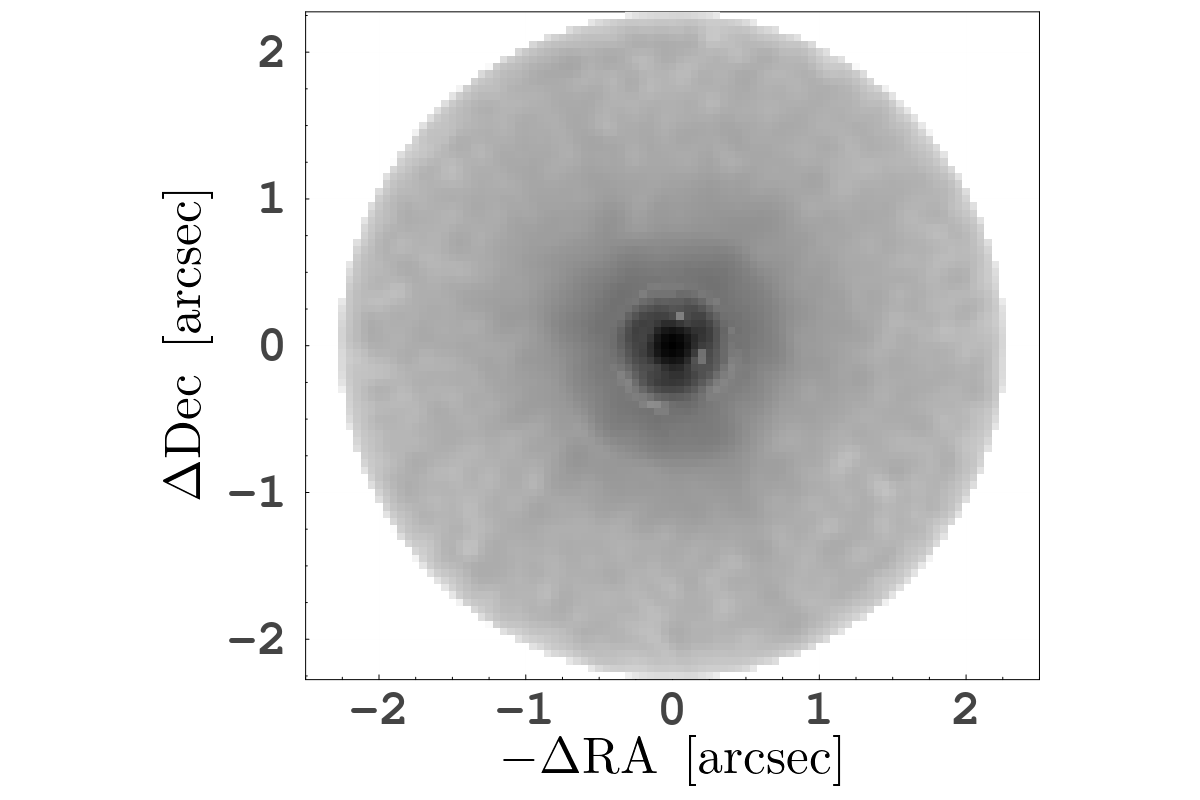}
\caption{VIS point-spread function recovered by \texttt{PSFEx} in the centre of the field of view. Pixel intensities are scaled on a logarithmic
  stretch. The image sampling is $0\farcs05$ per pixel.
  The negative pixels close to the centre are artefacts caused by the  oversampling, limited number of stars, and  regularisation scheme applied by \texttt{PSFEx} \citep[see][]{bertin11}.
  }
\label{fig:vispsf}
\end{figure}

\subsection{Masking}

\citet{EROLensData} describe the semi-automated generation of masks for the field, aiming to exclude spurious
detections and light contamination near the haloes or diffraction spikes of stars, as well as stellar ghost images.
We apply these masks to all of our catalogues and furthermore exclude regions affected by very extended low-redshift galaxies, whose substructure (e.g., extended spiral arms) may otherwise be incorrectly deblended into smaller objects.

\section{Photometric redshifts}
\label{sc:photoz}

The computation of photometric redshifts (photo-$z$s) and the calibration of their redshift distributions
is conducted in two steps and broadly follows the pattern established for \Euclid WL cosmology analyses \citep{Masters2015, EuclidSkyOverview}, as well as precursor Stage-III WL surveys \citep[e.g., ][]{Hildebrandt2021,Myles2021}. Firstly, galaxies are assigned to subsamples on the basis of their best-estimated photometric redshift and secondly, the redshift distribution within each of these tomographic redshift bins is inferred by matching the target galaxies to a reference data set in their high-dimensional photometric space. The following subsections detail these two steps.

\subsection{Photo-$z$ point estimates}

We compute photometric redshifts for galaxies in the \Euclid and Subaru overlap area using the \texttt{Phosphoros}\footnote{\url{https://phosphoros.readthedocs.io/en/latest/}} package (Paltani et al., in prep.), developed for the \Euclid Science Ground Segment. \texttt{Phosphoros} is a fully Bayesian template-fitting code that shares many of the features of \texttt{LePhare} \citep{lephare1,lephare2}, on which it was initially modelled, but expands upon the functionality and flexibility of its predecessor (see the documentation and Paltani et al., in prep., for details).

Template-fitting approaches to photo-$z$ estimation are susceptible to biases and spurious peaks in redshift in the presence of systematic uncertainties, especially in the form of unknown zero-point calibration offsets, photometric measurement biases, template mis-specification, and errors in the passband throughput curves. Disentangling, measuring, and correcting for these various effects is at best a laborious task, and in most cases simply intractable. A pragmatic alternative that is often employed \citep[e.g.,][]{COSMOS2020} is to use the subset of the data with known redshifts to identify systematic differences between the predicted fluxes from their best-matching templates (at fixed redshift) and their measured values, and to then apply calibration offsets to the data. In this approach, errors in the templates (for example) get absorbed into the photometric calibration, improving the measured redshifts for objects represented by the spectroscopic subset, but leaving unknown biases for those that are not. Nevertheless, some confidence can be gained by examining the photometric offsets as a function of redshift \citep[e.g., ][]{Hartley2022}. Errors in the templates and filter curves introduce redshift-varying offsets, while true calibration errors produce constant offsets. To guard against residual biases, especially in galaxies that are not represented among the spectroscopic objects, a systematic uncertainty of between a few to ten percent of an object's flux is typically added in quadrature to the measured flux uncertainties in each band.

In the case of our A2390 data we have very limited spectroscopic information, consisting mostly of cluster member galaxies, and the additional complication of moderately-high Galactic reddening. The reddening in particular can be problematic in the case of structure on scales finer than the {\it Planck} map \citep{Planckdust2013} that we use when applying attenuation to the templates. Indeed, the cirrus emission that we subtract off (see Sect.\thinspace\ref{sc:Data:zirrus}) varies on very fine spatial scales, and we find that we are unable to obtain a stable value for the CFHT $u$-band photometric adjustment. As a result, we drop this band from our current work.
We also drop the \IE band during photo-$z$ measurement, on account of the lack of chromatic corrections in the current version of our catalogues.
As outlined in
\cite{EROData},
the very broad passband of \IE makes low-level detrending operations (such as flat-fielding) chromatic in nature, and thus require corrections to measured object photometry at a later stage. At this time, those necessary corrections are not in place for the ERO data, and so we expect significant colour-dependent biases in the photometric measurements of galaxies. Such biases have been confirmed by visual inspection of their multi-band spectral energy distributions (SEDs), with the \IE band often low with respect to the other red optical bands.
Since we have deep multi-band photometry from Subaru Suprime-Cam, we are able to proceed without the \IE photometric information for the computation of photo-$z$s, and will address the necessary corrections in the future.

With the three NISP bands and six Suprime-Cam bands, we proceed to the photometric zero-point adjustment. We gather spectroscopic redshifts via CDS,\footnote{\url{https://cds.unistra.fr//}} combining data sets from
\citet{Lamareille2006},
\citet{Nakamura2006},
\citet{Rines2018},
and \cite{sohn20}, with most objects contained in the \cite{sohn20} compilation. In total we acquire 330 spectroscopic redshifts (after excluding known active galactic nuclei), of which about $80\%$ are at or close to the cluster redshift.
The \texttt{Phosphoros} setup that we use for the systematic photometric adjustments and to measure photo-$z$ point estimates and probability distributions for the whole catalogue is summarised in Table \ref{tab:phosphoros}. The photometric adjustment factors that we derive through this process are listed in Table \ref{tab:zeropoints}, alongside the systematic fractional flux uncertainties that we apply during the photo-$z$ measurement run.

\begin{table}
  \centering
  \caption{Configuration parameters for the \texttt{Phosphoros} photo-$z$ package.}
\begin{tabular}{ll}
\hline\hline
 \rule{0pt}{2ex}
  Parameter & Value / Range \\[1pt]
  \hline
   \rule{0pt}{2ex}
   Template SED set & 31 COSMOS SEDs \\
                    & \citep{Ilbert2009} \\
   Redshift & $(0,10)$ \\
   Reddening, $E(B-V)$ & $(0,0.5)$ \\
   Reddening curve & \cite{Prevot1984} \\
                   & \cite{Calzetti2000} \\
   Luminosity prior & Tophat \\
         & $-24\le {M}_B \le 0$ \\
   Milky Way attenuation correction & \cite{Fitzpatrick99} \\
   IGM absorption & \cite{Inoue2014} \\
   \hline
\end{tabular}
\label{tab:phosphoros}
\end{table}

\begin{table}
  \centering
  \caption{Photometric magnitude zero-point adjustments (relative factor applied to the source fluxes) and fractional systematic flux uncertainties applied during photometric redshift computation.}
\begin{tabular}{ccc}
\hline\hline
 \rule{0pt}{2ex}
  Band & ZP adjustment & ZP uncertainty \\[1pt]
  \hline
   \rule{0pt}{2ex}
   \YE & 0.999 & 0.05 \\
   \JE & 1.028 & 0.05 \\
   \HE & 1.008 & 0.05 \\
   $B$  & 1.061 & 0.05 \\
   $V$  & 1.020 & 0.05 \\
   $R_\mathrm{c}$ & 0.952 & 0.05 \\
   $I_\mathrm{c}$  & 0.910 & 0.05 \\
   $i$  & 0.909 & 0.05 \\
   $z^\prime$  & 0.923 & 0.05 \\
   \hline
\end{tabular}
\label{tab:zeropoints}
\end{table}

\begin{figure}[htbp!]
\centering
\includegraphics[angle=0,width=1.0\hsize]{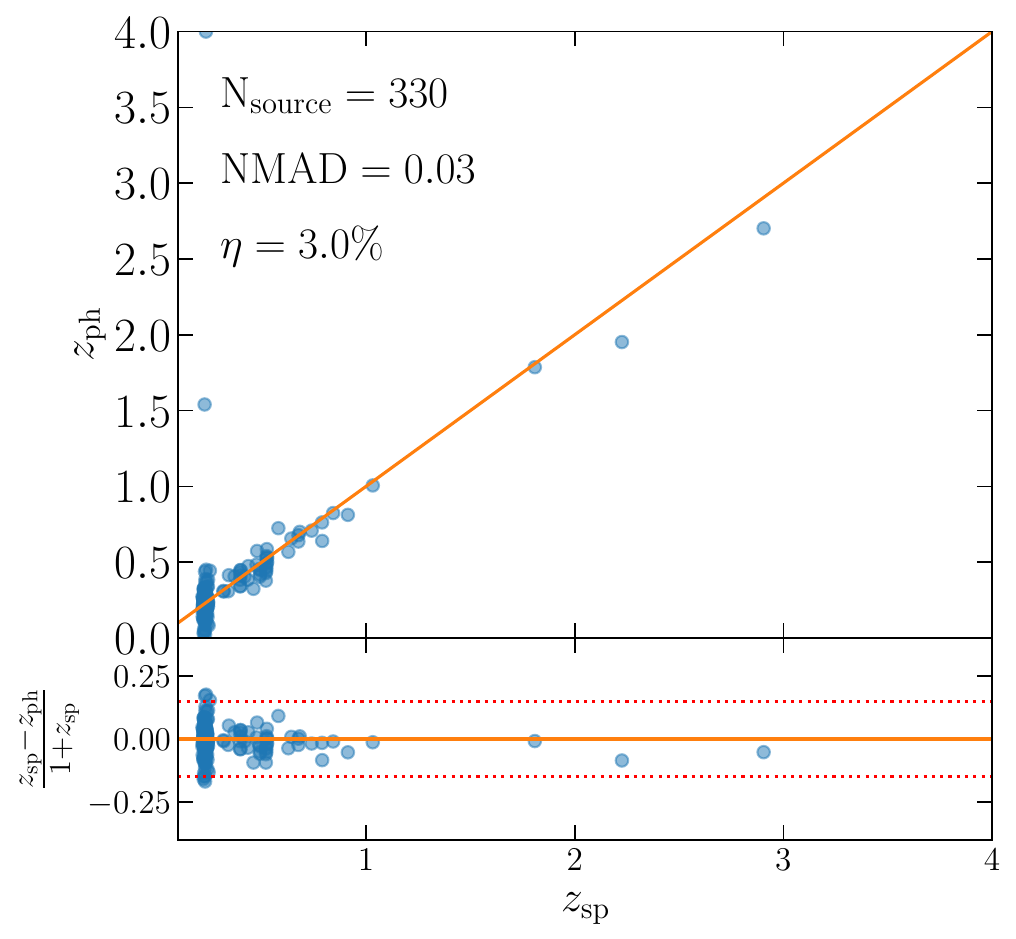}
\caption{Comparison of the best \texttt{Phosphoros}-derived photometric redshifts (\( z_\mathrm{ph} \)) with spectroscopic redshifts (\( z_\mathrm{sp} \)) for sources in the A2390 field. The top panel shows the direct comparison, while the bottom panel displays the redshift residuals defined as \( (z_\mathrm{sp}-z_\mathrm{ph}) / (1 + z_\mathrm{sp}) \). The normalised median absolute deviation (NMAD) and outlier rate (\( \eta \)) are indicated. Dashed lines at \( \pm 0.15 \) show the residual threshold.}
\label{fig:photoz}
\end{figure}

\texttt{Phosphoros} outputs both the redshift corresponding to the maximum of the multi-dimensional posterior distribution and the marginalised maximum posterior redshift.
To construct our
tomographic redshift subsamples we use the peak of the 1-D marginalised redshift distribution because it performs marginally better on the spectroscopic set of objects than the other point estimates we measure.
Figure~\ref{fig:photoz} shows the comparison of our measured photo-$z$ with the spectroscopic redshifts. The outlier rate, \( \eta = 3\% \), is defined as the fraction of sources with scaled residuals \( \Delta z = (z_\mathrm{spec} - z_\mathrm{phot}) / (1 + z_\mathrm{spec}) \) exceeding 0.15 in absolute value,
while the normalised median absolute deviation,
${\rm NMAD} =  0.03$, is defined as the normalised median absolute deviation of the residuals, ${\rm NMAD} = 1.4826 \times \mathrm{median}(|\Delta z - \mathrm{median}(\Delta z)|)$.
Both values are typical for a sample of reasonably-bright galaxies with deep optical to near-infrared broadband photometry. Although the spectroscopic sample is small, these numbers are encouraging given the manipulations to the images that were required to remove the cirrus emission.

The boundaries of our redshift bins are pre-defined, taking into account the expected overall distribution of objects in redshift, the desire to avoid (as far as possible) cluster member galaxies, and the dependence of photo-$z$ precision on the overall ${\rm S/N}_\mathrm{flux}$ of the photometry.
We define two magnitude ranges, $22<\IE<24.5$ (`bright sample') and $24.5<\IE<26.5$ (`faint sample'), with the intention of producing reasonably tight and well-separated bins for objects with more precise photo-$z$ (brighter objects), and broader bins for fainter, more difficult to measure objects that are more numerous at higher redshifts. We form six bins in redshift, four of which are used in the cluster mass measurement (avoiding the cluster itself and
galaxies at very high redshifts,
where template degeneracies are known to have an impact).
See Table \ref{tab:tomo_obj_num} for the number of objects in the different combinations of magnitude and photometric redshift bins.
\begin{table}[htbp!]
    \centering
    \caption{Number of objects in different magnitude and photometric redshift bins.}
    \smallskip
    \begin{tabular}{rrr}
          \hline\hline
 \rule{0pt}{2ex}
       Redshift bins & $22 < \IE <24.5$ &  $24.5 < \IE <26.5$  \\[1pt]
         \hline
          \rule{0pt}{2ex}
         (0.2,0.3] & 1041 & 677 \\
         (0.3,0.6] & 2760 & 3611 \\
         (0.6,0.9] & 4359 & 5706 \\
         (0.9,1.5] & 4134 & 11362 \\
         (1.5,2.8] & 2465 & 14504 \\
         (2.8,6]   & 195  & 1056 \\
         \hline
    \end{tabular}
    \label{tab:tomo_obj_num}
\end{table}

\subsection{\label{sec:nofz}Construction of the redshift distributions}

Building accurate redshift distributions, $n(z)$, for samples of galaxies is a topic that has received a great deal of attention in the WL
literature over recent years. The role that the mean redshift (in particular) plays in the cosmological interpretation of a measured WL signal has placed redshift measurements at the centre of studies of systematic uncertainties \citep[see ][ for a review]{Newman2022}. A broad consensus has emerged around two main methods for redshift inference: cross-correlation with a finely redshift-binned tracer sample \citep{Newman2008}; and matching the target galaxies in colour space to a known reference set of objects \citep{Lima2008,Masters2015}.
Since cross-correlation typically requires large sky areas to build sufficient constraining power, we will employ only the colour space matching.

The tool used by most recent cosmological analyses for calibrating the relation between redshift and the photometric space of a particular survey is a self-organising map (SOM), introduced by \cite{Masters2015} for the purpose of meeting the extremely tight redshift requirements in \Euclid.  An SOM is an unsupervised machine-learning method for dimensionality reduction that produces a two-dimensional array of nodes (or `cells') from the higher dimensional space, while preserving locality.
That is, neighbouring points in the high-dimensional space remain neighbours in the nonlinear projection to two dimensions.
Its value for redshift calibration is that the cells provide a partitioning of the photometric space that depends on the nature of the data itself.
That is, heavily occupied regions of colour space get split more finely (allowing for greater redshift fidelity), spurious objects tend to collect into small noticeable regions of the map and can be excluded, and troublesome cells, where degeneracies in redshift may occur, can also be predicted. But most important is the ability to compare the occupation of cells in the target galaxy sample with those in the reference data set that have known redshift information.
If a cell is devoid of spectroscopic calibrators, then the target galaxies in that same cell can be identified and excluded from the analysis, substantially reducing the bias arising from non-representativeness of the spectroscopic set. These empty cells can then be prioritised for future collection of spectroscopic redshifts in order to optimally use telescope time for improving redshift calibration; the rationale behind the
`Complete Calibration of the Color-Redshift Relation'
(C3R2) programme \citep{Masters2017}.

Among the assumptions of the use of an SOM for redshift calibration is that the photometric space of the calibrator spectroscopic objects and the target galaxies are well matched. In other words, the photometric calibration, measurement precision, and any systematic biases due to measurement methods should be shared by both the target and calibrating galaxy samples. In the Dark Energy Survey (DES), these necessary characteristics were achieved by a large programme of image injections \citep{Everett2022}, using the DES Deep Fields sub-survey \citep{Hartley2022} as the source of truth. In our task the requirement on the accuracy of our redshift distributions is not at the strict level of cosmological parameter inference, and neither is such a programme of image simulations practical. Instead, in the following sub-sections we perform an adaptation of the photometric space of our calibration data set, the COSMOS2020 catalogue, to our A2390 catalogue. We then argue that the colour spaces of the two catalogues are matched well enough for our present need, and finally we construct our $n(z)$ distributions.

\subsubsection{COSMOS photometric space adaptation}

The construction of the $n(z)$
for our WL source catalogues
requires a calibration data set that has matching photometry and is rich in redshift information, preferably for a complete (flux limited at the depth of the lensing catalogue) sample of galaxies. For \Euclid the photo-$z$ auxiliary fields \citep{EuclidSkyOverview} have been identified for this purpose.
Among them is the COSMOS field, arguably the key extragalactic deep field for WL
redshift calibration over the last decade.
We choose to base our $n(z)$
estimates on the information in this field, and in particular we use the COSMOS2020 catalogue \citep{COSMOS2020}. COSMOS2020 includes a very rich set of deep data across more than $30$ photometric bands, including a subset that is in common with Abell 2390, a set of photometric redshifts at an estimated precision of better than $2.5\%$ ($i<25$).
In addition,  the field contains an abundance of spectroscopic redshifts built up over many years.
In order to take advantage of this information, however, we must first unify the sets of photometry that we will use. In practice, this means generating fluxes for COSMOS2020 objects in the bands that are present in our A2390 field but not in COSMOS, and then ensuring that the photometric scatter is similar in the two catalogues.

The bands present in our A2390 catalogue and missing in the COSMOS2020 catalogue\footnote{There are multiple versions of the COSMOS2020 catalogue, depending on the photometric and photo-$z$ measurement methods desired. Only the combination of the classic \texttt{SExtractor} photometry with LePhare photo-$z$ provides the measurements and the zero-point calibration offsets in the Suprime-Cam bands that we require.} are the three \Euclid NISP bands (\YE, \JE, and \HE) and three of the Suprime-Cam bands ($R_\mathrm{c}, I_\mathrm{c}$, and $z^\prime$). Observations with the $z$ filter and older SuprimeCam MIT/LL
chips \citep[see][]{miyazaki02}  that form the $z^\prime$ throughput band were taken, but are not included in the COSMOS2020 catalogue. We generate photometry in the missing bands by means of a template-guided interpolation (for a full description see Tarsitano et al.{\thinspace}in prep.). More precisely, each COSMOS galaxy is fit using \texttt{Phosphoros} and the set of 31 COSMOS template SEDs \citep{Ilbert2009}, as though we were measuring photo-$z$,
but for this purpose we force the redshift to match that of the best photo-$z$ from the COSMOS2020 catalogue.

At this point we could simply integrate our desired filters over the template SEDs, but in practice this has two major disadvantages: (1) we would end up with quantised (rest-frame) colours that may not sample the SOM well, and (2) we may propagate errors over a wide range in wavelength. For example, photometric scatter in $B-V$ colour may result in a template fit that is a poor representation of the amplitude at long wavelengths and thus introduce a significant bias in the optical to NIR colours, even if the shape of the fitted SED at NIR wavelengths is appropriate. Instead, we performed a local weighted interpolation using colours measured from the template SED.
We identify the two broad bands that bracket a missing band in mean filter wavelength and compute the two colours that involve these filters and the missing band\footnote{That is, the colours, bluer band minus missing band, and missing band minus redder band.} using the best-fit template. These two colours, combined with the catalogued fluxes of the bracketing bands provide two predictions for the flux of the missing band. We then take a weighted average, where the weight corresponds to the inverse distance in mean filter wavelength. The initial flux errors are then propagated from the uncertainties on the neighbouring bands used.

\begin{figure}[htbp!]
\centering
\includegraphics[angle=0,width=1.0\hsize]{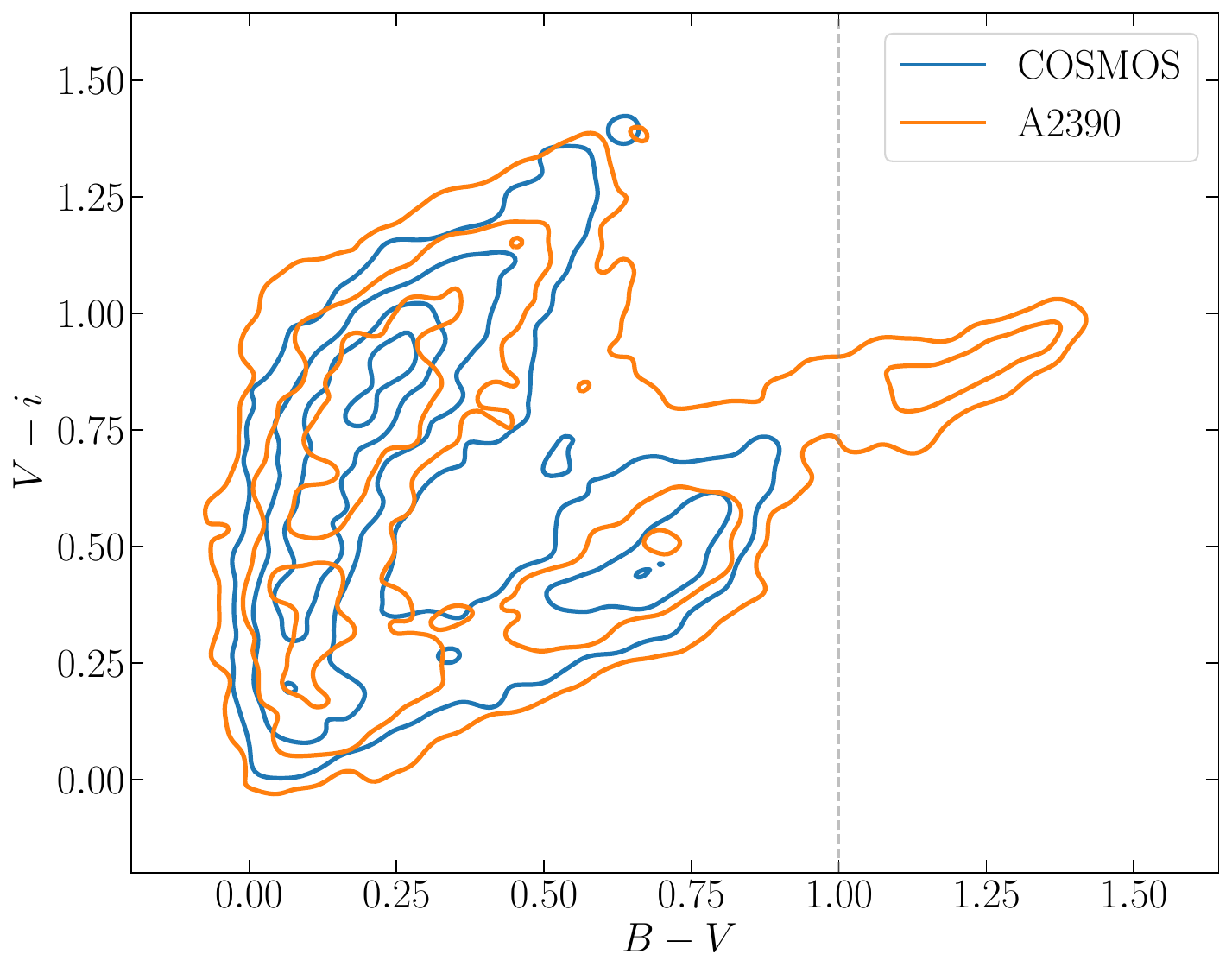}
\caption{Galaxy number density in the $B-V$, $V-i$ colour-colour space for our A2390 catalogue and the COSMOS2020 catalogue with matched bands. Contours show the relative number density, with both blue and orange distributions sharing the same contour levels. The grey dashed line shows the selection cut. }
\label{fig:colour_matching}
\end{figure}

Figure \ref{fig:colour_matching} shows an example two-colour space for our A2390 sample and the band-matched COSMOS sample. The redshift of A2390 places the strong $4000\thinspace$\AA\, break feature almost mid-way between the mean wavelengths of the $B$ and $V$ SuprimeCam filters. As a result, the red cluster member galaxies form a clearly-visible sequence, with very red values of $B-V$ colour. We remove these objects from the weak lensing source catalogue,
keeping only  objects with
$B-V<1$. The two distributions are otherwise similar in terms of the location of the main two sequences, but differ slightly in the number density distribution. We expect such differences due to the different line-of-sight cosmic structure of the two fields, including the bluer cluster members that were not removed by our $B-V$ colour cut.

\subsubsection{\label{sc:nofz}Final SOM-based $n(z)$}

\begin{figure}[htbp!]
\centering
\includegraphics[angle=0,width=1.0\hsize]{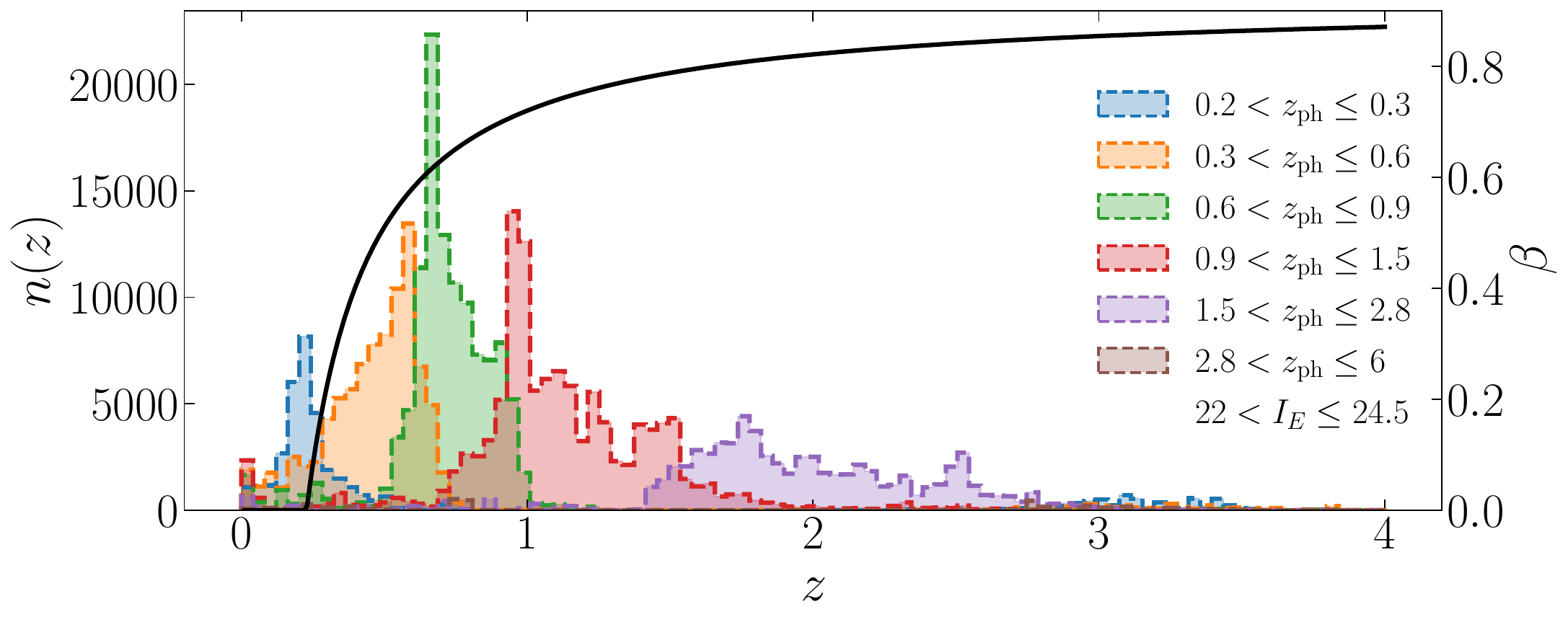}
\includegraphics[angle=0,width=1.0\hsize]{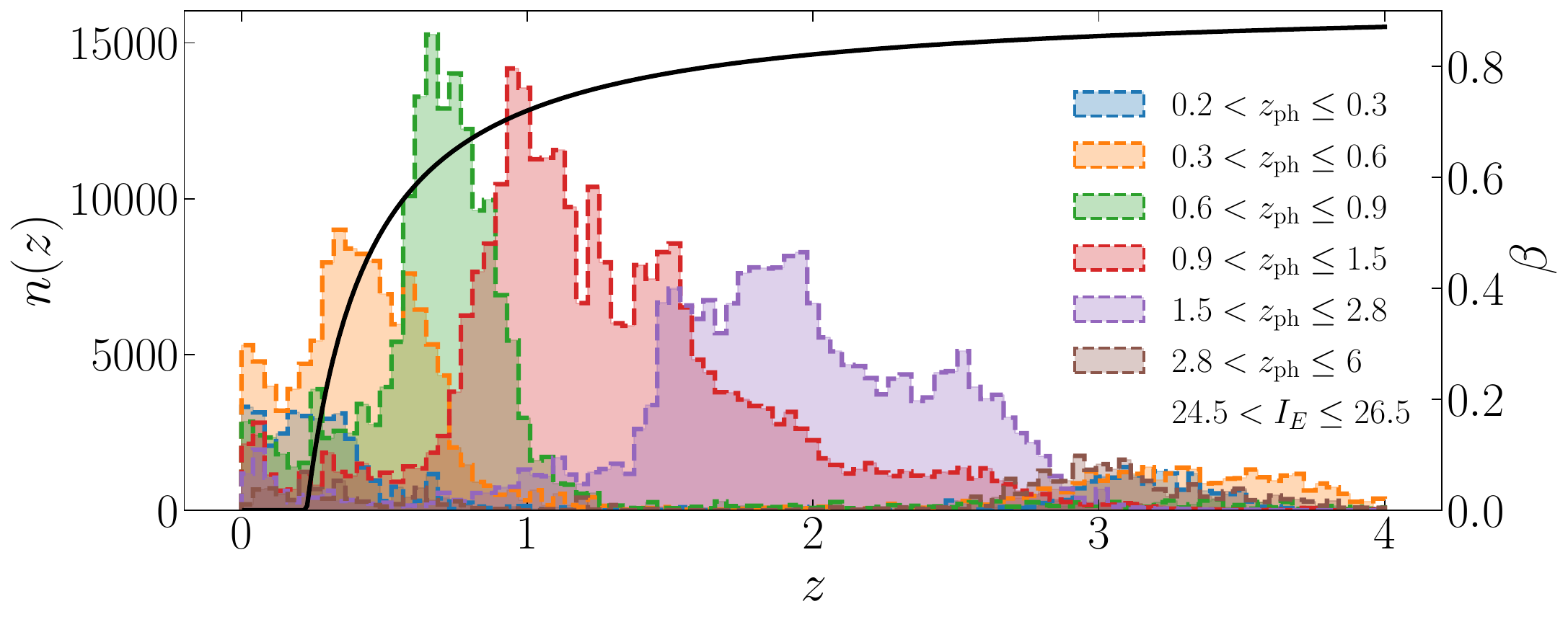}
\caption{Redshift distributions for each of our redshift and magnitude bins using the \texttt{SE++} shear weights. The upper panel shows the distributions for the magnitude range $22<\IE<24.5$, and the lower panel for the range $24.5<\IE<26.5$. The black curve shows the geometric lensing efficiency $\beta(z)$, with the corresponding axis plotted on the right.}
\label{fig:nz}
\end{figure}

A single SOM is trained using the photometry of the target galaxy sample from the A2390 field, and then used for all redshift bins. The redshift calibrator objects from COSMOS2020 are assigned to redshift bins in the same way as the target galaxies for the A2390 field.
Doing so helps guard against biases that may arise from how the specific photometric noise realisation of an object moves it between redshift bins. Correctly following how noise moves objects between bins in this way, as well as between cells of the SOM, ensures that our $n(z)$ estimates are as well calibrated as they can be (Roster et al., in prep.).

The $n(z)$ for a redshift bin is built by assigning the target galaxies and calibrator objects to their best-matching SOM cell. The calibrator objects form a normalised mini-$n(z)$ distribution for each cell, which is then multiplied by the sum of the shear weights of the target galaxies assigned to the cell. The final $n(z)$ for a photometric redshift bin is then simply the sum of all cells. Clearly, if a given cell lacks any assigned calibrator objects then the $n(z)$ for that cell cannot be computed. The target objects assigned to that cell must therefore be excluded from our analysis. Similarly, objects with zero shear weight also drop out of the sample. Following this process, 35\,479 objects are available for the weak lensing mass estimation of A2390, with a further small fraction excluded depending on the shape measurement method used (see Sect. \ref{sc:Shapes}).

We show the final $n(z)$ distributions in Fig.\thinspace\ref{fig:nz}, where the upper panel shows the distributions for our brighter magnitude range and the lower panel the fainter range. Two characteristics of the distributions are immediately obvious: the bins are broader for the faint magnitude subset, as expected; and the lowest and highest redshift bins have a substantial overlap. The latter effect is common in photo-$z$ and known to be largely due to a confusion between the two strong break features in galaxy SEDs (the Balmer/$4000\thinspace$\AA\, break at low-$z$ and the Lyman break at high-$z$).
Indeed, the COSMOS2020 photo-$z$s also suffer from this effect to a degree, despite the exquisite photometry. At our faintest magnitudes the scatter in COSMOS2020 photo-$z$s remains reasonably small ($\sigma_z \simeq 0.04$), but the fraction of outliers rises due to this issue. However, this is not a concern for our study since we drop the lowest and the highest redshift bins, which are most affected by the cross-contamination, from our main WL analysis (see Sect.\thinspace\ref{sc:wl_results}).

\section{\label{sc:Shapes}Weak lensing shape measurements}

Weak lensing analyses require accurate measurements of galaxy shapes.
The primary method designated to be used for this task in  {\Euclid}'s first main data release (DR1) is the new forward modelling method \texttt{LensMC} \citep{congedo24}.
To demonstrate its performance on early \Euclid data, we
employ \texttt{LensMC}
in this ERO analysis, as detailed in Sect.\thinspace\ref{sc:Shapes:LensMC}.
However, \texttt{LensMC} has not  been used in published works analysing real imaging observations so far. Therefore, we decided to compare the  \texttt{LensMC}-based analysis
to weak lensing constraints obtained using other shape measurement algorithms, thereby providing an empirical cross-check.
For this, we in particular employ a pipeline based on the \texttt{KSB+} formalism   \citep{kaiser1995,luppino1997,hoekstra1998}, which has been applied to similar data sets in the past, as detailed in Sect.\thinspace\ref{sc:Shapes:KSB}.
As explained in  Sect.\thinspace\ref{sc:Shapes:SourceExtractor} and Appendix \ref{appendix:SE++-details},
we additionally obtain shape estimates using
\texttt{SourceXtractor++} \citep[][abbreviated as \texttt{SE++}]{bertin22,kuemmel22},
in addition to its use
for photometric measurements (see Sect.\thinspace\ref{sc:Data:photometric_catalog}).
Because of different selections, these methods yield different number densities of weak lensing source galaxies\footnote{\label{footnote:ngal}For the quoted number densities, the area corresponds to observed sky area (including masked regions), while we do not count objects located within masked areas in the VIS image.}, as summarised in Table \ref{tab:shear_cats} and Figure \ref{fig:count_joint}.
We present a first comparison of the resulting shear estimates via a matched catalogue analysis in Sect.\thinspace\ref{sc:Shape_Comparison}.
A more quantitive comparison is provided in Sect.\thinspace\ref{sec:WLconstraintsanduncertainties} via the inferred  weak lensing mass estimates. In contrast to a matched catalogue analysis the latter properly accounts for the impact of shape weights and avoids a  potential risk to  compromise the calibration of one method by imposing additional selections from the other methods.

Given the limited space, we show illustrative figures related to the shear catalogue creation in this section for the \texttt{KSB+} method only.
The PSF model employed for the other methods was already
presented in Sect.\thinspace\ref{sc:PSFEXmodel}. Further plots related to the
 \texttt{SourceXtractor++} analysis  and the
 \texttt{LensMC} shear calibration
are provided in Appendices
 \ref{appendix:SE++-details}
and  \ref{appendix:lensmc}, respectively.

\begin{figure}[htbp!]
\centering
\includegraphics[angle=0,width=1.0\hsize]{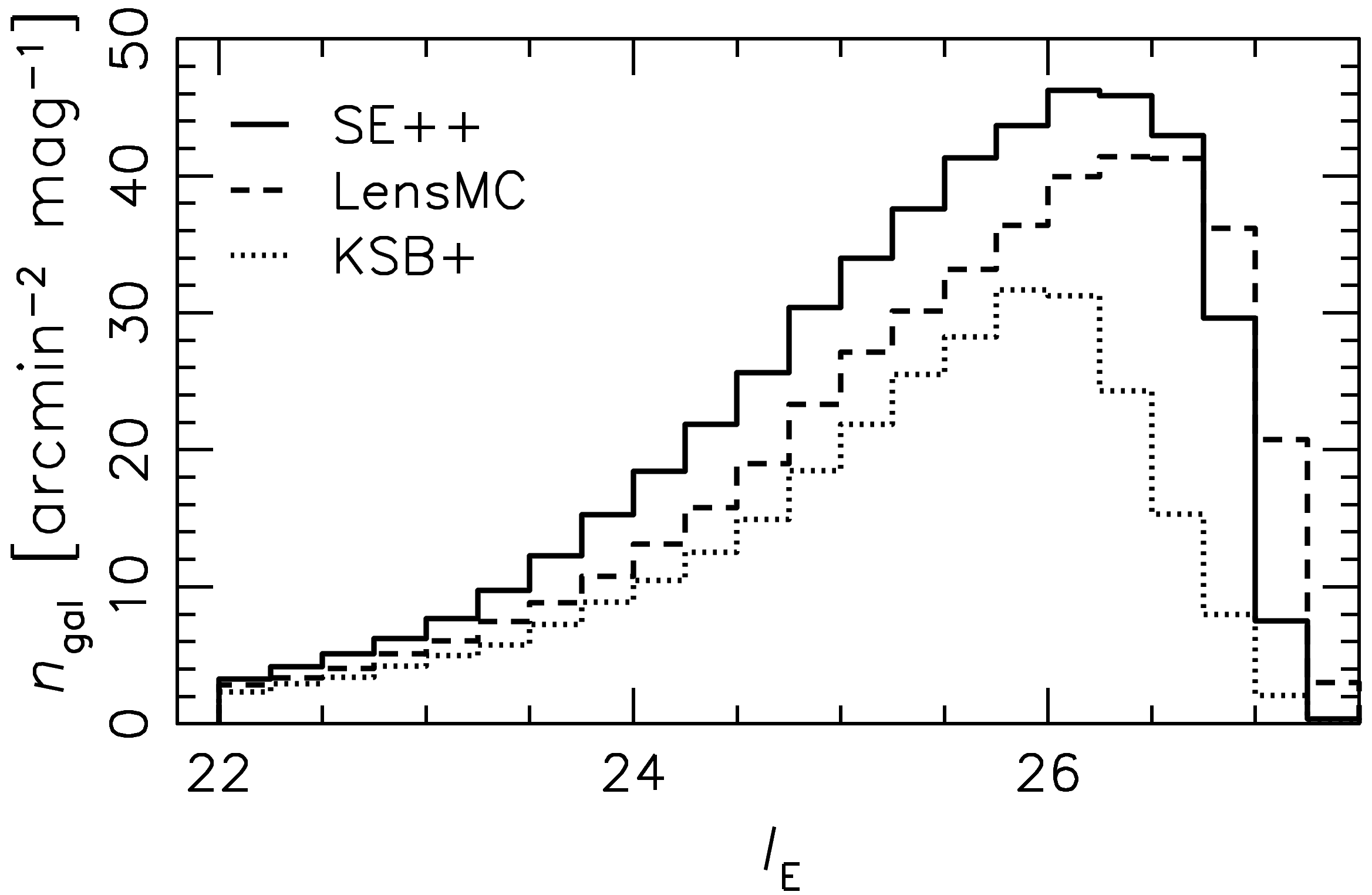}
\caption{Number density of objects in the \texttt{LensMC} (dashed),  \texttt{SE++} (solid), and \texttt{KSB+} (dotted) weak lensing source catalogues, computed
within the central
$0\fdg555 \times 0\fdg555$
of the VIS stack, after applying shear selection cuts and removal of objects in masked areas, but without applying photometric redshift selections.}
\label{fig:count_joint}
\end{figure}

\begin{table}[htbp!]
    \centering
    \caption{Overview over the shear catalogues, including the sections providing detailed descriptions and the source densities $n_\mathrm{gal}$ in the total catalogue, as well as in the \IE ranges of our analysis and the nominal EWS.}
    \smallskip
    \begin{tabular}{ccccc}
   \hline
   \hline
 \rule{0pt}{2ex}
     Method & Section & \multicolumn{3}{c}{$n_\mathrm{gal} [\mathrm{arcmin}^{-2}]$} \\
     & (\& Appendix) & Total  & \mbox{$22<\IE <26.5$} & \mbox{$\IE <24.5$}\\[1pt]
         \hline
          \rule{0pt}{2ex}
      \texttt{lensMC} & \ref{sc:Shapes:LensMC} \& \ref{appendix:lensmc} & 110 & 82 & 22\\
       \texttt{KSB+} & \ref{sc:Shapes:KSB}  & 74 & 65 & 18\\
        \texttt{SE++} & \ref{sc:Shapes:SourceExtractor}  \& \ref{appendix:SE++-details}  & 126 & 102 & 29\\
         \hline
    \end{tabular}
    \label{tab:shear_cats}
\end{table}

\subsection{\label{sc:Shapes:LensMC}\texttt{LensMC}}

\texttt{LensMC}
is a forward modelling shape measurement method that accounts for the PSF convolution and samples the posterior distribution of galaxy parameters via a
Markov chain Monte Carlo analysis \citep[][hereafter \citetalias{congedo24}]{congedo24}.
The method was designed specifically to meet the stringent requirements of \Euclid and Stage IV WL surveys, both in terms of cosmic shear accuracy and computational scalability to measure 1.5 billion galaxies.
It is now the designated shape measurement method for the \Euclid data release 1 (DR1).

Since \citetalias{congedo24}
\texttt{LensMC}
has gone through substantial testing on these data, as well as early science operation data.
Examples of changes are the correction of background gradients (in addition to the median) at the scale of the postage stamp extracted around the target galaxy or group of neighbouring galaxies (512 pixels in size).
This helps to further mitigate any residual gradients (left over by the image reduction) that would impact our lensing measurements.
Additionally, we now have much better control of outliers that are taken care of by
robust sigma clipping.
This helps to control the impact of residual cosmic rays or unmasked detector features.
Bright star masks are also included in the measurement.
\texttt{LensMC}
takes in the
\texttt{SE++}
photometric catalogue and uses the estimated world coordinates, object IDs, segmentation map, and \texttt{PSFEx} model to measure object shapes, positions, sizes, fluxes, magnitudes, $\chi^2$ values, signal-to-noise ratios, and parameter errors in a bulge+disc forward-modelling approach.
As described in \citetalias{congedo24},
objects are grouped with a friends-of-friends algorithm with a scale of $\ang{;;1}$ and are jointly measured if belonging to the same group.
\texttt{LensMC} flags objects (and therefore assigns them zero weight)
if they are too close to masked pixels, when the segmentation map IDs are not consistent with the objects, or in case of general failures.
We defined a selection function based on the measured total flux-averaged half-light radius. After checking the distributions, we implemented the star-galaxy separation by selecting objects that have a half-light radius larger than $\ang{;;0.09}$.
At the same time we removed faint galaxies with very large (and often non-physical) size estimates, which can occur for very noisy objects.
The magnitude-dependent selection that we adopted for this keeps
objects with a half-light radius less than {$[-\ang{;;0.1875}\thinspace(\IE-24)+\ang{;;1.85}]$}. This selection excludes all objects with a  half-light radius larger
than $\ang{;;1.85}$ at \mbox{$\IE=24$}, but this limit increases steadily as the galaxies become larger and brighter.
Shear weights are defined as in \citetalias{congedo24}.
All objects that were flagged or excluded by the selection above were assigned zero weight because they were deemed unsuitable for the WL analysis.

\citetalias{congedo24} conducted tests on WL
image simulations based on the \Euclid Flagship simulations \citep{EuclidSkyFlagship} mimicking data from the VIS instrument \citep{EuclidSkyVIS}.
However, these simulations only considered weak reduced shears ($|g|=0.02$), as would be adequate in the
cosmic shear regime.
In order to conduct first tests of
\texttt{LensMC} in the cluster shear regime, we analysed an additional set of image simulations with input shears  \mbox{$|g|\le 0.2$}, as detailed in  Appendix \ref{appendix:lensmc}.
Based on these tests
we found
that shear biases behave largely linearly for \texttt{LensMC} in this extended regime, which is why a standard linear multiplicative bias correction is sufficient for our study.
Through
 these tests we also
identified a dependence of the estimated multiplicative bias
on details of the PSF model, including its sampling.
For future \Euclid WL studies, the \Euclid Science Ground Segment is developing and calibrating a physical, forward-modelling, super-resolution model of the \Euclid PSF, which was, however, not yet available for this
analysis.
We therefore employed the  {\tt PSFEx} PSF model described in Sect.\thinspace\ref{sc:PSFEXmodel}  for this ERO analysis, sampled at the native VIS pixel scale, with a refined multiplicative shear bias correction as detailed in Appendix \ref{appendix:lensmc}.
For the current study, we considered a conservative 3\% uncertainty for the multiplicative bias correction to account for  potential differences between the data and calibration simulations regarding the PSF model and source population properties.
Following \citet{li23} the shear calibration for the \Euclid DR1 will apply a vine-copula remapping
to ensure matching source populations (Jansen et al., in prep.).
Together with the improved PSF models,
as well as corrections for the impact of complex galaxy morphologies \citep{csizi25},
that work will enable a much tighter shear calibration,
which was however
not yet available at the time of this ERO analysis.

We note that our current analysis does not account for the impact of the SED dependence of the PSF \citep{cypriano10,eriksen18}. As detailed in Sect.\thinspace\ref{sc:Shapes:KSB}, we estimate that this adds an additional 1.2\% systematic uncertainty to the multiplicative shear calibration, which we add in quadrature, yielding a total multiplicative shear bias uncertainty of 3.2\%.
This uncertainty is fully sufficient for our single-target study, which is
dominated by statistical uncertainties (see Sect.\thinspace\ref{sec:WLconstraintsanduncertainties}).

As in \citetalias{congedo24},
we carried out a number of validation checks, including testing the reduced $\chi^2$ distribution, which is a very useful diagnostic of the stability of the measurement. This distribution peaks at 1.0 with
a small residual positive skewness, as expected for real data in the case of
adequate modelling and error estimation.
Validating the distributions also informed the selection we applied to the catalogue, as discussed
above.

Figure \ref{fig:count_joint} shows the number counts of objects in the  catalogues from \texttt{LensMC} and the other shape measurement methods after applying shape selections and removing objects in masked areas. The total source density$^{\ref{footnote:ngal}}$ in the correspondingly filtered \texttt{LensMC}  catalogue amounts to  110 $\mathrm{arcmin}^{-2}$, of which  22 $\mathrm{arcmin}^{-2}$ have $\IE<24.5$, and 82 $\mathrm{arcmin}^{-2}$ fall into the interval $22<\IE<26.5$ employed in our WL analysis.
We note that the \texttt{LensMC} shape catalogue extends noticeably beyond the depth limit \mbox{$\IE<26.5$} imposed by the photometric redshift analysis (see Sect.\thinspace\ref{sc:photoz}).

\subsection{\label{sc:Shapes:KSB}\texttt{KSB+}}
We also generated a
WL catalogue using the \texttt{KSB+} formalism
    \citep{kaiser1995,luppino1997,hoekstra1998}, employing the implementation from \citet{erben2001} as detailed in \citet{schrabback10}.
This pipeline
is also used
for shape measurements in cluster WL
analyses by \citet{schrabback18,schrabback18b,schrabback21b},
\citet{thoelken18},
and \citet{zohren22}.
We employ the
correction for multiplicative
WL
shear estimation bias
derived by \citet[][hereafter \citetalias{hernandez20}]{hernandez20} for this \texttt{KSB+} implementation, accounting for the bias dependence on the signal-to-noise ratio
${\mathrm{S}/\mathrm{N}}_\mathrm{KSB}$,
which is measured including the \texttt{KSB+} weight function \citep[see][]{erben2001,schrabback07}.
\citetalias{hernandez20}
tune their image simulations
such that they closely resemble deep {\it Hubble} Space Telescope (HST)
WL data with a resolution of 0\farcs1 (PSF FWHM) based  on
observations from the Cosmic Assembly Near-IR Deep Extragalactic Legacy Survey \citep[CANDELS,][]{grogin11,koekemoer11},
including realistic clustering.
They also explored an alternative scenario matching the properties of
WL data from the Very Large Telescope (VLT) High Acuity Wide field K-band Imager (HAWK-I),
 with a PSF FWHM of 0\farcs4. Here \citetalias{hernandez20} find that multiplicative shear biases shift by less than $0.9\%$ compared to the HST-like setup, suggesting a low sensitivity of the calibration to the exact simulation details.
Accordingly, we
expect
that this calibration is also applicable to the {\it Euclid} ERO observations, which have a resolution approximately at the geometric mean of the scenarios explored by \citetalias{hernandez20}.
Based on their analysis,
\citetalias{hernandez20} estimate a residual systematic uncertainty of their derived multiplicative shear bias calibration of $1.5\%$.
Similarly to the \texttt{LensMC} analysis we inflate this uncertainty to 3\% to account for
differences in both the PSF shapes
and the source background selection compared to  \citetalias{hernandez20}.

We note that \citetalias{hernandez20} find no indications for significant  nonlinear shear biases for reduced shears up to $|g|<0.4$ for this \texttt{KSB+} implementation, allowing us to safely ignore nonlinear corrections at the accuracy requirements of this ERO analysis.
Interestingly, this differs from the results obtained by \citet{Jansen24}, who find a significant nonlinear bias component for the {\tt galsim} \citep{rowe15} \texttt{KSB+} implementation, suggesting a dependence on the detailed \texttt{KSB+} implementation differences.

\begin{figure}[htbp!]
\centering
\includegraphics[angle=0,width=1.0\hsize]{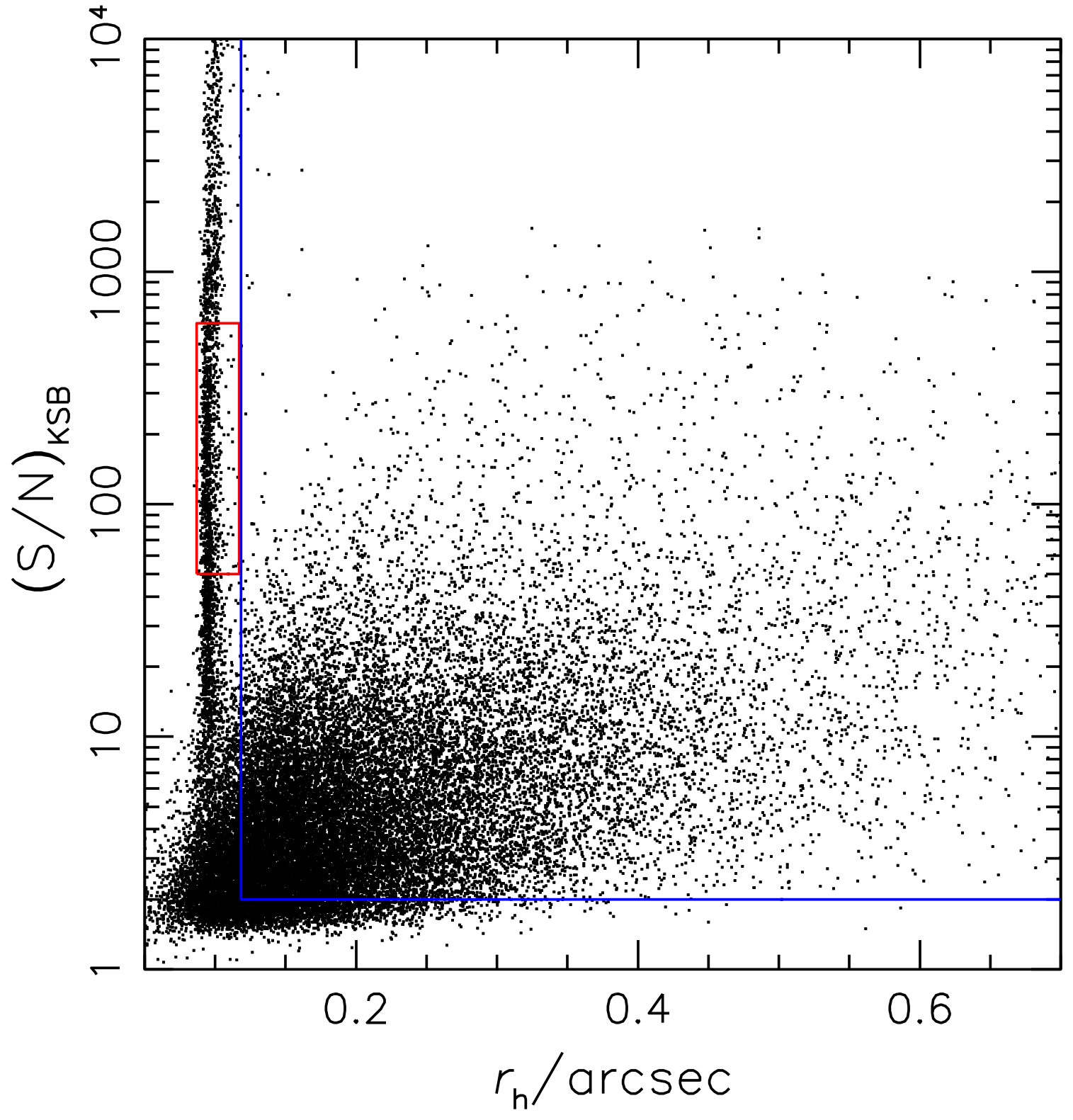}
\caption{Distribution of the \texttt{KSB+} signal-to-noise ratio
${\mathrm{S}/\mathrm{N}}_\mathrm{KSB}$
versus half-light radius $r_\mathrm{h}$ for objects in the unfiltered \texttt{KSB+} catalogue.
The red box and blue lines indicate the pre-selection regions for the stars that are employed in the PSF modelling and for the galaxies, respectively. For clarity only a random subset of 20\% of catalogue entries is displayed. Stars and noisy or poorly resolved
galaxies are further removed from the shear catalogue via cuts in photometric redshift,
magnitude, the \texttt{SExtractor} signal-to-noise ratio, and
additional \texttt{KSB+} selections  (see \citetalias{hernandez20}).}
\label{fig:rh_lnSN}
\end{figure}

\begin{figure*}[t!]
\centering
\includegraphics[angle=0,width=0.45\hsize]{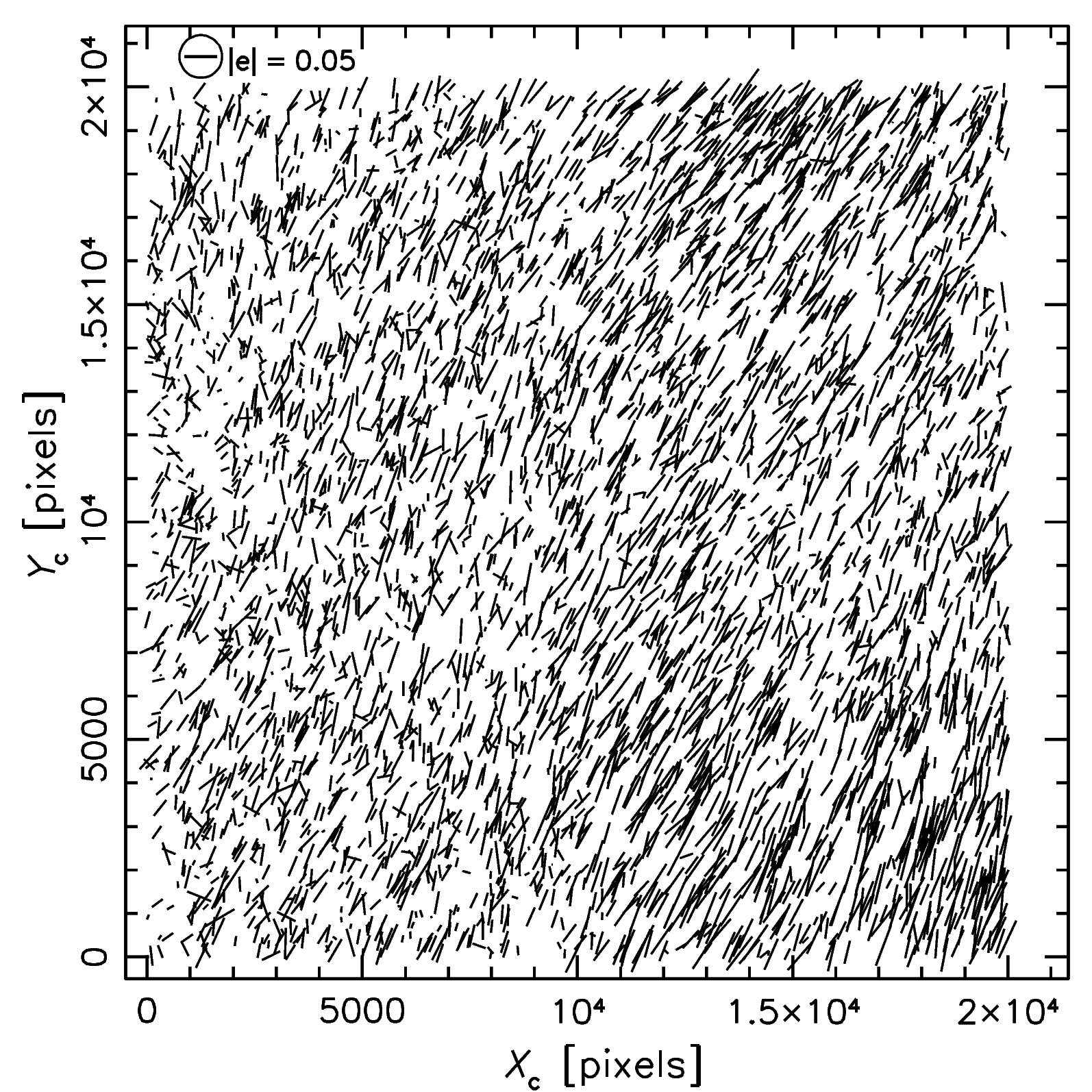}
\includegraphics[angle=0,width=0.45\hsize]{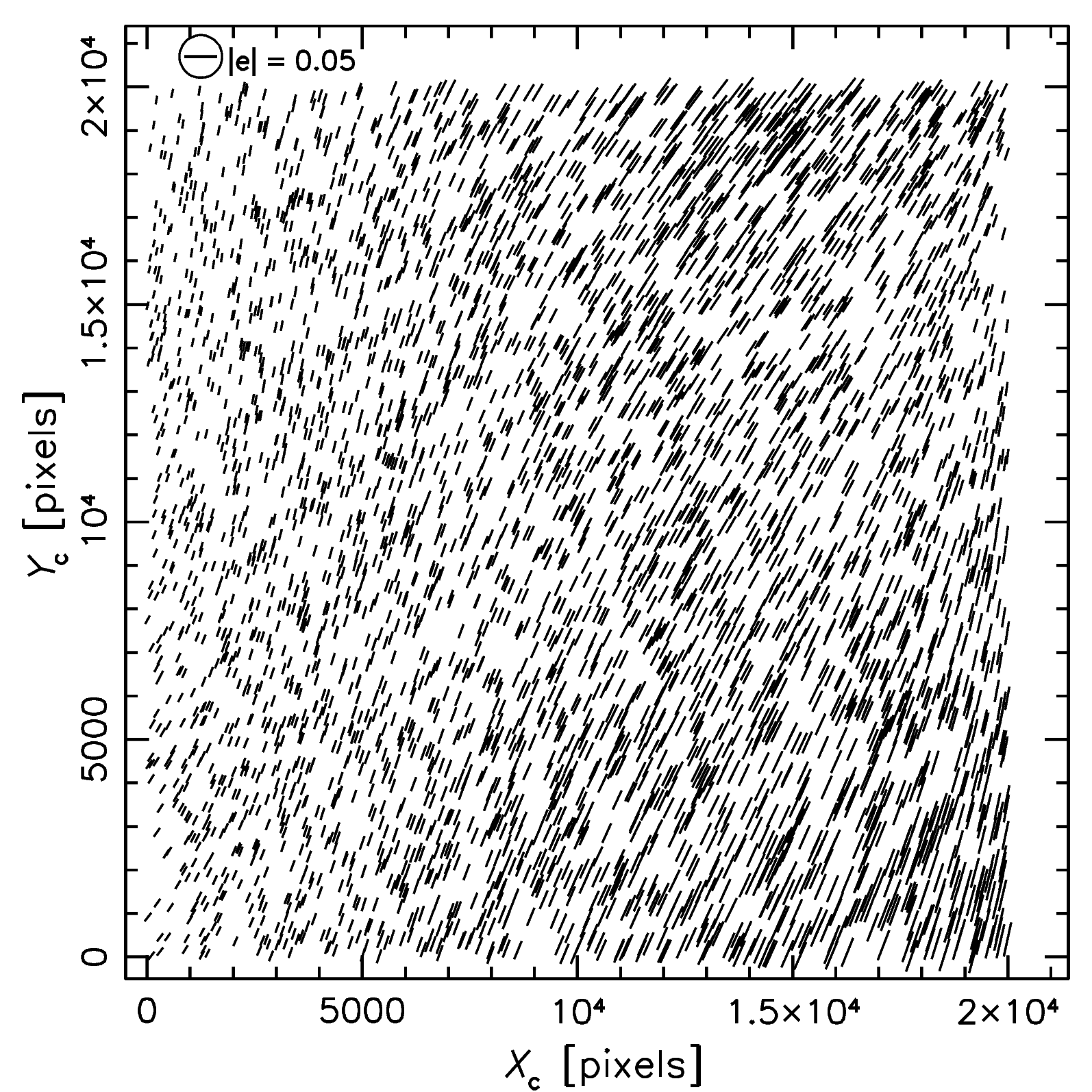}
\caption{Spatial variation of the PSF polarisation $\epsilon_\alpha$  measured using a \texttt{KSB+} Gaussian filter scale \mbox{$r_\mathrm{g}=0\farcs16$} (left panel) and its 2D third-order spatially interpolated model (right panel). Here we consider the central
$20\,000\times 20\,000$
pixels of this VIS stack (extending slightly beyond the primary WL region of interest, see Sect.\thinspace\ref{sc:Data:ground_based_data}), where the depicted coordinates $(X_\mathrm{c}, Y_\mathrm{c})$ relate to the pixel positions in the stack $(X_\mathrm{s}, Y_\mathrm{s})$ as  \mbox{$(X_\mathrm{c}, Y_\mathrm{c})=(X_\mathrm{s}-10799, Y_\mathrm{s}-10499$}). The whisker in the top left indicates a reference polarisation with \mbox{$|e|=0.05$}.}
\label{fig:ksb_ellipticity}
\end{figure*}

Figure\thinspace\ref{fig:rh_lnSN} shows the distribution of measured objects in the unfiltered \texttt{KSB+} catalogue as a function of the half-light radius $r_\mathrm{h}$ and
${\mathrm{S}/\mathrm{N}}_\mathrm{KSB}$.
In the figure the red box indicates the cuts that are used to select stars for the PSF modelling, where we exclude not only faint and noisy stars, but also brighter stars
to avoid the impact of nonlinear effects such as the  brighter-fatter effect \citep{guyonnet15}.
The selected stars have a median half-light radius
$r_\mathrm{h,median}^*=0\farcs09529\pm 0\farcs00007$
as measured by \texttt{analyseldac} \citep{erben2001}, as well as
median values of the
flux radius the and
full-width-at-half-maximum  (FWHM)
as measured by \texttt{SExtractor} \citep{bertin96}
of
$r_\mathrm{f,median}^*=0\farcs1210\pm 0\farcs0001$  and $\mathrm{FWHM}_\mathrm{median}^*=0\farcs1585\pm 0\farcs0002$, respectively.

The selected stars are used to obtain local estimates of PSF parameters, such as the components  of the \texttt{KSB+} PSF polarisation $e_\alpha$, measured as a function of the \texttt{KSB+} Gaussian filter scale $r_\mathrm{g}$.
We find that the  PSF properties as measured with \texttt{KSB+}  vary fairly smoothly across the relevant part of the VIS  image stack, which is why we employ a simple third-order polynomial interpolation for the current study (see Fig.\thinspace\ref{fig:ksb_ellipticity}).
We  note that the overall level of PSF ellipticity is quite low. That is, when measured with a Gaussian filter scale of \mbox{$r_\mathrm{g}=0\farcs16$} (as employed for typical compact galaxies),
the root-mean-square (r.m.s.) of the polarisation model amounts  to only 2.7\% when combining both polarisation components and averaging over the area depicted in Fig.\thinspace\ref{fig:ksb_ellipticity}.

The blue lines in Fig.\thinspace\ref{fig:rh_lnSN} indicate lower limits  $r_\mathrm{h,min}=0\farcs1404$ and  ${\mathrm{S}/\mathrm{N}}_\mathrm{KSB,min}=2$ employed in the galaxy selection.
The source selection furthermore includes cuts in \texttt{KSB+} parameters (see \citetalias{hernandez20}),
as well as
\mbox{$\mathrm{S/N}_\mathrm{flux}>8$},
which is well-bracketed by the scenarios tested by
\citetalias{hernandez20}.
Figure\thinspace\ref{fig:count_joint} compares the number counts of objects in the  catalogues from \texttt{KSB+} and the other shape measurement methods after applying the corresponding shape selections and removing objects in masked areas.
The total source density$^{\ref{footnote:ngal}}$ in the correspondingly filtered \texttt{KSB+}  catalogue amounts to  74 $\mathrm{arcmin}^{-2}$, of which  18 $\mathrm{arcmin}^{-2}$ have \mbox{$\IE<24.5$} and 65 $\mathrm{arcmin}^{-2}$ fall into the interval \mbox{$22<\IE<26.5$} employed in our WL analysis.
Compared to the other shape catalogues the source density is somewhat lower for the \texttt{KSB+} catalogue.
This is due to a combination of several factors, including a more   conservative removal of both objects with close neighbours and galaxies that are  noisy or poorly resolved.
Large galaxies are also removed within the employed \texttt{KSB+} pipeline if they are not well covered by the internal postage stamp cutout.

Figure\thinspace\ref{fig:ksb_mag} shows the dispersion $\sigma_{\epsilon,\alpha}$ of the measured ellipticity estimates from all objects in the fully filtered \texttt{KSB+} galaxy catalogue, split into bins of $\IE$. Combining both ellipticity components $\epsilon_1$ and $\epsilon_2$, we fit these values with a  third-order polynomial interpolation
(smooth curve in
Fig.\thinspace\ref{fig:ksb_mag})
in order to define an empirical shape weight $w_i=[\sigma_\epsilon(\IE)]^{-2}$
 \citep[e.g.,][]{schrabback18b}.

\begin{figure}[htbp!]
\centering
\includegraphics[angle=0,width=1.0\hsize]{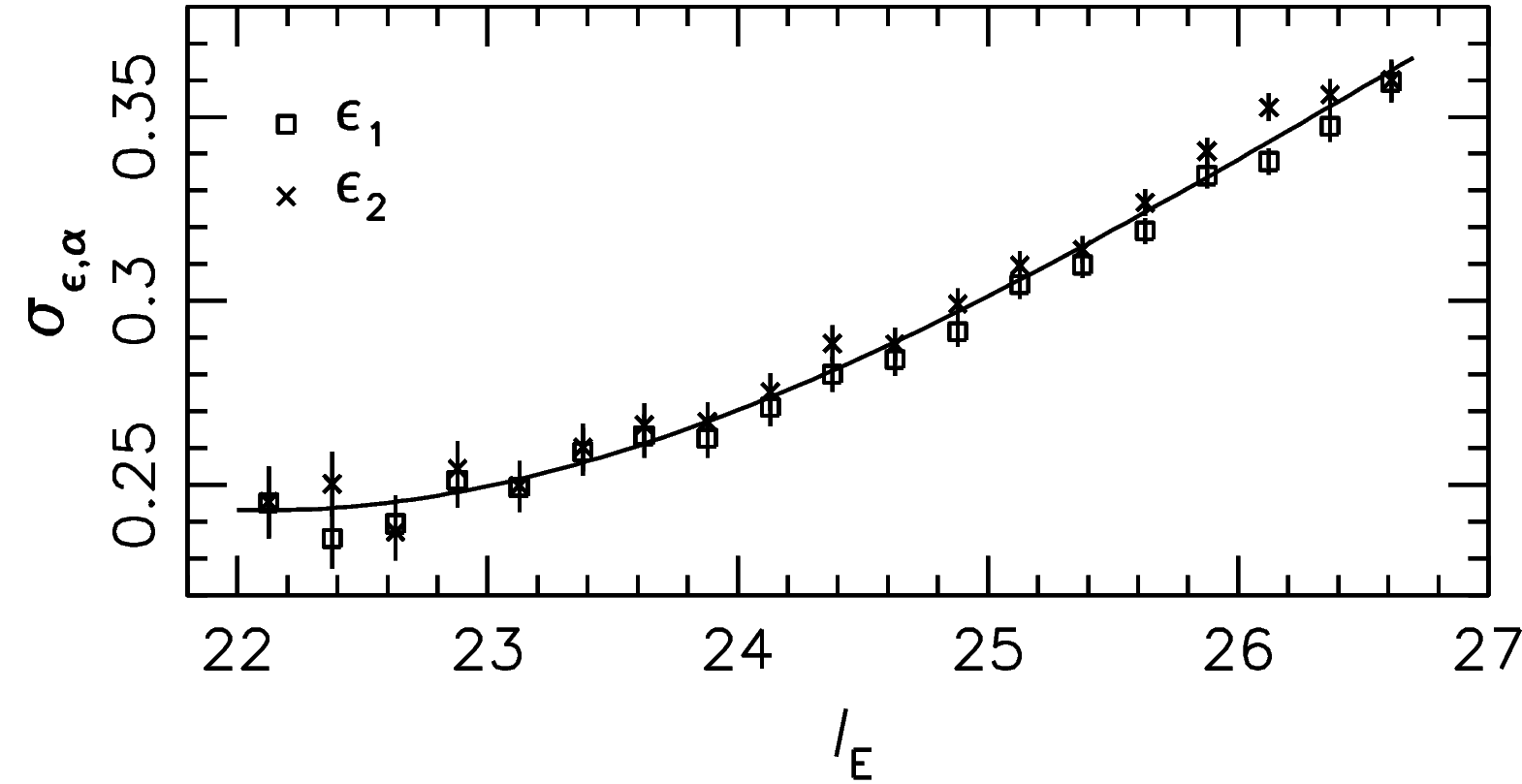}
\caption{Dispersion of the fully-corrected \texttt{KSB+} ellipticity estimates as a function
$\IE$,
shown for both ellipticity components $\epsilon_1$ and $\epsilon_2$. The smooth curve shows the best-fit third-order polynomial interpolation function, which is used for the computation of the empirical shape weight.}
\label{fig:ksb_mag}
\end{figure}

In this work we neglect the impact
of the SED dependence of the PSF \citep{cypriano10,eriksen18} for all shape catalogues.
To assess the impact of this, we employ the formalism from \citet[][see their equation A7]{cypriano10},
for which we compute the required size ratio of the PSF and the galaxies via the corresponding flux radius from \texttt{SExtractor}, averaged over all selected galaxies in the \texttt{KSB+} catalogue. The resulting effective shift in the multiplicative shear bias depends on the difference in the FWHM of the effective PSFs
for
stars and galaxies.
When assuming the corresponding estimates by \citet{cypriano10} for the
average galaxy population versus a typical disk or halo stars, we obtain a shift in the multiplicative bias by 1.2\% or 0.8\%, respectively.
We use the larger one of these values as estimate for the resulting multiplicative bias uncertainty, which we add in quadrature to the 3\% estimate discussed above, yielding a joint uncertainty of 3.2\%.

\subsection{\label{sc:Shapes:SourceExtractor}\texttt{SourceXtractor++}}
In our work \texttt{SE++} is not only employed for photometric measurements  (see Sect.\thinspace\ref{sc:Data:photometric_catalog}),
but also as one of the methods to obtain PSF-corrected galaxy shapes for the WL analysis.
Differing from the photometric analysis, this separate  \texttt{SE++} run employs a single S\'ersic profile, directly providing an ellipticity estimate, as well as best-fit sizes for star/galaxy discrimination.
Recent tests done as part of the Euclid Morphology Challenges \citep{bretonniere23,merlin23} indicate that \texttt{SE++} can not only recover accurate multi-band photometry, but also morphological parameters including ellipticity, orientation, S\'ersic index, and half-light radius.
Likewise, earlier tests conducted by \citet{mandelbaum15} and \citet{martinet19} demonstrated that \texttt{SExtractor}'s model-fitting engine can reach an accuracy that is sufficient for Stage III WL surveys.
We note however  that
\texttt{SE++} has not yet gone through the same level of testing on WL image simulations as the other two shape measurement methods employed in this study.
We therefore regard the constraints from the other methods as primary results, where the comparison to the   \texttt{SE++} constraints provides a data-driven validation for the   \texttt{SE++} implementation.

Like for the other two shape measurement methods, galaxy ellipticities are better measured for high-surface-brightness extended sources. Noise and PSF blurring induce more uncertain ellipticity, which must be properly accounted for in a weighting scheme.
The recovered
uncertainties on the
$\epsilon_1$ and $\epsilon_2$
ellipticity components are added in quadrature to a constant shape noise floor
$\sigma_{\rm \epsilon,0}\equiv0.25$
in order to down-weight poorly constrained galaxy shapes.
 Appendix \ref{appendix:SE++-details} provides further details on our employed \texttt{SE++} implementation.

For the different shape catalogues the number counts of objects, after applying shape selections and removing objects in masked areas, are compared in  Fig.\thinspace\ref{fig:count_joint}.
The total source density$^{\ref{footnote:ngal}}$ in the correspondingly filtered  \texttt{SE++}  catalogue  amounts to  126 $\mathrm{arcmin}^{-2}$, of which  29 $\mathrm{arcmin}^{-2}$ have \mbox{$\IE<24.5$} and 102 $\mathrm{arcmin}^{-2}$ fall into the interval \mbox{$22<\IE<26.5$} employed in our WL analysis.

\subsection{\label{sc:Shape_Comparison}Raw shear profile comparison}
As a consistency test we conduct a first comparison of the shear signal probed by the three shear catalogues in this subsection.
The catalogues differ in their depth and  weighting, which affects the expected shear signal and will be accounted for in  the full analysis presented in Sect.\thinspace\ref{sc:wl_results_mass}.
For this first test presented here we limit the catalogues to the joint sample of sources that have non-zero shape weights for all three shape measurement methods.
We then compute the tangential
component
\begin{equation}
    g_\mathrm{t}  = - g_1 \cos{2 \phi} - g_2 \sin{2\phi}\label{eq:gt}\;,
\end{equation}
and the cross-component
\begin{equation}
    g_\times = + g_1 \sin{2\phi} - g_2 \cos{2 \phi}  \;,
\end{equation}
of the reduced shear estimates,
where $\phi$ denotes the azimuthal angle with respect to the cluster centre (see Sect.\thinspace\ref{sc:wl_results_mass} regarding the choice of centre).
We compare their averages as a function of
the physical (not co-moving) radial distance $r$ from the cluster centre
for the three shear catalogues in Fig.\thinspace\ref{fig:matched_profile}, finding very good agreement both for the tangential component and the cross component. The latter is consistent with zero for all three shape catalogues, as expected in the case of accurately removed instrumental signatures.
As a further check we split the matched source sample into bins of galaxy size,
likewise finding  good agreement  between the three catalogues in terms of the resulting reduced shear profiles.

\begin{figure}[htbp]
\centering
\includegraphics[angle=0,width=1.0\hsize]{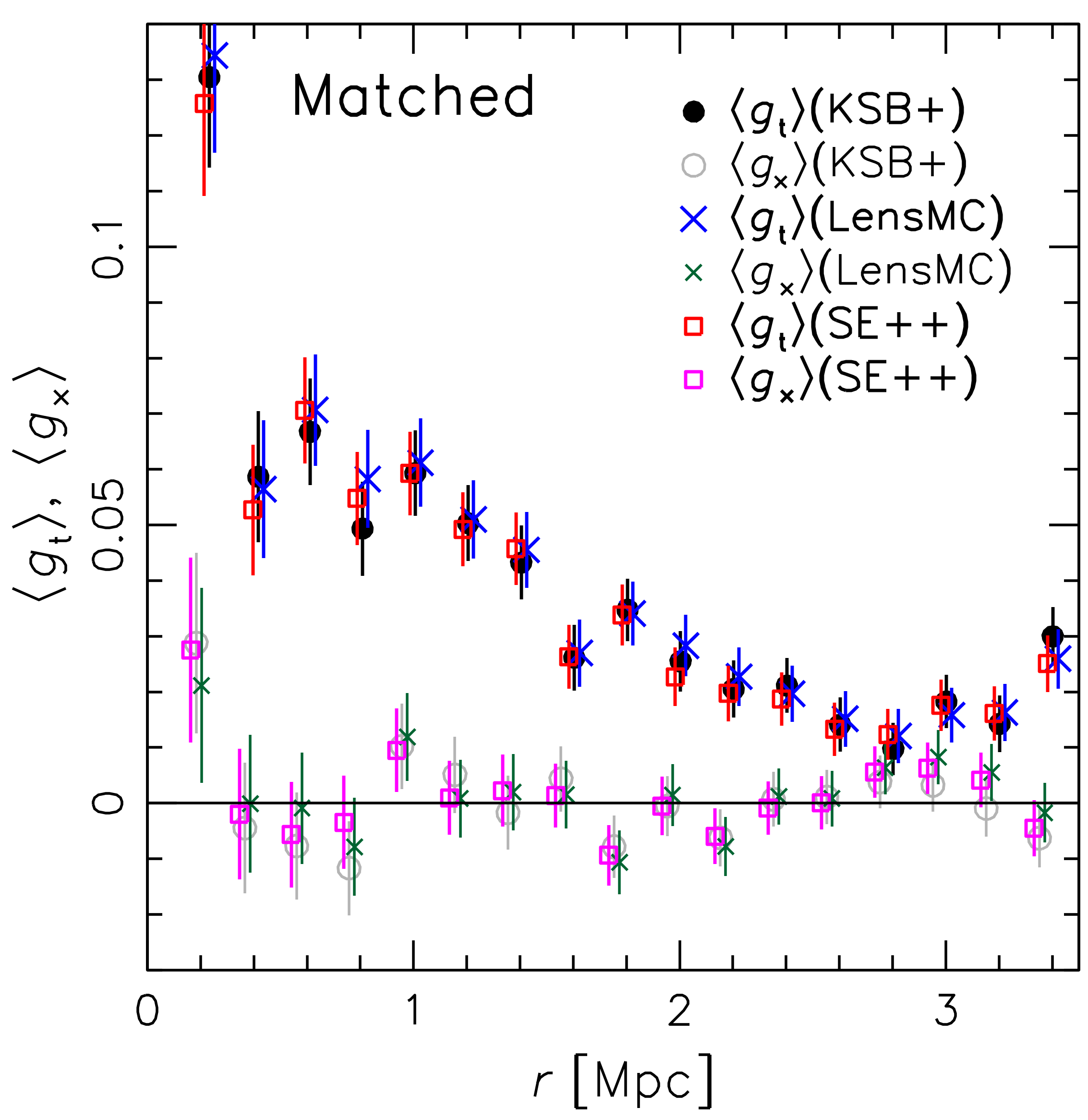}
\caption{Matched-sample comparison of the raw reduced shear profiles (prior to the background selection and contamination correction) obtained from the three shape catalogues, showing both the tangential component $\langle g_\mathrm{t}\rangle$ and the cross-component  $\langle g_\times\rangle$.
In the average computation we only include sources that have non-zero shape weights as computed by all three methods.
The error-bars include shape noise and are therefore  correlated between the methods, explaining why the tangential reduced shear profiles obtained by the three methods differ from each other less than the error-bars.  For the \texttt{KSB+} catalogue $\langle g_\mathrm{t}\rangle$ is plotted at the correct position, while the other data points have been
offset
along the $x$-axis for clarity.
}
\label{fig:matched_profile}
\end{figure}

\section{\label{sc:cluster_member_contamination}Quantifying and correcting for cluster member contamination}

Cluster members constitute an excess population compared to random galaxy populations that are used to estimate the source redshift distribution (see Sect.\thinspace\ref{sc:nofz}). Accordingly, it is necessary to accurately estimate
the radius-dependent  contamination of the selected source galaxies by cluster members, in order to correct the estimated shear signal for the dilution caused by this contamination. Here we determine this correction based on the number density profile (Sect.\thinspace\ref{sc:source_number_density_profile}), which, however, needs to be corrected for the impact of source obscuration (Sect.\thinspace\ref{sc:obscuration_correction}) and lensing magnification  (Sect.\thinspace\ref{sec:magnification}).
Similarly to \citet{kleinebreil25} we conduct this contamination analysis in the same photometric redshift bins as used for the redshift calibration, since we expect a dependence of the contamination level on the source photo-$z$ and magnitude.

\subsection{\label{sc:source_number_density_profile}Source number-density profile}
For all combinations of photo-$z$ and magnitude bins (see Sect.\thinspace\ref{sec:nofz})
we calculate the radial source number-density profile in geometric annuli,
which mirror the binning of the shear measurement (see Sects.\thinspace\ref{sc:Shape_Comparison} and \ref{sc:wl_results_mass}). Here, we take the masked fraction $f_\mathrm{masked}$ of each annulus into account and compute the corresponding number densities and Poisson errors
\begin{align}
    n_\mathrm{gal} &= \frac{N_\mathrm{gal}}{\pi\left(r_2^2-r_1^2\right)\left(1-f_\mathrm{masked}\right)}\;,\\
    \sigma\left(n_\mathrm{gal}\right) &= \frac{\sqrt{N_\mathrm{gal}}}{\pi\left(r_2^2-r_1^2\right)\left(1-f_\mathrm{masked}\right)}\;,
\end{align}
where $r_1$ and $r_2$ are the inner and outer radii of the annulus, $N_\mathrm{gal}$ is the number of sources in a given bin, and $f_\text{masked}$ is the geometric masked fraction.
We show the resulting number density profiles in Fig.\thinspace\ref{fig:nd_raw}.
The bins with low photometric redshifts $z_\mathrm{ph}$ show a pronounced increase in source number density towards the cluster centre, which is expected due to cluster member contamination.
On the other hand, some of the high-redshift bins
show a noticeable drop below $\SI{1}{Mpc}$, which is likely due to the optical obscuration caused by bright cluster galaxies and possibly magnification (see Sects.\thinspace\ref{sec:obscuration} and \ref{sec:magnification}). Contamination generally appears to be stronger for the bright sample (\mbox{$22.0<\IE<24.5$}), and obscuration seems to affect the faint sample ($24.5<\IE<26.5$) more strongly.

\subsection{\label{sc:obscuration_correction}Estimation of and correction for source obscuration}
\label{sec:obscuration}
Bright cluster galaxies can cause significant optical obscuration and impede the ability to detect background galaxies along and close to their lines of sight.
In order to extract an accurate cluster member contamination model from source number density profiles,
 we therefore need to take the obscuration effects of the lens into account. We modify the image injection pipeline described in \citet{kleinebreil25} for this task.

We generate galaxy images with \texttt{GalSim} \citep{rowe15} based on galaxy properties that we draw randomly from the Flagship catalogue \citep{EuclidSkyFlagship}.
In particular, we generate double-Sérsic profile models based on the Flagship galaxy properties (which model the disc and the bulge of a galaxy), and convolve the resulting image with a \textit{Euclid}-like optical PSF following \citet{tewes19} and \citet{Jansen24}. We subsequently inject the simulated galaxy stamps into the real VIS observations at random positions and at a density of
$\SI{3}{arcmin^{-2}}$
within the Flagship VIS magnitude interval
\mbox{$20<\IE<24.5$}
per injection run.
This is equal to $10\%$ of the nominal
number density of the
EWS.
Towards fainter magnitudes we inject sources based on the Flagship magnitude distribution with the same fraction, leading to an overall galaxy injection density of
$\SI{29.4}{arcmin^{-2}}$.

Following the source injection we re-run the object detection using \texttt{SourceExtractor} \citep{bertin96},
employing identical settings as used for the generation of the WL source catalogue.
Since we know the injected objects' positions
we are able to calculate their detection probability (as the ratio of the numbers of re-detected and injected galaxies). We show a high-resolution map of the resulting detection probability in Fig.\thinspace\ref{fig:pdet_map}.

\begin{figure}[t]
    \centering
    \includegraphics[width=\linewidth]{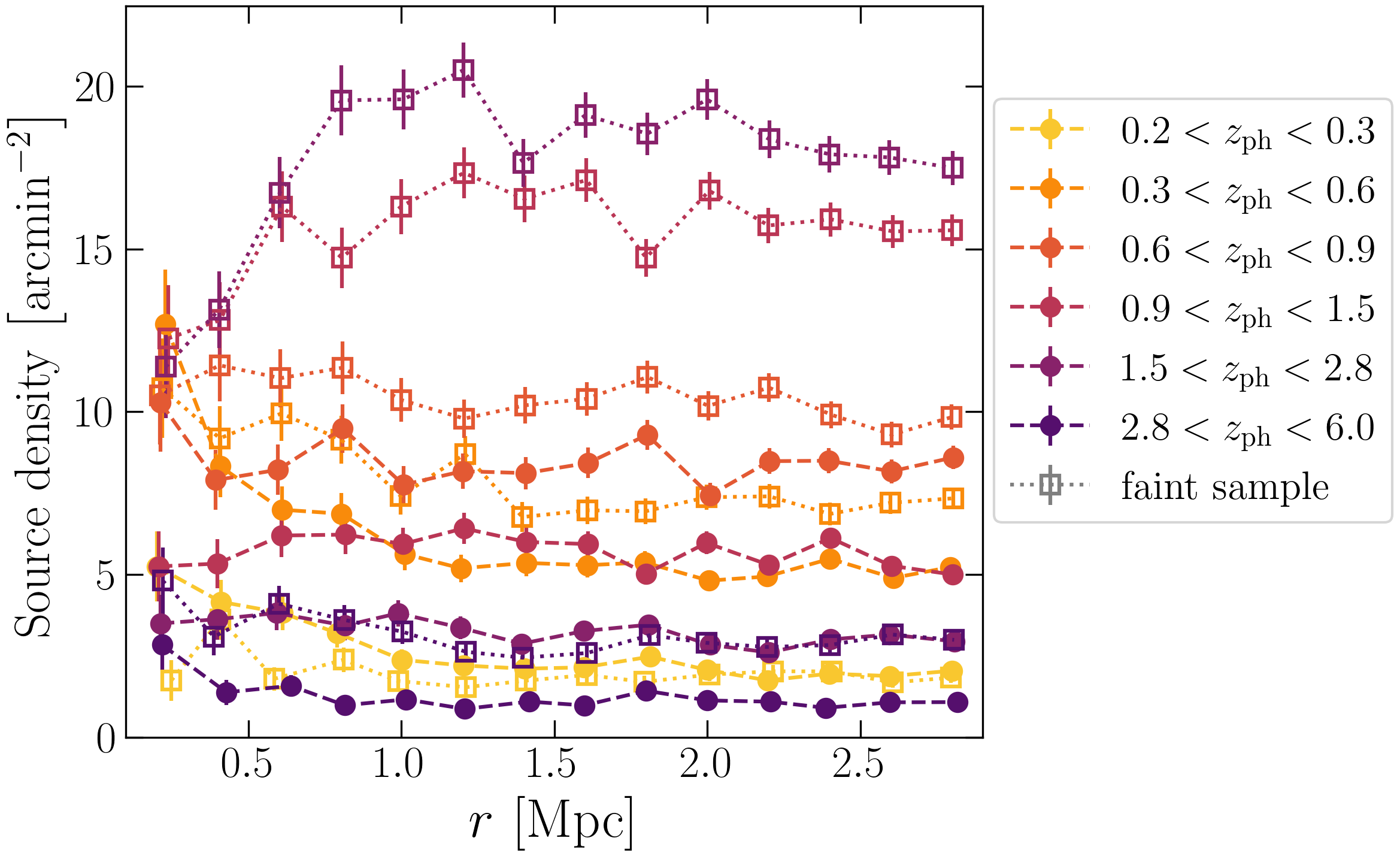}
    \caption{Number density profiles of WL sources in bins of photo-$z$
    (colour-coded)  and magnitude, where the solid and open symbols correspond to the bright and faint sample, respectively.
    Error-bars indicate Poisson uncertainties. In this figure, we account for masked areas, but not for obscuration.
    }
    \label{fig:nd_raw}
\end{figure}

\begin{figure}[t]
    \centering
    \includegraphics[width=\linewidth]{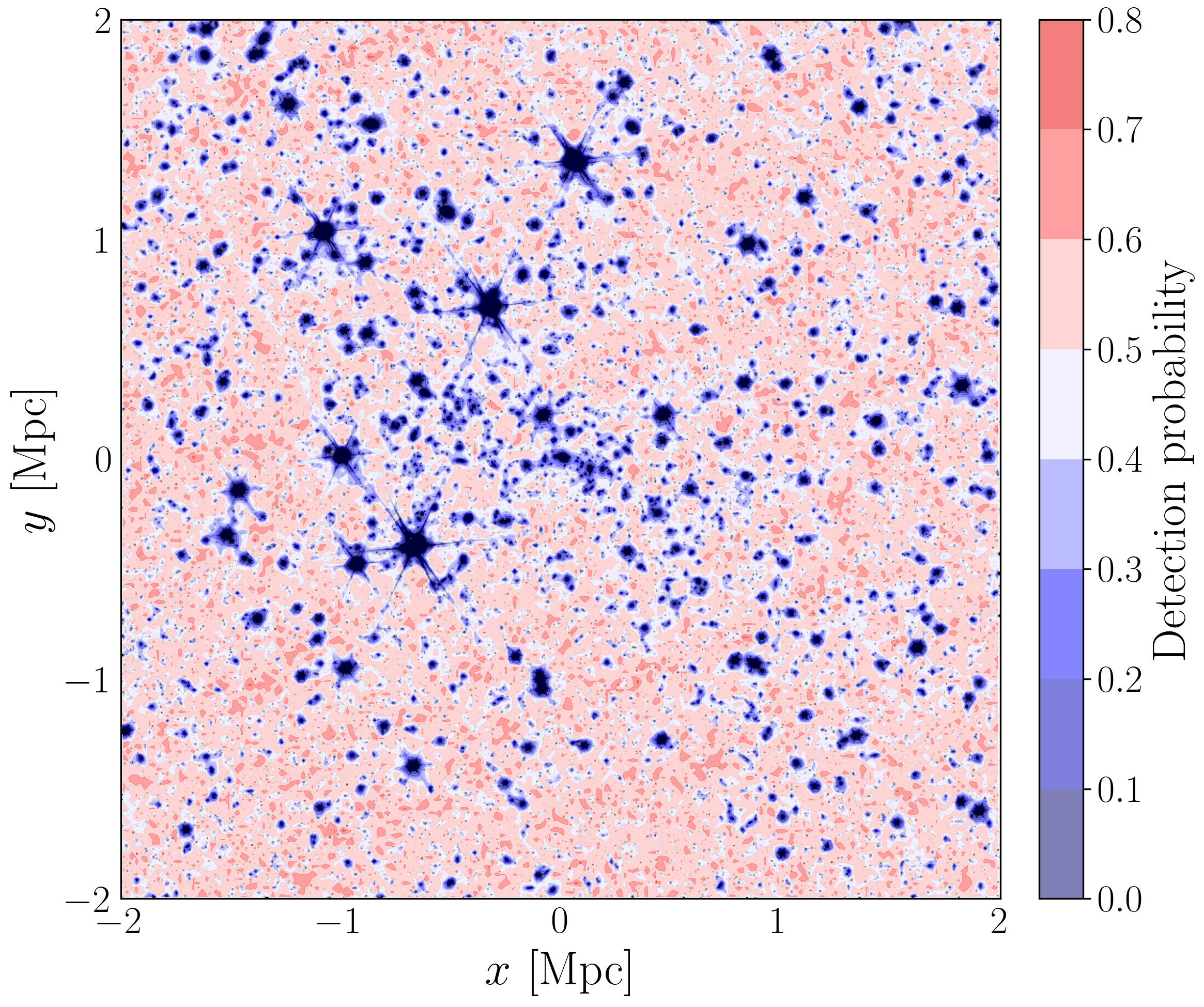}
    \caption{Overlaid with the monochrome VIS image of Abell 2390, the colour coding shows the
    detection probability map of the whole injection input sample in the inner $\SI{4}{Mpc}\times\SI{4}{Mpc}$, with no selection in input magnitude, photo-$z$, or other quantities applied. The image injection pipeline captures cluster galaxies and foreground stars including their diffraction spikes.}
    \label{fig:pdet_map}
\end{figure}

We inject $10^7$ galaxy images in total, which we split into three injection areas (or `batches') to compensate for the increasing geometric area of equidistant annuli. The first batch of injection runs covers the full $\SI{8}{Mpc}\times\SI{8}{Mpc}$ of the observed cluster field and consists of $422$ individual runs that provide a total of $\SI{4e6}{}$ injected galaxy image stamps into the VIS image. The second batch covers an intermediate area of the central $\SI{4}{Mpc}\times\SI{4}{Mpc}$, where we again inject $\SI{4e6}{}$ image stamps over 1686 injection runs. The third batch only covers the inner $\SI{2}{Mpc}\times\SI{2}{Mpc}$  of the cluster, where we inject $\SI{2e6}{}$ image stamps over 3373 injection runs. Each injection run in these three batches has the same galaxy injection density of
$\SI{29.4}{arcmin^{-2}}$.

The effect that foreground cluster galaxies have on the detection probability of background sources is expected to depend on the brightness of the background sources.
Therefore, we use the same split into magnitude bins that we employ for the WL analysis (\mbox{$22.0<\IE<24.5$} and \mbox{$24.5<\IE<26.5$}, see Sects.\thinspace\ref{sec:nofz} and \ref{sc:wl_source_bin_selection_beta}).
Given a potential additional dependence on size and signal-to-noise ratio, we additionally mimic the corresponding selections of the \texttt{KSB+} analysis using   \texttt{SExtractor}-measured quantities.
In the injection analysis we do not find a significant additional dependence of the detection probability on the input
photo-$z$
from the Flagship catalogue. We therefore  do not apply a further
sub-selection into photometric redshift bins.
Figure\thinspace\ref{fig:pdet_profile} shows the resulting detection probability profile for the two magnitude bins,
normalised to the mean of the values of the three outermost bins, which we regard as the field value. Here, the high-resolution VIS mask has been taken into account.
The density of re-detected injections decreases rapidly below a cluster-centric distance of $\SI{1}{Mpc}$, especially in the fainter magnitude bin, which we attribute to the increasing spatial density of cluster galaxies. The innermost annulus shows a sharp drop in detection density, likely due to the brightest cluster galaxy (BCG).
We note an additional intermediate dip at \mbox{$r\simeq 2.5 \,\mathrm{Mpc}$}.
We speculate that this may be caused by Galactic cirrus present in this area.
To test this hypothesis we  created a rudimentary cirrus mask and observe that the dip in detection density in this radial range corresponds well to the mask's radial profile (compare to Fig.\thinspace\ref{fig:pdet_profile}).

\begin{figure}[t]
    \centering
    \includegraphics[width=\linewidth]{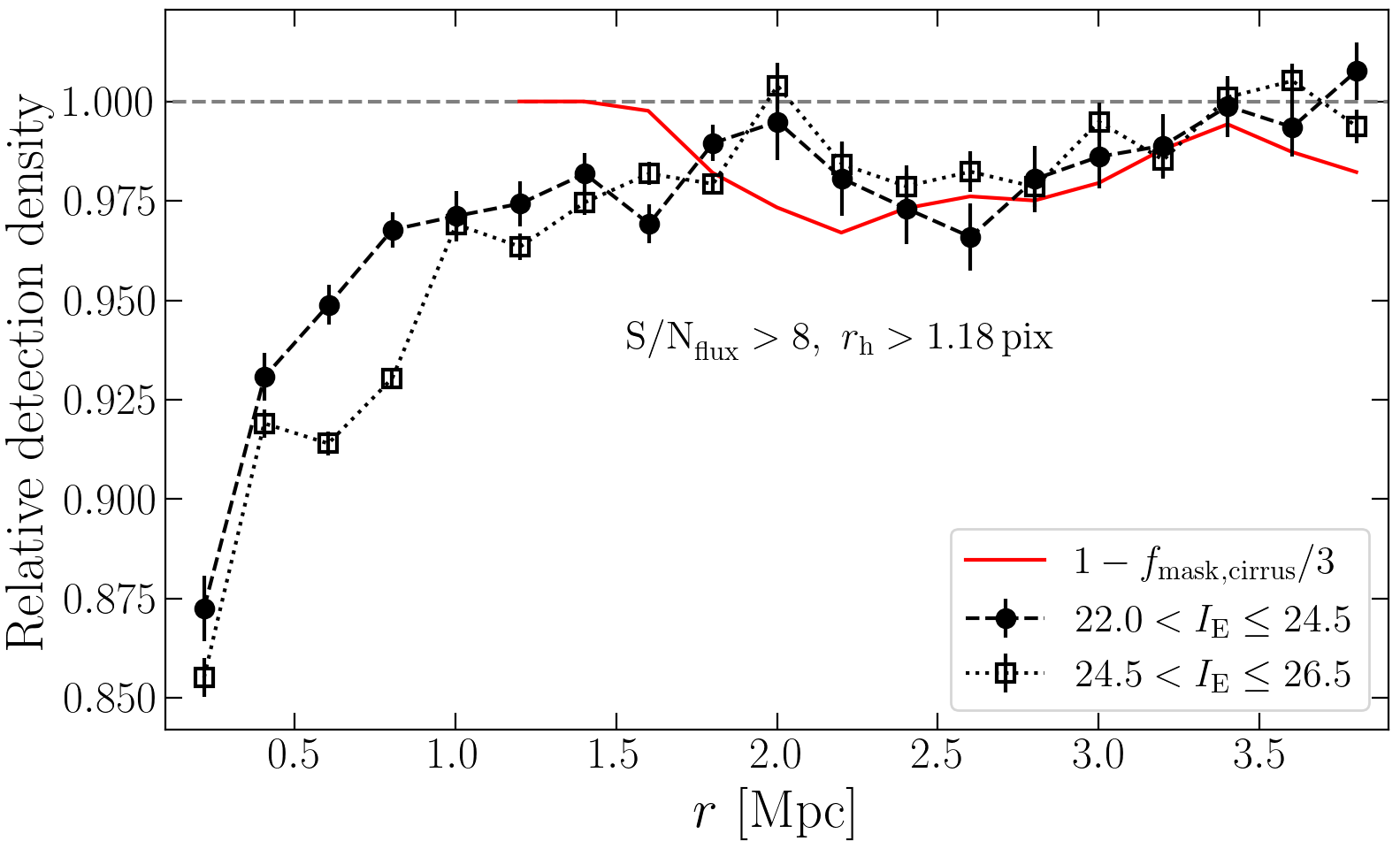}
    \caption{Relative spatial density of re-detected galaxies with a Flagship redshift in the range $0.3<z_\text{obs}<2.8$
    that fulfil the WL selection cuts.
    We show the contribution of a rudimentary cirrus mask to the unmasked area of each annulus, scaled by $\frac{1}{3}$, to visually match the corresponding dip in detection density.  }
    \label{fig:pdet_profile}
\end{figure}

\subsection{\label{sc:accounting_for_magnification}Accounting for magnification}
\label{sec:magnification}
Massive cluster haloes cause redshift-dependent lensing magnification, which affects the observed source number density in several ways \citep[e.g.,][]{schrabback18}.
Firstly, it increases the observed separation between objects, thereby reducing the number density of sources.
Moreover, the observed galaxy images increase in size. This can change the source selection if an analysis employs a size cut, which is the case in this work.
The size change may additionally affect the lensing weights derived from shape measurements.
We ignore this second-order correction here, but suggest that its impact is investigated for future studies that analyse larger cluster samples and have tighter accuracy requirements.
Lastly, the brightness of the sources increases because of magnification, changing the number of galaxies in a given magnitude bin.

We compute an approximate correction for the combined impact that magnification has on the source number density profile with the help of the Flagship catalogue.
Note that we do not follow the  often employed simplified treatment of magnification via the slope of the number counts \citep[e.g.,][]{broadhurst95}. Instead, we  apply an artificial magnification
to a full Flagship-based  galaxy population in order to be able to account for the range of effects listed in the previous paragraph.

In particular, we select $10^7$ Flagship galaxies randomly and divide them into tomographic redshift bins based on the `observed' redshift $z_\mathrm{obs}$ (including peculiar velocities) in the Flagship catalogue.
We assume  a reference Navarro--Frenk--White \citep[NFW, ][]{navarro97}  density profile with $M_\mathrm{200c}=1.5\times 10^{15}\, M_\odot$
and $c_{200\mathrm{c}}=4$
(consistent with earlier studies of the cluster, see \citealt{applegate14}, \citealt{okabe16}, \citealt{herbonnet20}, and Sect.\thinspace\ref{sc:comparison_earlier})
to compute the radius-dependent magnification for each galaxy in the Flagship sample following
\citet{schrabback18} as
\begin{equation}
\label{eq:mu}
  \mu(z_\mathrm{s},r)\simeq 1+2\,\frac{\beta(z_\mathrm{s})}
    {\beta_0}\kappa_0 (r)\;,
\end{equation}
where we employ  the `true' redshift  (without peculiar velocity) as the source redshift $z_\mathrm{s}$.
The geometric lensing efficiency
\begin{equation}
\label{eq:beta}
\beta(z_\mathrm{s})=\frac{D_\mathrm{ls}}{D_\mathrm{s}} H(z_\mathrm{s}-z_\mathrm{l}) \;,
\end{equation}
 depends on the
angular diameter distances between the observer and the source $D_\mathrm{s}$, and the lens and the source $D_\mathrm{ls}$, respectively. Here, $H(z_\mathrm{s}-z_\mathrm{l})$ denotes the Heaviside step function, which is equal to one for sources with $z_\mathrm{s}>z_\mathrm{l}$ and vanishes otherwise.
In Eq.\thinspace(\ref{eq:mu}) $\beta_0$ is an arbitrary fiducial lensing efficiency, which we calculate at the central redshift of each tomographic bin, and $\kappa_0(r)$ is the corresponding convergence at this redshift.
Magnification  changes galaxy sizes and magnitudes according to
\begin{align}
   r_\mathrm{h,eff}^\mathrm{mag} &= r_\mathrm{h,eff}\,\sqrt{\mu(z_\mathrm{true},r)}\;,\\
    \IE^\mathrm{mag} &= \IE - 2.5\log_{10}\mu(z_\mathrm{true},r)\;.
\end{align}

We approximately mirror the \texttt{KSB+} source selection in terms of cuts in (lensed) half-light radius and
$\mathrm{S}/\mathrm{N}_\mathrm{flux}$,
and then count the galaxies in the un-lensed and the lensed sample, for both the bright and  faint magnitude bins. We additionally include a weight \mbox{$W=1/\mu(z_\mathrm{true})<1$} for each galaxy in the magnified sample to account for the magnification of the observed sky area. The ratio of the two number counts then yields an estimate for the impact that magnification has on the source number densities.
We show the resulting radial profiles for the different bin combinations in Fig. \ref{fig:magnification}.

\begin{figure}
    \centering
    \includegraphics[width=\linewidth]{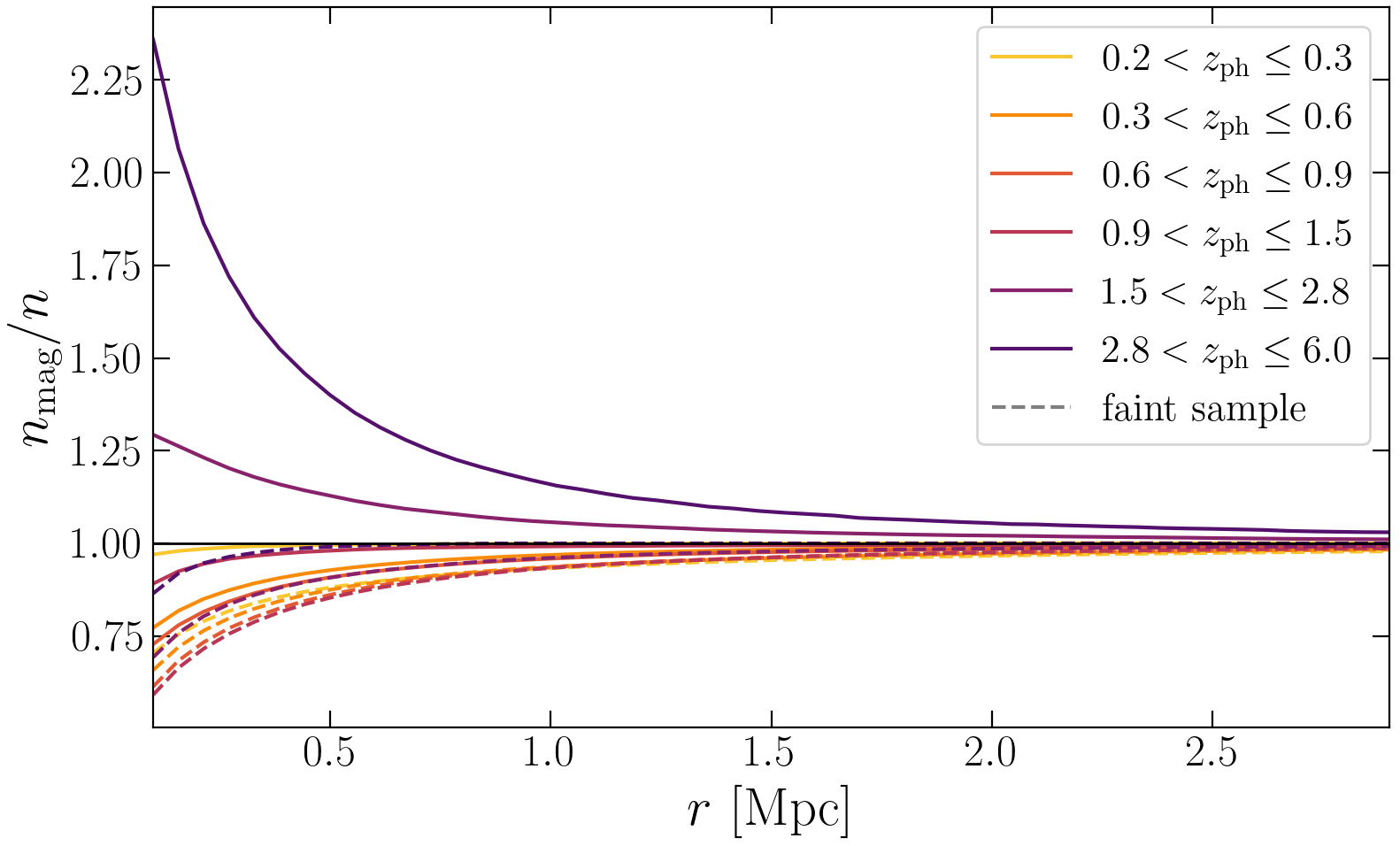}
    \caption{Impact of magnification on source number densities for our setup and an $M_{200\mathrm{c}}=1.5\times 10^{15}\, M_\odot,\ c_{200\mathrm{c}}=4$ NFW halo. The solid and dashed lines show the bright (\mbox{$22.0<\IE<24.5$}) and  faint (\mbox{$24.5<\IE<26.5$}) samples, respectively.}
    \label{fig:magnification}
\end{figure}

\subsection{\label{sec:contamination_correction}Cluster member contamination}
We use the estimated detection probability profiles from Sect.\thinspace\ref{sec:obscuration}
to boost the radial source density profiles on a bin-by-bin basis.
We decide against a model fit of the detection probability profiles,
because we can capture the intrinsic obscuration fingerprint of the cluster (and Galactic cirrus) in this way. We show the resulting detection bias-corrected source number density profile in Fig.\thinspace\ref{fig:nd_corr}, normalised to the mean value of the three outermost bins. We fit an exponential profile,
\begin{equation}
    \frac{n(r)}{n_\text{outer}}=1+A\,\exp\left(1-\frac{r}{r_\text{S}}\right)\,,
\end{equation}
 to the corrected source number density profiles, where
a separate amplitude $A$ is fit for each magnitude and photometric redshift bin combination, while  the scale radius $r_\text{S}=\SI{350(39)}{kpc}$ is jointly constrained from all bin combinations.
This model is a direct measure of the boost factor that we apply to the shear measurement. We show the resulting contamination
$1-n_\text{outer}/n(r)$
for the \texttt{KSB+} source sample in Fig.\thinspace\ref{fig:contamination}.

\begin{figure}
    \centering
    \includegraphics[width=\linewidth]{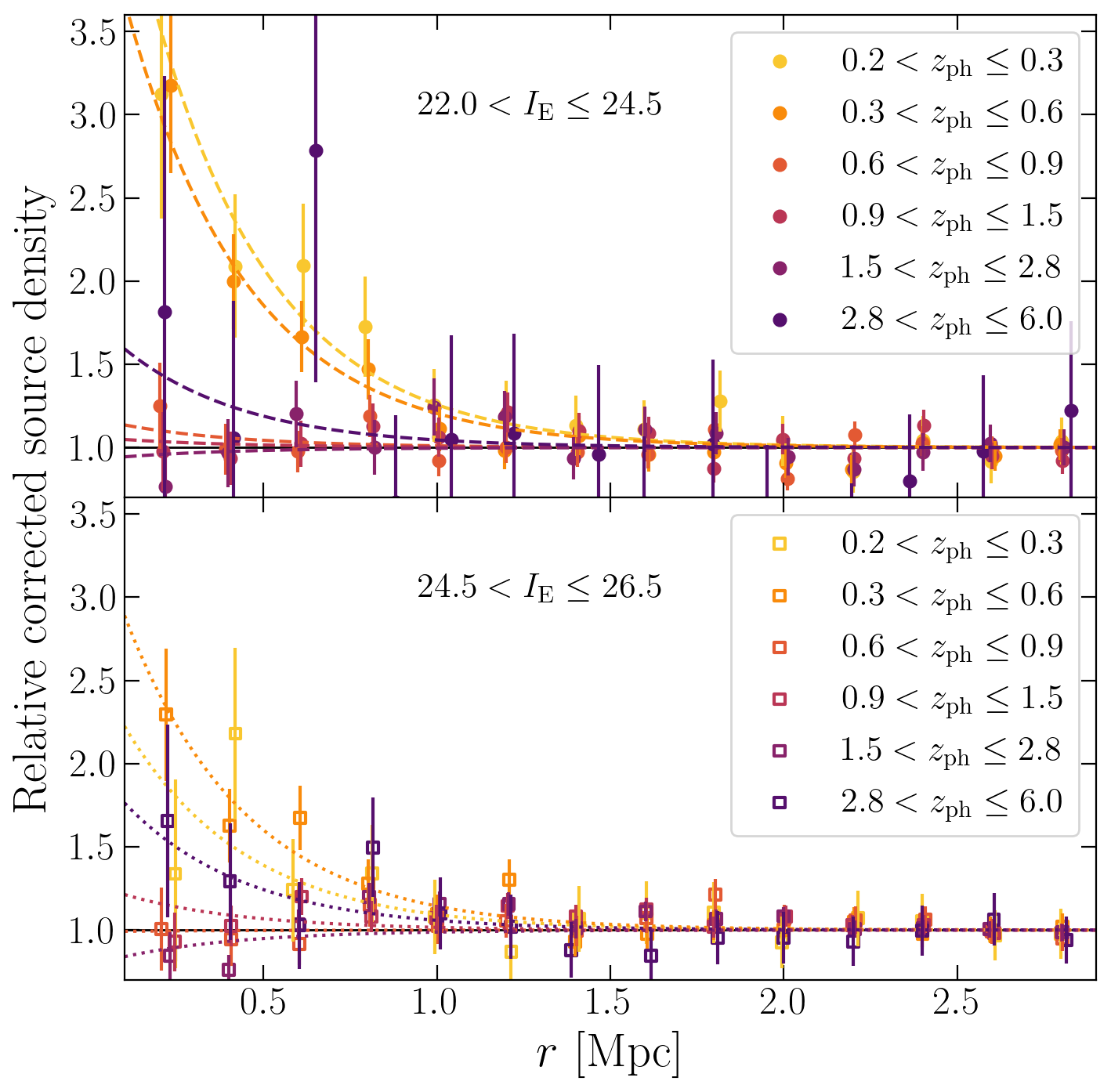}
    \caption{Source number density profiles after correction for the impact of source obscuration and magnification,  scaled to the corresponding mean value of the three outermost annuli.
   The increase over the baseline is a direct measure of the boost factor that we need to apply to compensate for cluster member contamination. We show an exponential model fit as dashed (\mbox{$22.0<\IE<24.5$}) and dotted (\mbox{$24.5<\IE<26.5$}) curves.}
    \label{fig:nd_corr}
\end{figure}

\begin{figure}
    \centering
    \includegraphics[width=\linewidth]{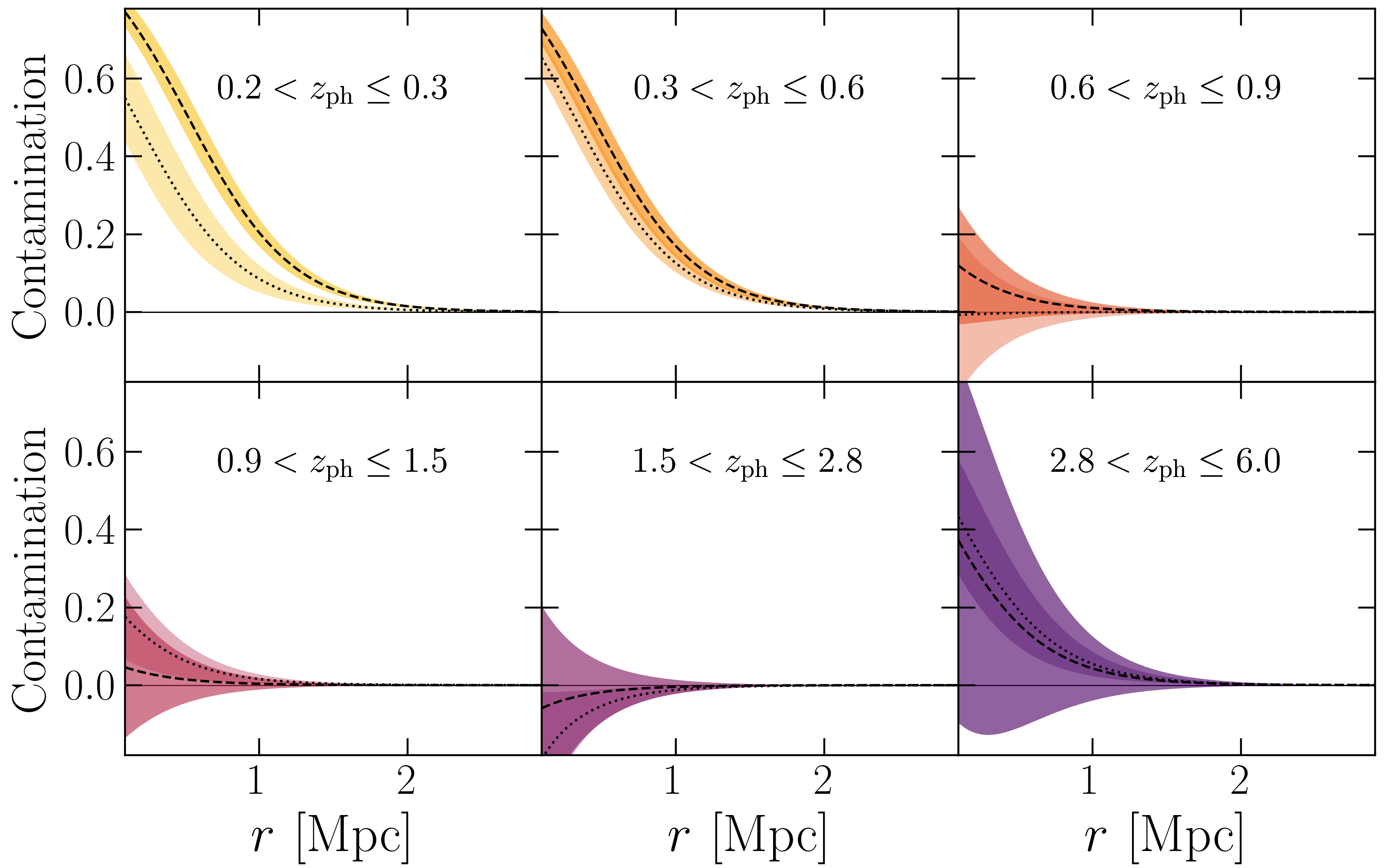}
    \caption{Contamination profiles for the \texttt{KSB+} lensing sample based on the fits of the obscuration- and magnification-corrected radial source density profiles for the different tomographic redshift and magnitude bin combinations. Dashed lines correspond to the bright sample (\mbox{$22.0<\IE<24.5$}), while dotted lines show the profiles for the faint sample (\mbox{$24.5<\IE<26.5$}). The shaded areas represent the $\pm 1\sigma$ uncertainties of the model fit.}
    \label{fig:contamination}
\end{figure}

\section{\label{sc:wl_results}Weak lensing results}

\subsection{\label{sc:wl_source_bin_selection_beta}Source bin selection and average lensing efficiencies}
In the contamination analysis presented in Sect.\thinspace\ref{sc:cluster_member_contamination} we considered a total of six photometric redshift bins. For the main WL results presented here
we drop both the bin with $0.2<z_\mathrm{ph}<0.3$, which includes the cluster redshift,
and the highest photometric redshift bin   $2.8<z_\mathrm{ph}<6.0$.
The latter bin is removed
since it does not contribute significant constraining power, containing only few sources, while
suffering  from both
cross-contamination in the $n(z)$ calibration (see Sect.\thinspace\ref{sec:nofz})
and
large uncertainties
regarding the contamination model (see Fig.\thinspace\ref{fig:contamination}).
We report the average geometric lensing efficiency $\langle \beta \rangle$, as well as $\langle \beta^2 \rangle$
for the remaining four photometric redshift bins and both magnitude bins in Table \ref{tab:beta}, taking the shape weights of the different shear catalogues into account.
Galaxies for which the source redshift distribution cannot be calibrated (see Sect.\thinspace\ref{sc:nofz}) are dropped  both for this computation and for the analyses presented in the following subsections.
\begin{table}[htbp!]
\caption{Average lensing efficiency values $\langle\beta\rangle$ and  $\langle\beta^2\rangle$ computed for the different magnitude and photo-$z$
bin combinations using the shear weights $w$ from \texttt{LensMC}, \texttt{SE++}, and \texttt{KSB+}, respectively.}
\smallskip
\label{tab:beta}
\smallskip
\addtolength{\tabcolsep}{-0.4em}
\begin{tabular}{cccccccc}
\hline
\hline
 \rule{0pt}{2ex}
$z_\mathrm{ph}$ & $\IE$ &  \multicolumn{2}{c}{$w_\mathrm{LensMC}$} & \multicolumn{2}{c}{$w_\mathrm{SE++}$} & \multicolumn{2}{c}{$w_\mathrm{KSB+}$} \\
& &  $\langle\beta\rangle$ & $\langle\beta^2\rangle$ &  $\langle\beta\rangle$ & $\langle\beta^2\rangle$ &  $\langle\beta\rangle$ & $\langle\beta^2\rangle$ \\[1pt]
\hline
 \rule{0pt}{2ex}
(0.3,0.6] & (22.0,24.5] & 0.448 & 0.244 & 0.448 & 0.244 & 0.452 & 0.244 \\
(0.3,0.6] & (24.5,26.5] & 0.467 & 0.303 & 0.469 & 0.305 & 0.461 & 0.293 \\
(0.6,0.9] & (22.0,24.5] & 0.610 & 0.392 & 0.609 & 0.391 & 0.621 & 0.399 \\
(0.6,0.9] & (24.5,26.5] & 0.574 & 0.371 & 0.573 & 0.371 & 0.582 & 0.375 \\
(0.9,1.5] & (22.0,24.5] & 0.709 & 0.522 & 0.709 & 0.522 & 0.719 & 0.529 \\
(0.9,1.5] & (24.5,26.5] & 0.718 & 0.540 & 0.717 & 0.539 & 0.720 & 0.540 \\
(1.5,2.8] & (22.0,24.5] & 0.797 & 0.648 & 0.797 & 0.648 & 0.804 & 0.654 \\
(1.5,2.8] & (24.5,26.5] & 0.785 & 0.637 & 0.785 & 0.636 & 0.788 & 0.640 \\

\hline
\end{tabular}
\end{table}

\subsection{\label{sc:wl_results_reconstruction}Weak lensing convergence reconstruction}

\begin{figure*}
    \centering
    \includegraphics[width=0.32\linewidth]{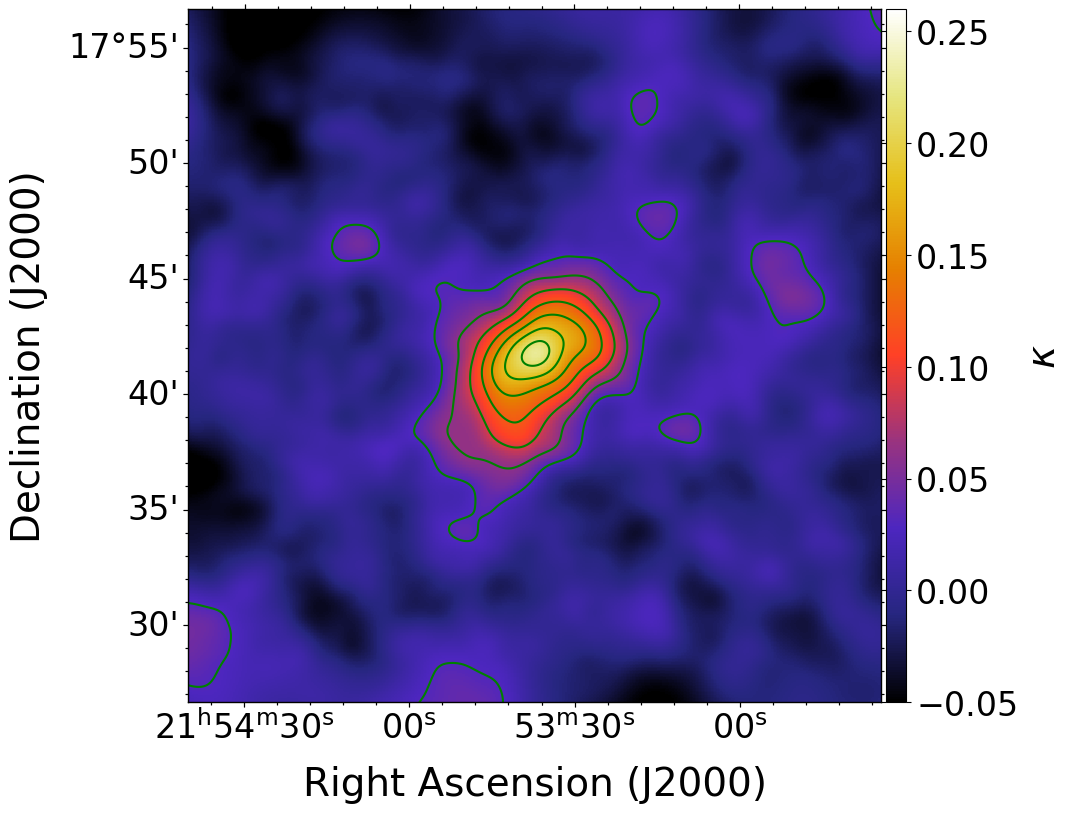}
         \includegraphics[width=0.32\linewidth]{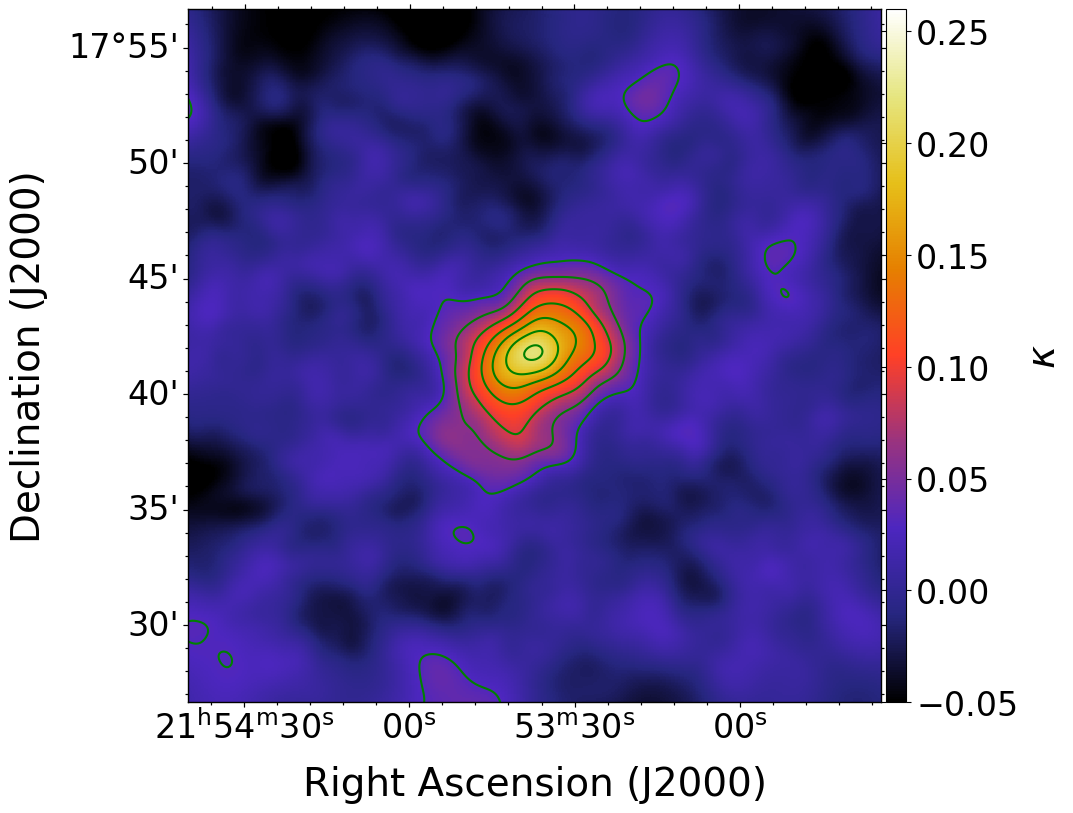}
             \includegraphics[width=0.32\linewidth]{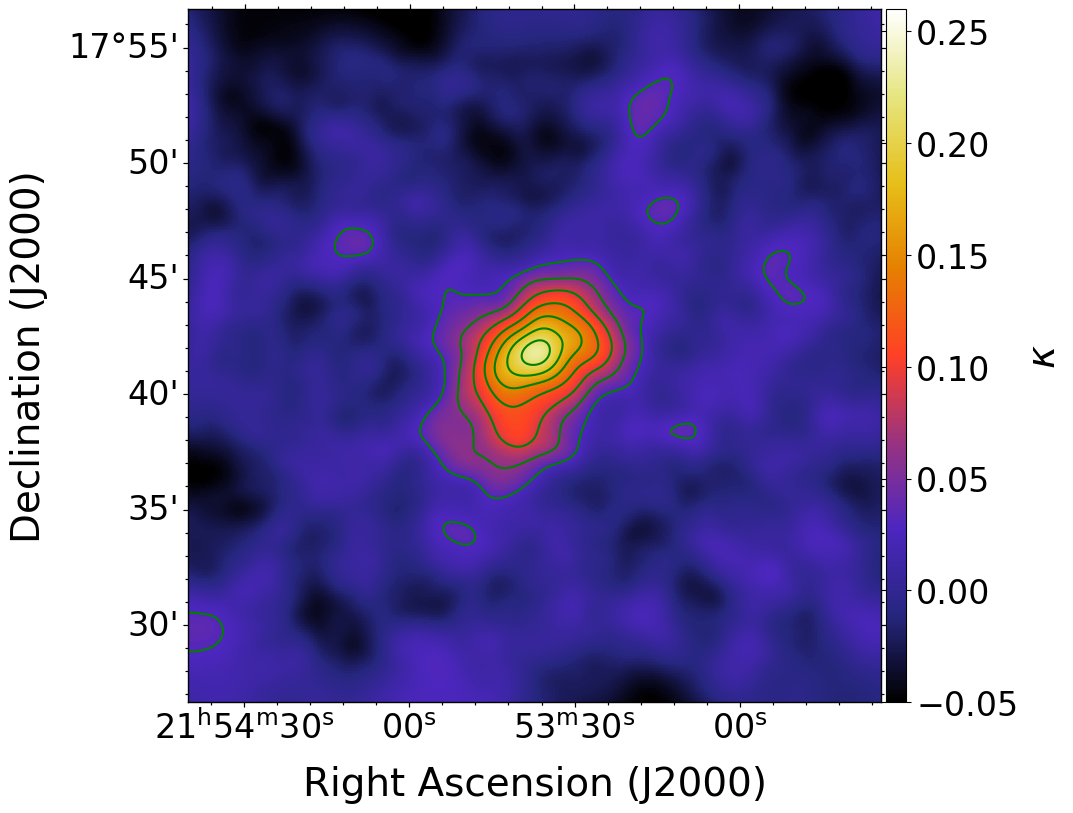}
    \includegraphics[width=0.32\linewidth]{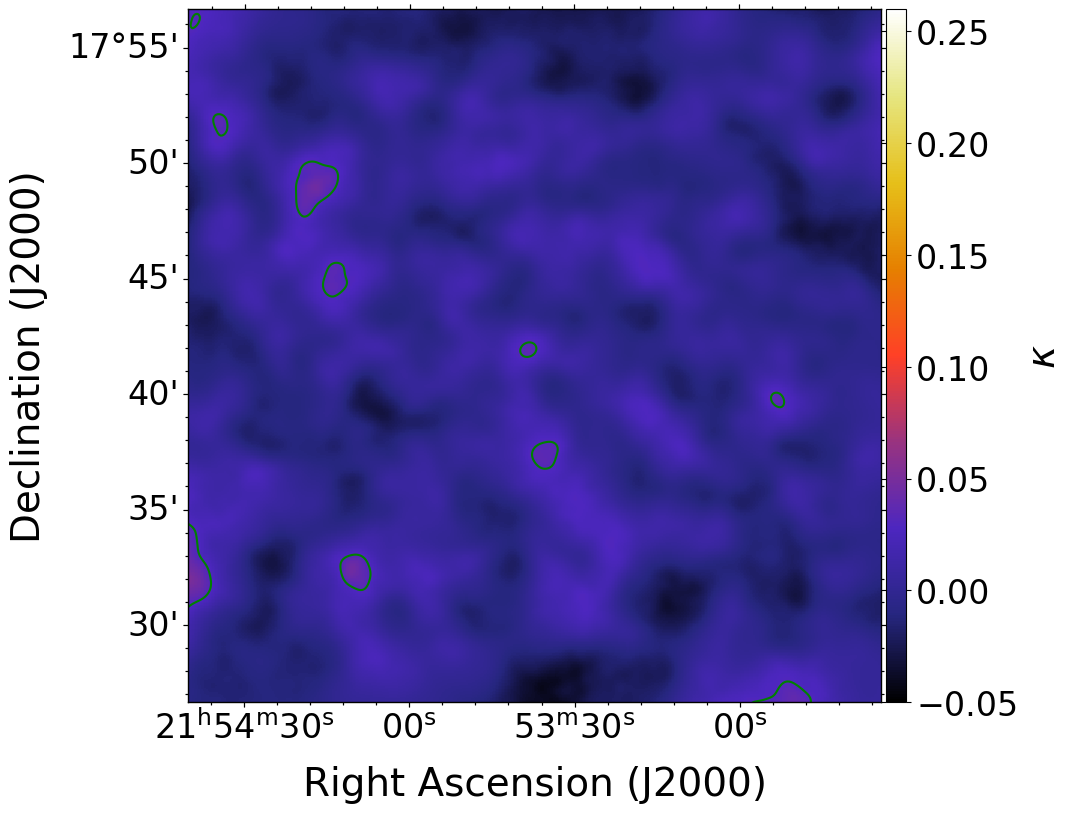}
    \includegraphics[width=0.32\linewidth]{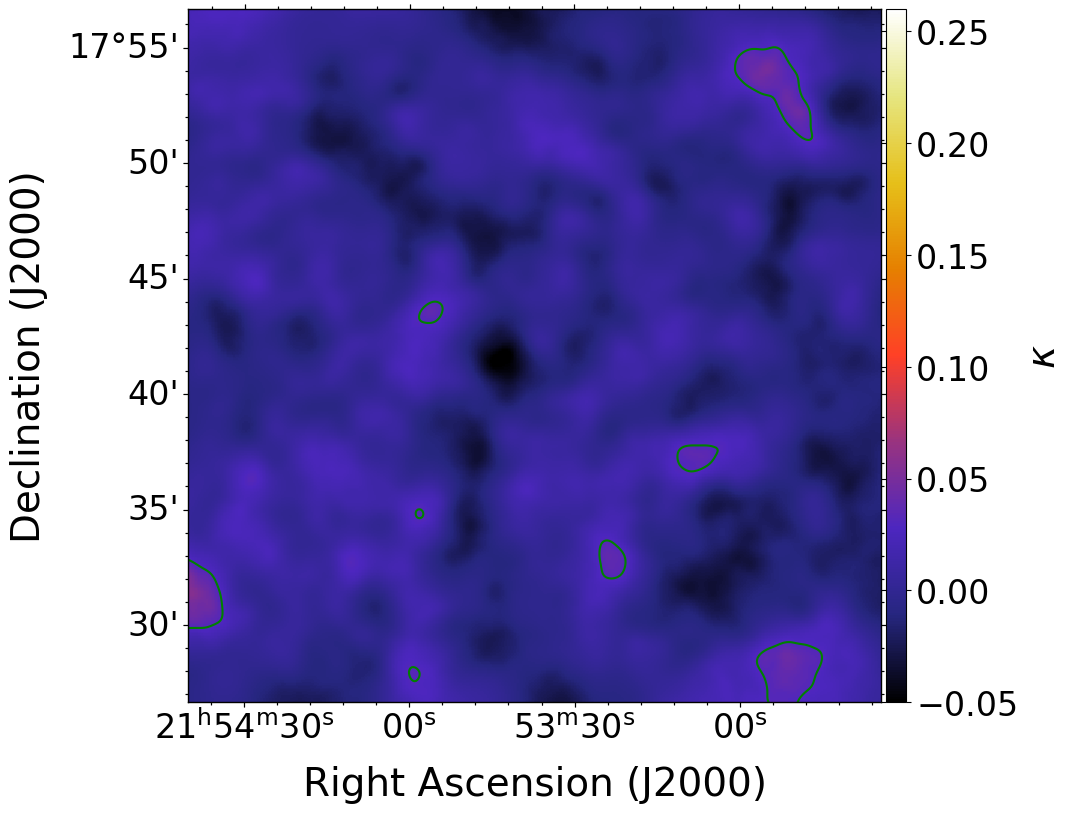}
    \includegraphics[width=0.32\linewidth]{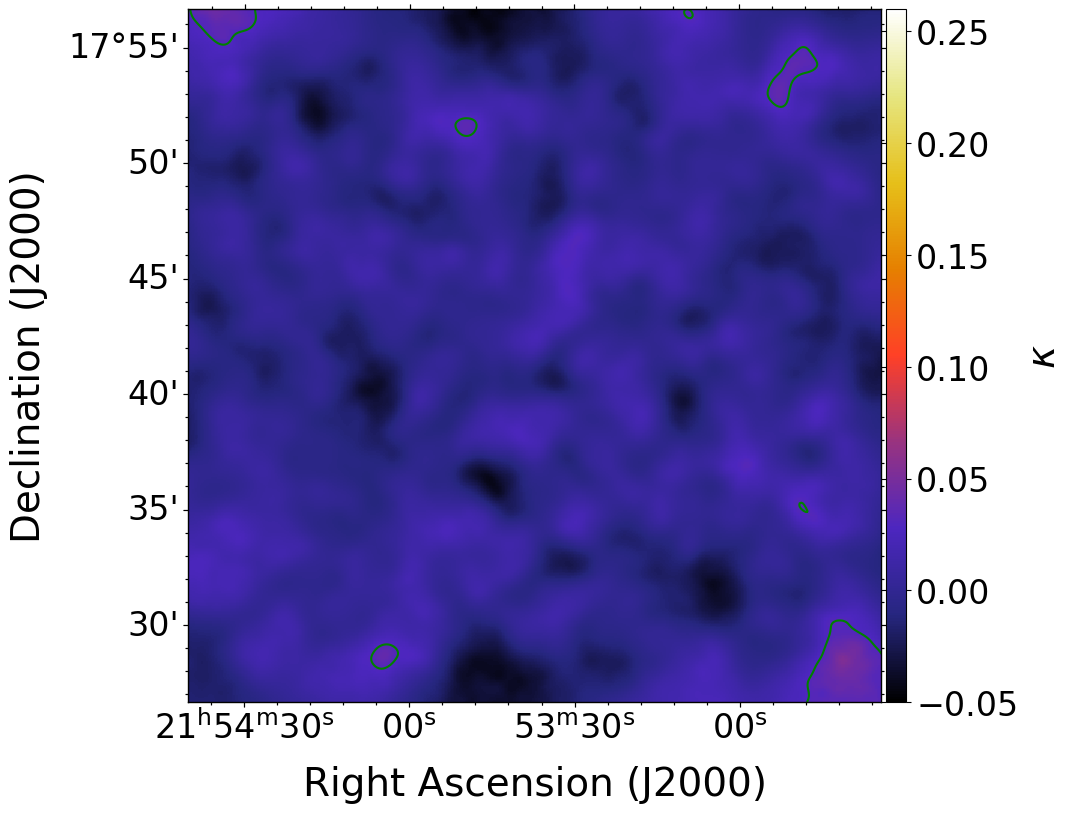}
   \caption{Convergence $\kappa$ reconstructions based on the \texttt{LensMC} (left), \texttt{KSB+} (middle), and \texttt{SE++} (right) shear catalogues, showing the $E$ mode (top) and $B$ mode (bottom). Contours are spaced in steps of  \mbox{$\Delta\kappa=0.03$}  starting at \mbox{$\kappa=0.03$}.  }
\label{fig:massrecon}
\end{figure*}

\begin{figure*}
    \centering
        \includegraphics[width=0.75\linewidth]{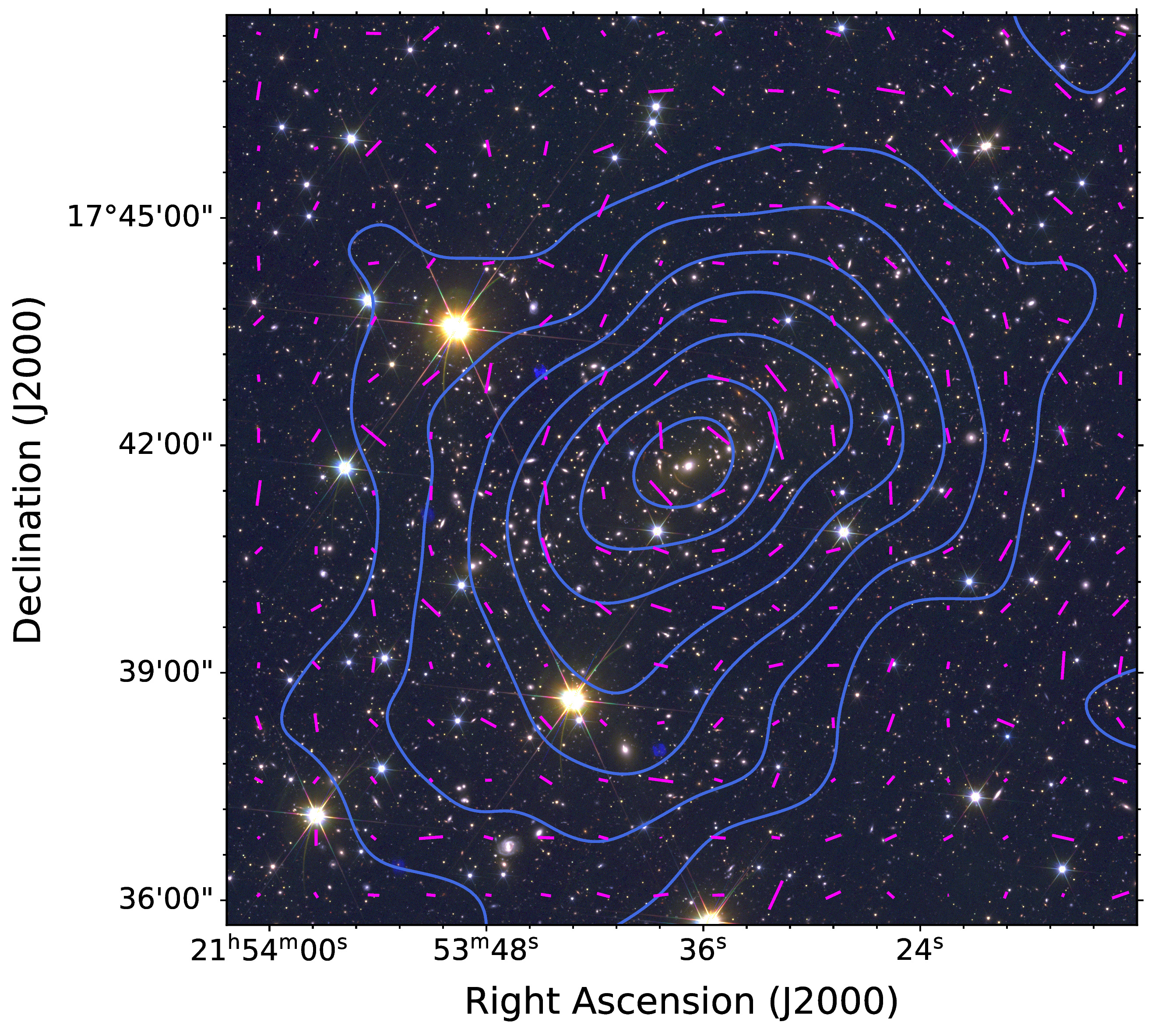}
    \caption{Overlay of a {\it Euclid}
    $\HE \YE \IE$ colour image of A2390 with the Wiener-filtered $E$-mode $\kappa$ reconstruction shown as contours  in steps of  $\Delta\kappa=0.03$, starting at $\kappa=0.03$, as derived from the \texttt{LensMC} shape measurements.
    For illustration the magenta whiskers show the estimated shear field binned on a coarse grid without applying smoothing (a finer grid is used to compute the $\kappa$ reconstruction, see Appendix \ref{appendix:wiener_filtered}).
    }
    \label{fig:image_mass_overlay_lensMC}
\end{figure*}

The
WL convergence $\kappa=\Sigma/\Sigma_\mathrm{crit}$ is defined as the ratio
of the
surface mass distribution $\Sigma$
and the critical surface mass density
\begin{equation}
  \label{eqn:sigmacrit}
  \Sigma_{\mathrm{crit}} = \frac{c^2}{4\pi G}\frac{1}{D_{\mathrm{l}}\, \beta }\;,
\end{equation}
which is given in terms of the  gravitational constant $G$,
the light speed $c$,
the angular diameter distance  to the lens $D_{\mathrm{l}}$,
and
the geometric lensing efficiency $\beta$ (see Eq.\thinspace\ref{eq:beta}).

Since the
WL convergence $\kappa$ and shear $\gamma$ are second-order derivatives of the lensing potential \citep{bartelmann01}, reconstructions of the $\kappa$ distribution can be estimated from the shear field up to an integration constant known as the mass-sheet degeneracy
\citep{falco85,schneider95}.
For the reconstruction we employ a Wiener-filtered algorithm following \citet{mcinnes09} and \citet{simon09}, as described in more detail in Appendix \ref{appendix:wiener_filtered}.
In the reconstruction we set  the mean convergence in the field to zero to fix the mass-sheet degeneracy.
This
can lead to noticeable $\kappa$ underestimations for small field sizes \citep[e.g.,][]{schrabback21b}, but is less of an issue for the wide field covered by the \Euclid observations and complementary ground-based data. Also note that we do not use the $\kappa$ reconstructions shown in Figs.\thinspace\ref{fig:massrecon} and \ref{fig:image_mass_overlay_lensMC}
(as well as Figs.\thinspace\ref{fig:image_mass_overlay_KSB} and \ref{fig:image_mass_overlay_SEpp}) for quantitative constraints, but rather only for illustration purposes.

When fully corrected for shape measurement and selection biases, the galaxy ellipticity estimates provide unbiased estimates for the reduced shear, which is linked to the (unobservable) shear $\gamma$ and convergence $\kappa$ as
\begin{equation}
 \label{eqn:reduced shear}
g=\frac{\gamma}{1-\kappa}\, .
\end{equation}
Given the high cluster-mass scale and lensing efficiency, the distinction between shear and reduced shear cannot be neglected for our study. For the quantitative mass constraints presented in Sect.\thinspace\ref{sc:wl_results_mass} we correctly model the reduced shear from the redshift-dependent shear and convergence.
For the convergence reconstructions
we apply an approximate correction, scaling first the individual boost-corrected reduced shear estimates from the different tomographic redshift bins (with \mbox{$0.3< z_\mathrm{ph}<2.8$}) and both magnitude bins to the same average lensing efficiency, in order to combine them into a single catalogue of sources that have  non-zero shape weights and photometric redshift calibration weights. We then conduct the convergence reconstruction iteratively, applying a conversion from reduced shear to shear based on the convergence map of the previous iteration.

Figure\thinspace\ref{fig:massrecon} shows the convergence reconstructions over the area covered by the \Euclid and photometric data for the
three different shear catalogues  in the top panels ($E$ mode). The bottom panels show
the corresponding $B$-mode
reconstructions,
for which galaxy ellipticities have been rotated by $45^\circ$, providing an estimate for the level of noise and potential residual systematics in the reconstruction \citep[e.g.,][]{massey07}.
Based on the comparison of the $E$-mode and $B$-mode reconstructions,
Fig.\thinspace\ref{fig:massrecon} shows
that the cluster is detected with high significance for all three shear catalogues.
All reconstructions show very similar morphologies,
with a significant elongation approximately along the south-east to north-west direction. This can be compared to the \Euclid optical+NIR colour image of the cluster via the overlay presented in Fig.\thinspace\ref{fig:image_mass_overlay_lensMC}
for the \texttt{LensMC} analysis, and in
 Figs.\thinspace\ref{fig:image_mass_overlay_KSB} and
\ref{fig:image_mass_overlay_SEpp} in Appendix \ref{app:overlays} for the other shear catalogues.
This shows that the $\kappa$ contours trace the distribution of cluster galaxies well for all three reconstructions, especially in the inner cluster region. These figures also show that the central peaks of the reconstructions closely coincide with the BCG.

\subsection{\label{sc:wl_results_mass}Weak lensing mass constraints}

\subsubsection{\label{sc:wl_shear_profile_fitting}Shear profile fitting}

In order to derive WL mass constraints for the cluster
we employ the sources of both magnitude bins and the four tomographic redshift bins with
\mbox{$0.3< z_\mathrm{ph}<2.8$}
that have  non-zero weights both in the shape analysis and photo-$z$
calibration.
We compute tangential reduced shear profiles separately for each
photometric redshift and magnitude bin combination, but  jointly fit all bins
with a single spherical NFW  \citep{navarro97} mass model, for which we compute tangential reduced shear profile predictions according to \citet{brainerd00}.
Here we
account for the
correct $\langle\beta\rangle$ for each bin combination
and also apply
corrections for the finite width of their
redshift distributions
based on the estimated
$\langle\beta^2\rangle$
following  \citet{seitz97}, \citet{hoekstra00}, and \citet{applegate14}.

For the mass constraints we employ the radial fit range
\mbox{$0.5\, \mathrm{Mpc}\le r \le 3.3 \,\mathrm{Mpc}$}. This avoids increased systematic uncertainties that occur both in the core region, where the contamination correction has the largest impact, and at very large scales, where deviations from a single NFW model are expected due to neighbouring structures \citep[e.g.,][]{grandis24}.

For noisy
WL data the centring of shear profiles poses a significant challenge for accurate mass measurements.
For example, when centre proxies such as the X-ray centroid,
the peak in the
 Sunyaev-Zeldovich-effect
\citep[SZE,][]{sunyaev70} signature,
or the BCG candidate are used, significant offsets can occur \citep[e.g.,][]{schrabback18}.
This can lead to biased mass constraints even when corrections for isotropic miscentring distributions are applied \citep{sommer24},
although  \citet{sommer25} find that this problem can be mitigated by
modelling the miscentring with respect to the centre of mass.
Likewise, substantial biases occur when centring shear profiles directly on noisy reconstructed
WL
convergence peaks \citep{sommer22}.
However, for our study this is not an issue, since the mass centre of the cluster is very well constrained, both by the WL data themselves (see Sect.\thinspace\ref{sc:wl_results_reconstruction}), and by strong lensing constraints of the core region.
In particular, the joint strong and weak lensing analysis presented in our companion paper
(Diego et al.~in prep.)
precisely constrains the cluster mass centre to the position
$\mathrm{RA(J2000)}=\ang{328.40494}, \mathrm{Dec(J2000)}=\ang{17.69460}$
located  about $6^{\prime\prime}$
southeast from the BCG.
This provides a centre for our shear profile analysis with negligible uncertainty, which is used throughout our analysis unless noted differently.

When fitting NFW models for individual clusters,
WL
studies
often employ a concentration--mass relation \citep[e.g.,][]{herbonnet20,schrabback21b} or a fixed  concentration \citep[e.g.][]{vonderlinden14b} given the limited constraining power and radial fit range of the data. Here we also follow this approach and assume a fixed concentration \mbox{$c_\mathrm{200c}=4$} consistent with \citet[][]{vonderlinden14b}, leaving the mass  \mbox{$M_\mathrm{200c}$} as the only free parameter (constraints on both mass and concentration derived from joint strong and weak lensing modelling are presented in Diego et al.~in prep.).
As usual, we employ over-density masses \mbox{$M_{\Delta c}$}, which are measured in
spheres with radius \mbox{$r_\mathrm{\Delta c}$}, within which the average density is equals $\Delta$ times the critical density of the Universe at the cluster redshift.

\begin{figure*}[htb]
\centering
\includegraphics[angle=0,width=0.32\hsize]{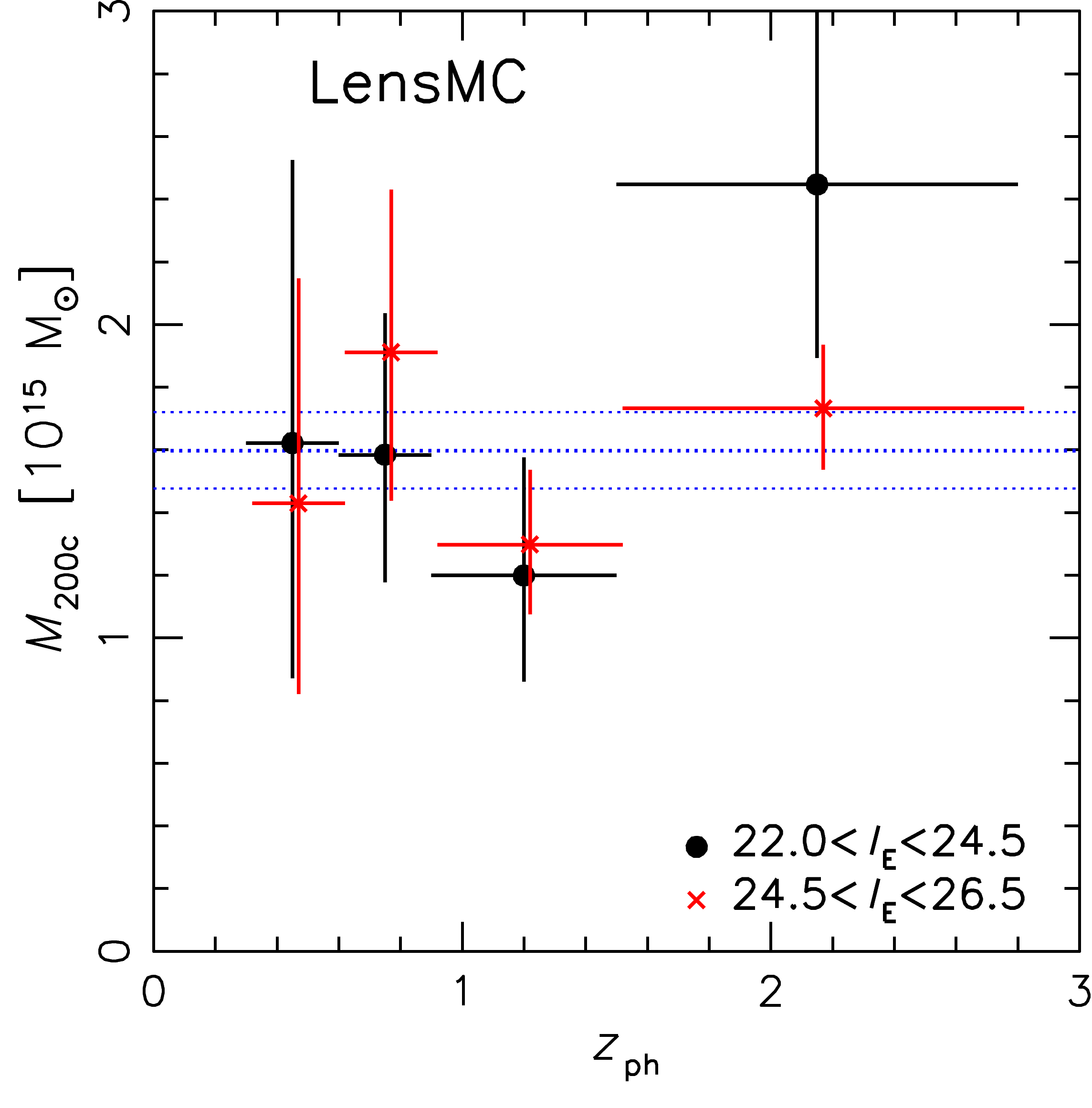}
\includegraphics[angle=0,width=0.32\hsize]{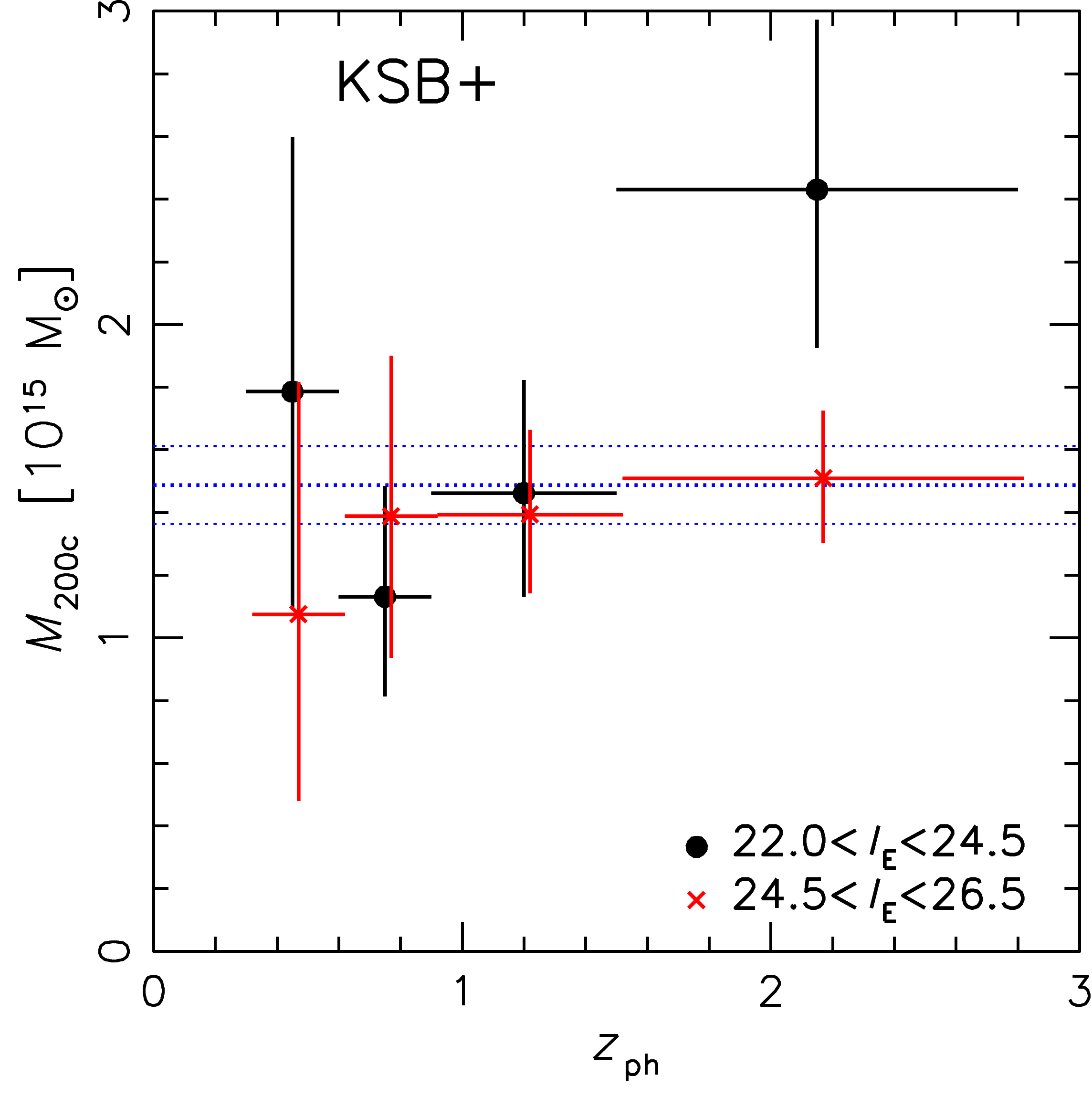}
\includegraphics[angle=0,width=0.32\hsize]{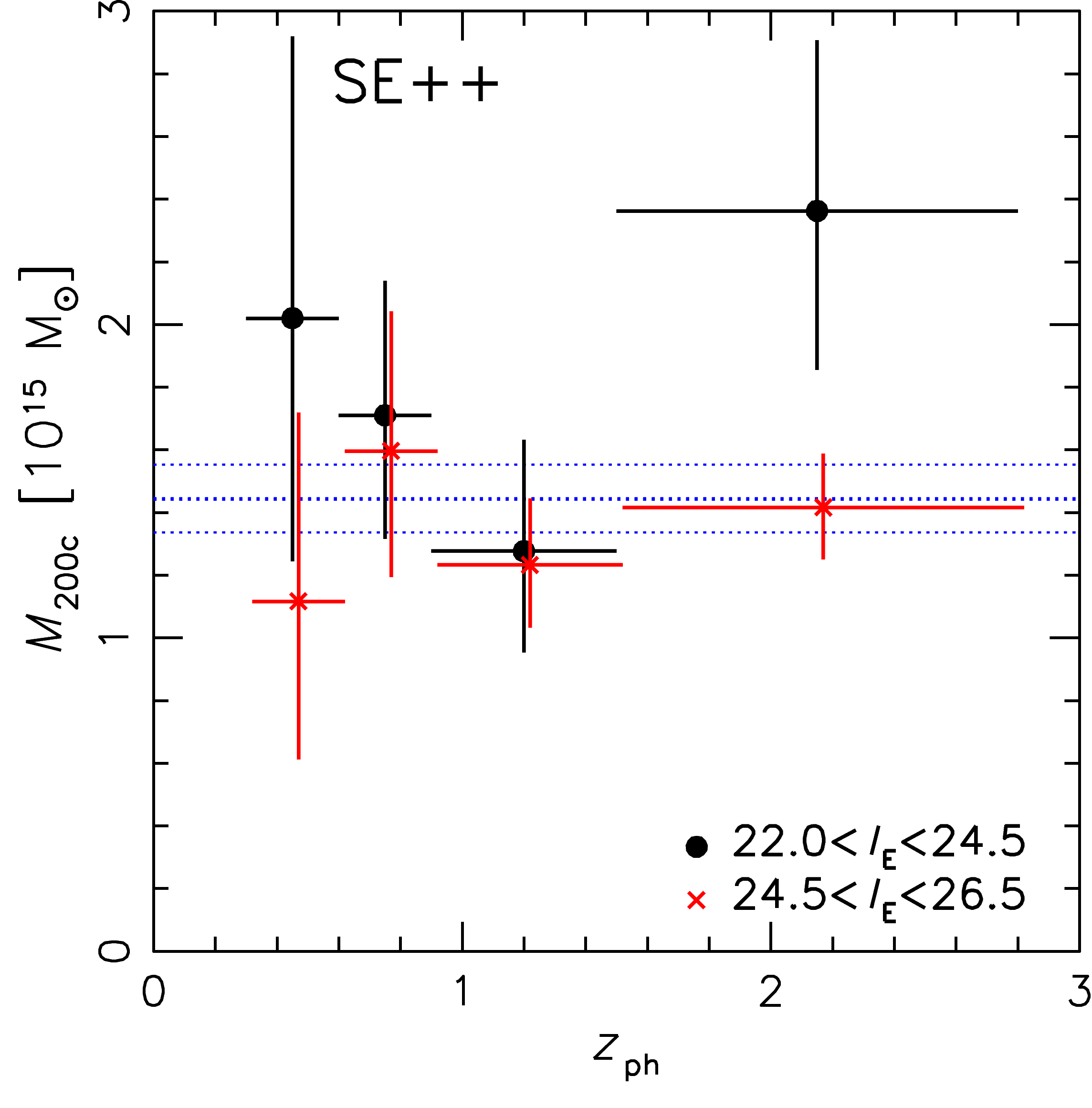}
\caption{WL constraints on $M_\mathrm{200c}$ based on the \texttt{LensMC} (left), \texttt{KSB+} (middle), and \texttt{SE++} (right) shear estimates. The dashed blue lines show the average and $1\thinspace\sigma$ uncertainty range, while the symbols show the individual  constraints from the different magnitude and photometric redshift bin combinations.  The black circles show the bright magnitude bins and are plotted at the bin centre with horizontal errorbars indicating the bin width. The red crosses show the results from the faint magnitude bin and have been offset by $\Delta z_\mathrm{ph}=0.02$ for clarity. All results assume a fixed concentration $c_\mathrm{200c}=4$, with plotted mass uncertainties corresponding to shape noise only.}
\label{fig:mass_as_fct_of_bin}
\end{figure*}

\begin{figure*}[htb]
\centering
\includegraphics[angle=0,width=0.32\hsize]{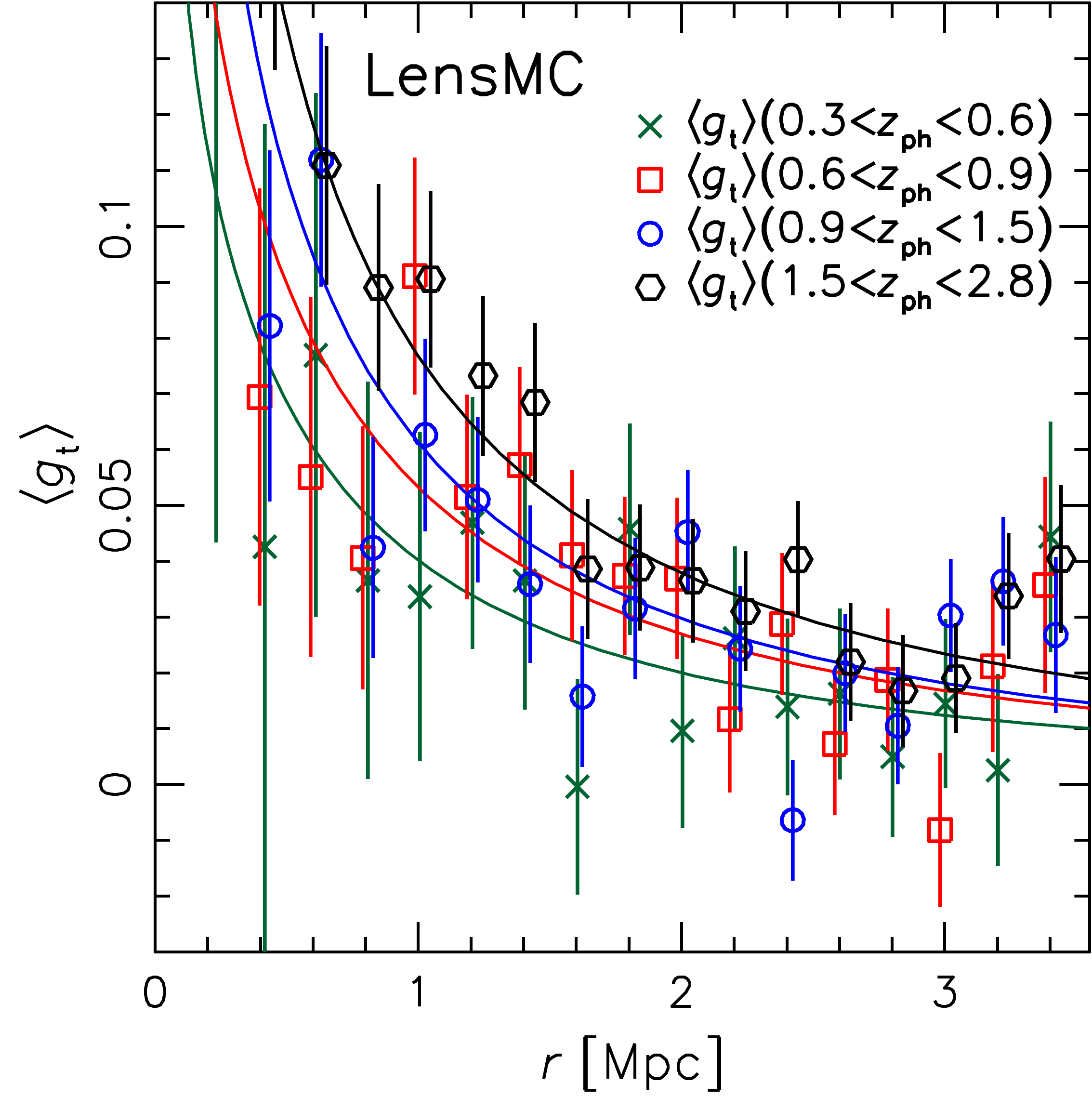}
\includegraphics[angle=0,width=0.32\hsize]{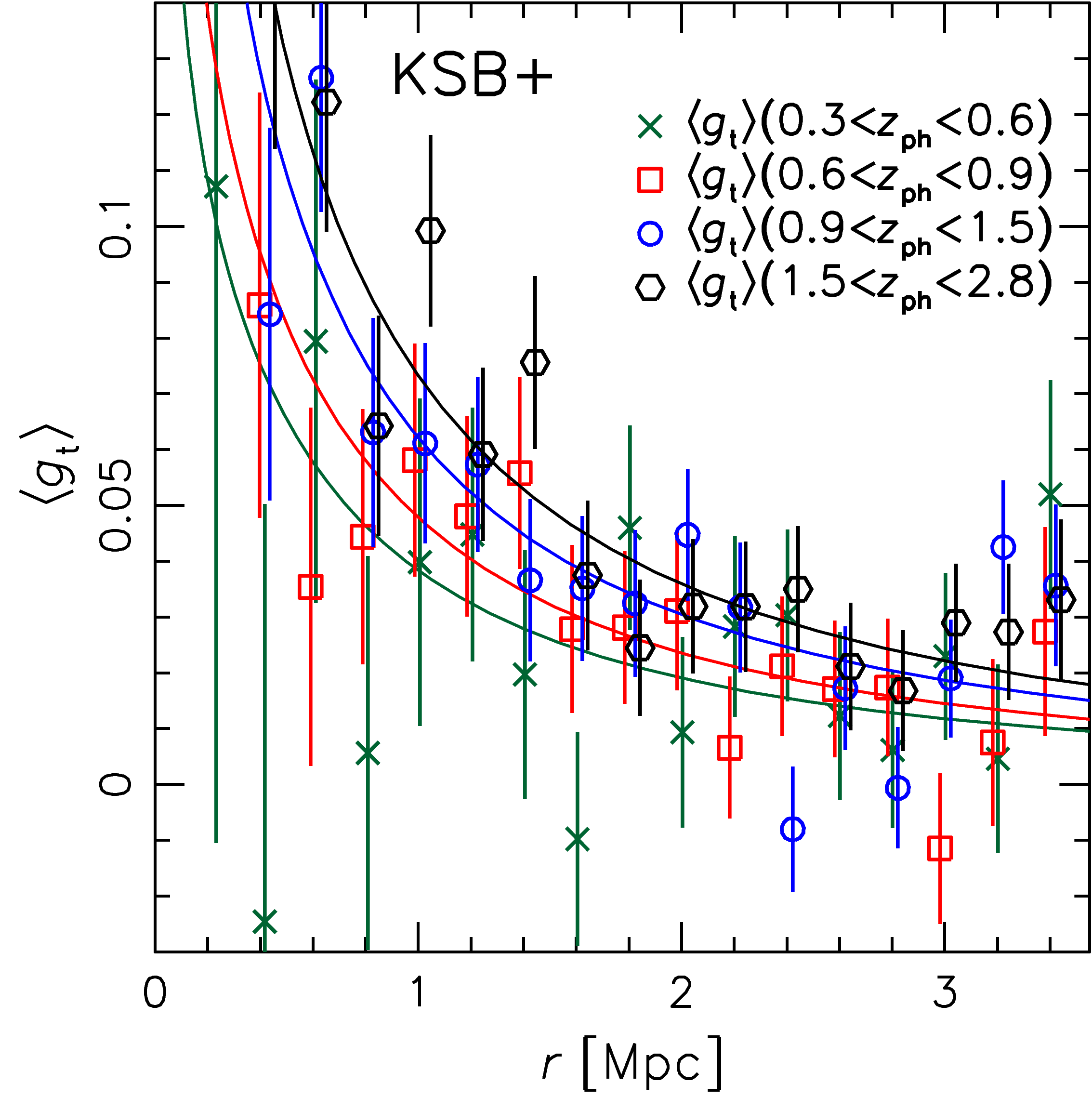}
\includegraphics[angle=0,width=0.32\hsize]{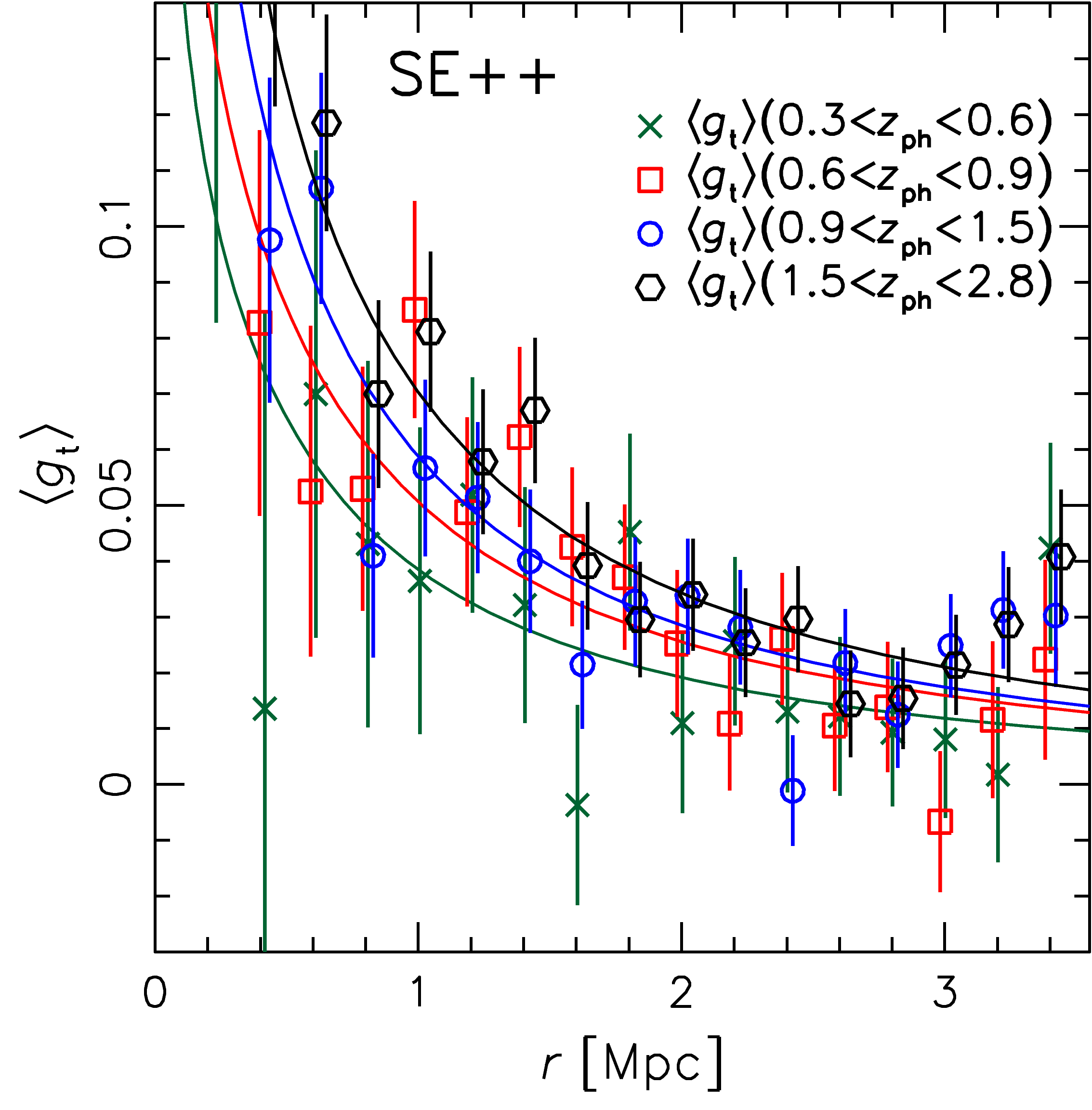}
\caption{Magnitude-bin-combined contamination-corrected tangential reduced shear
$\langle g_\mathrm{t}\rangle (r)$
profiles of A2390 based on the \texttt{LensMC} (left), \texttt{KSB+} (middle), and \texttt{SE++} (right) shear estimates. For this figure, the individual noisy
$\langle g_\mathrm{t}\rangle (r)$
profiles of the two magnitude  bins have been combined (slightly scaled to their average mean $\langle\beta\rangle$) to yield a single
$\langle g_\mathrm{t}\rangle (r)$
profile for each tomographic redshift bin.
For each set of data points the curve plotted in the same colour shows the correspondingly averaged  NFW reduced shear profile model
for the jointly constrained best-fit cluster mass,
assuming a fixed concentration $c_\mathrm{200c}=4$.
The data and model for the tomographic bin with \mbox{$0.3<z_\mathrm{ph}<0.6$} is shown at the correct position, while the
points and
models of the other tomographic bins have been shifted consistently
along the $x$-axis
for clarity.}
\label{fig:combined_shear_profile_tomo}
\end{figure*}

\begin{figure*}[htb]
\centering
\includegraphics[angle=0,width=0.32\hsize]{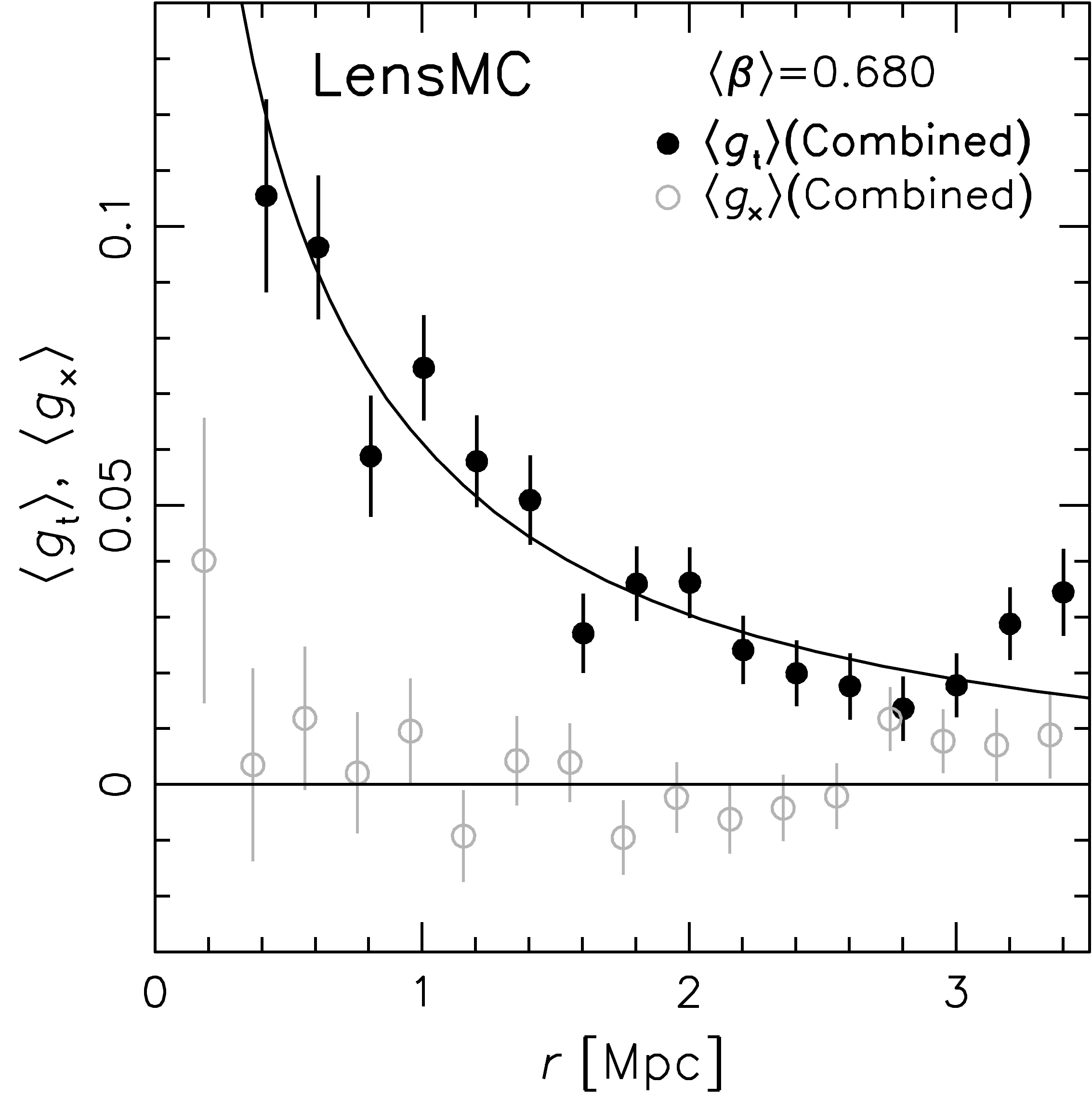}
\includegraphics[angle=0,width=0.32\hsize]{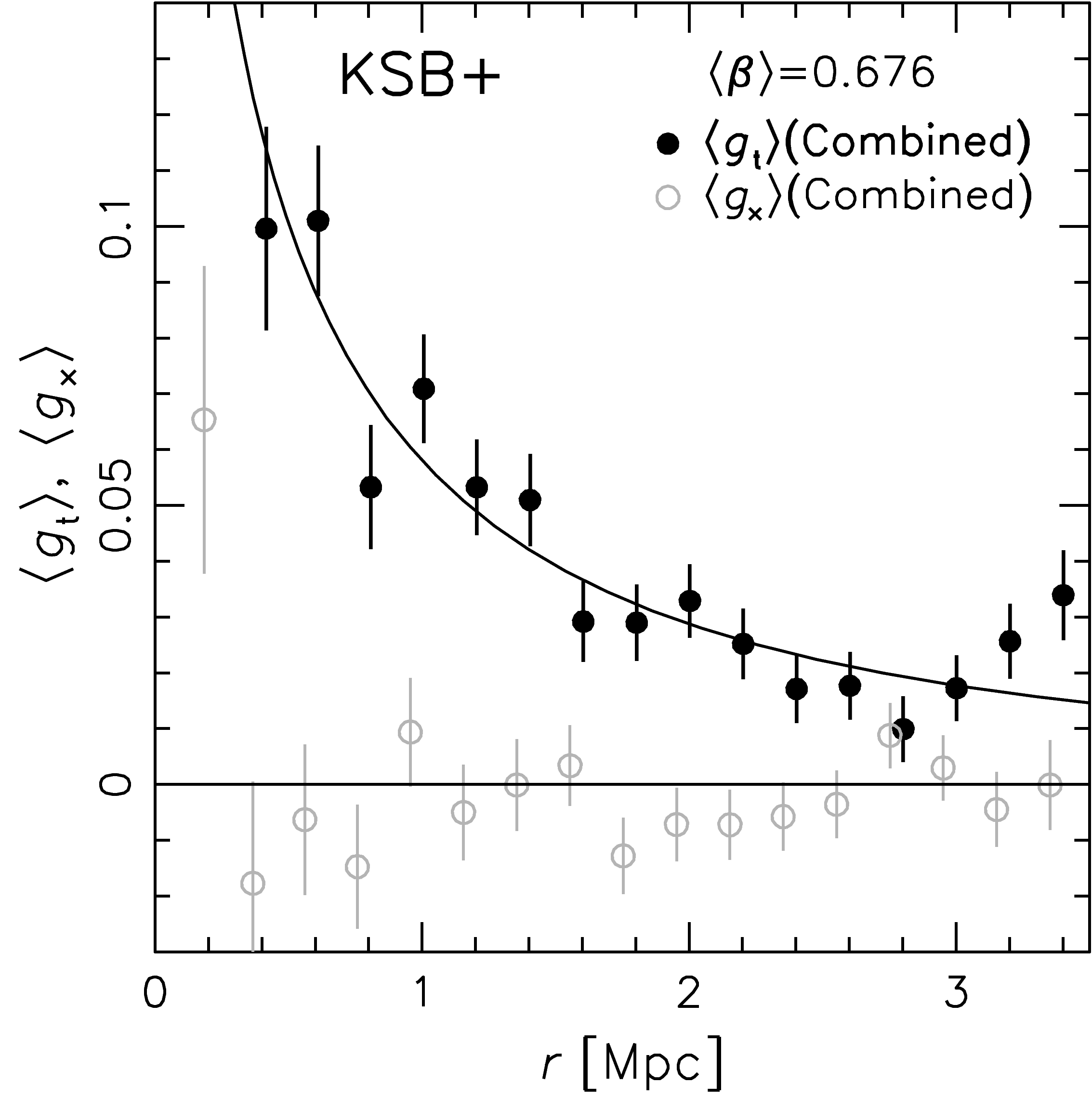}
\includegraphics[angle=0,width=0.32\hsize]{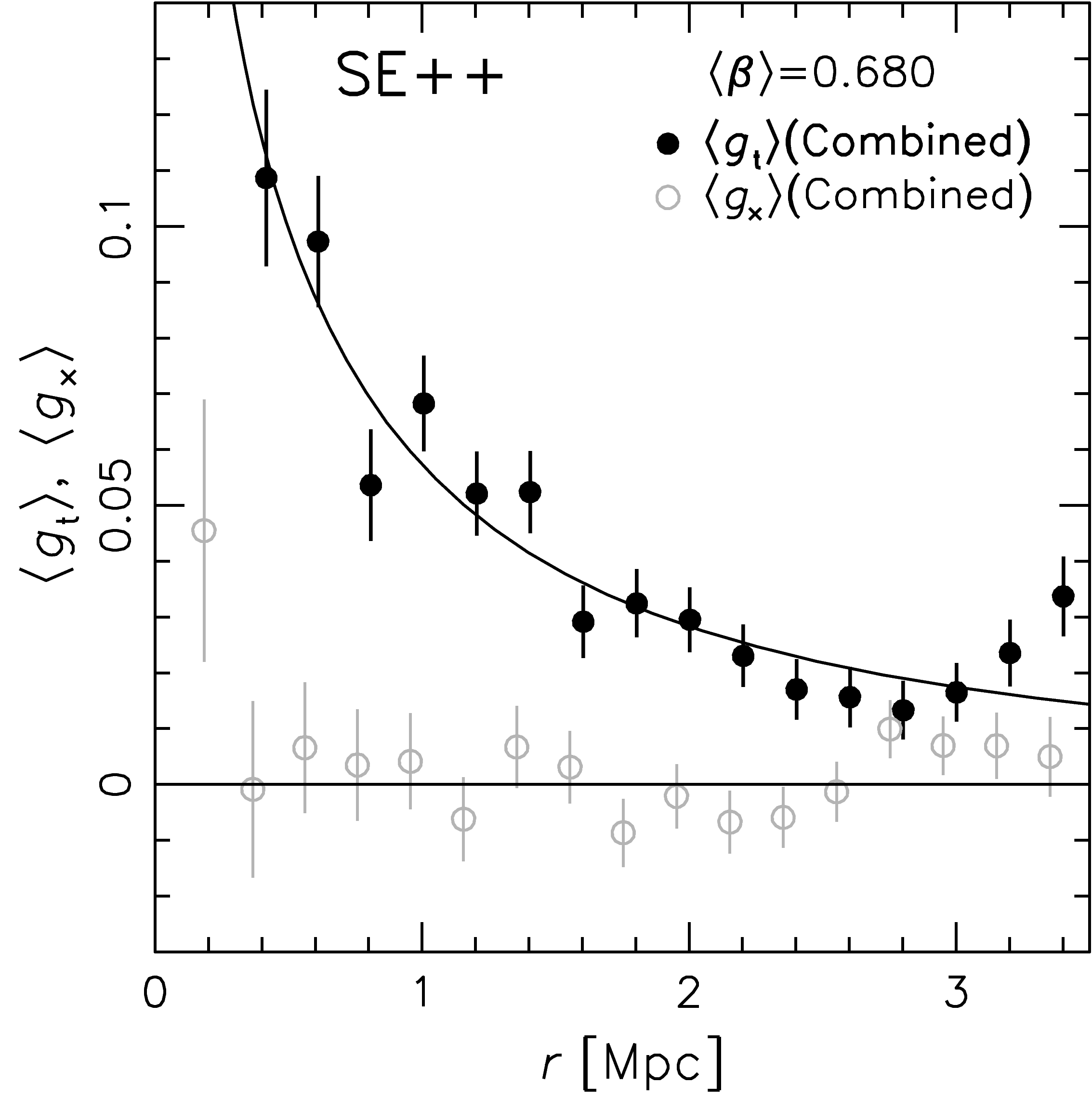}
\caption{Combined contamination-corrected reduced shear profiles of A2390 based on the \texttt{LensMC} (left), \texttt{KSB+} (middle) and \texttt{SE++} (right) shear estimates. For this figure, the individual noisy reduced shear profiles of the different magnitude and photometric redshift bins have been rescaled to the effective mean $\langle\beta\rangle$ and combined, including all tomographic bins with \mbox{$0.3< z_\mathrm{ph}<2.8$}. The curves show the correspondingly averaged best-fit NFW model prediction  for the tangential component assuming a fixed concentration $c_\mathrm{200c}=4$.}
\label{fig:combined_shear_profile}
\end{figure*}

\subsubsection{\label{sec:WLconstraintsanduncertainties}Resulting mass estimates and uncertainties}

\begin{table}[htbp!]
\caption{Weak lensing mass constraints for Abell 2390 assuming a spherical NFW mass model with fixed concentration $c_\mathrm{200c}=4$, where the first and second uncertainties listed correspond to the statistical uncertainties from shape noise and large-scale structure projections, respectively (see Sect.\thinspace\ref{sec:WLconstraintsanduncertainties} for systematic uncertainties).}
\smallskip
\label{tab:results}
\smallskip
\begin{tabular}{ccc}
\hline
\hline
  \rule{0pt}{2ex}
Shape catalogue & $M_\mathrm{200c}$ [$10^{14}M_\odot$] &  $M_\mathrm{500c}$ [$10^{14}M_\odot$] \\[1pt]
\hline
 \rule{0pt}{2ex}
\texttt{LensMC}  & $16.0\pm 1.2\pm 1.7$ & $11.1\pm 0.8\pm 1.2$ \\
\texttt{KSB+}  & $14.9\pm 1.2 \pm 1.6$ & $10.3\pm 0.9\pm 1.1 $\\
\texttt{SE++}  & $14.5\pm 1.1 \pm 1.6 $ & $10.0\pm 0.7 \pm  1.1$\\
\hline
\end{tabular}
\end{table}

We summarise the resulting mass constraints
for the three shape catalogues in Table \ref{tab:results},
including
also
derived mass constraints for $\Delta=500$. In addition to the shape noise caused by the intrinsic galaxy shapes, statistical uncertainty in cluster WL
is also caused by large-scale structure projections \citep[e.g.,][]{hoekstra01}. We estimate these by computing Gaussian shear field realisations for the broad combined redshift distribution of our tomographic bins following appendix B in \citet{simon12} and modelling the nonlinear matter fluctuation power spectrum according to \citet{takahashi12} assuming our reference cosmology. As a result we find that large-scale structure projections actually provide a larger contribution ($11\%$ relative uncertainty) to the statistical error budget for our data set compared to shape noise ($8\%$ relative uncertainty).

Compared to the combined $ 14 \%$ statistical uncertainty the
systematic uncertainty is small for our study. The largest contribution originates from the shear calibration,
where the \mbox{$ 3.2\%$} calibration uncertainties estimated for the \texttt{KSB+} and \texttt{LensMC} catalogues
(see Sect.\thinspace\ref{sc:Shapes}) translate into
$4.8\% $ mass uncertainties.
While the employed \texttt{SE++} implementation has not yet gone through the same level of
testing on WL image simulations, the excellent agreement, especially with the \texttt{KSB+} catalogue (see
also Sect.\thinspace\ref{sc:Shape_Comparison}), empirically demonstrates a similar level of shear calibration uncertainty for \texttt{SE++}.

A further source of systematic mass uncertainty is provided by the uncertainty of the contamination correction (see Sect.\thinspace\ref{sec:contamination_correction}). We estimate this
by varying the  contamination correction parameters within their fit uncertainties and refitting the WL cluster mass.
Here we find that the residual uncertainty of the contamination correction leads to a $0.7\%$ systematic mass uncertainty only, which is negligible compared to the other uncertainties.

Finally,
uncertainties in the procedure to calibrate the true source redshift distribution (see Sect.\thinspace\ref{sec:nofz}) will further contribute to the systematic mass uncertainty.
A full quantification of this uncertainty requires a realistic end-to-end simulation of the redshift calibration data and calibration procedure. Efforts in this direction are underway within the Euclid Consortium (see Roster et al., in prep.), but this is far beyond the scope of our current paper.
However,   to provide at least an approximate estimate of the level of uncertainty,  we have bootstrapped the calibration sample from COSMOS2020 and computed the dispersion of the estimated mean redshifts in the different combinations of magnitude and photometric redshift bins. For the bin combinations included in the WL mass analysis, this dispersion is in the range of $\sigma(\langle z\rangle)=0.006$--0.031.
Even if we conservatively assume that these shifts are maximally correlated between the different bin combinations, their joint impact still shifts the estimated cluster masses by less than 0.4\%. This weak sensitivity to redshift errors is thanks to the fact that most of the constraining power comes from the higher-redshift tomographic bins (\mbox{$0.9<z_\mathrm{ph}\le 1.5$} and \mbox{$1.5<z_\mathrm{ph}\le 2.8$}), where the geometric lensing efficiency $\beta$ depends only weakly on source redshift, given the fairly low cluster redshift (see Fig.\thinspace\ref{fig:nz}).
We note that this uncertainty estimate does not include  the impact of potential systematic uncertainties in the calibration  data itself.
Regarding such systematic uncertainties, similar previous deep weak lensing studies
employing a deep and well-calibrated photometric redshift catalogue
as reference sample,
found that catastrophic redshift outliers between galaxies at very low redshifts and very high redshifts are the primary concern
\citep{schrabback10,schrabback18,raihan20}.
For the current study, we circumvent this problem by not including the lowest and the highest photometric redshift bins, which are most affected by  catastrophic outliers (compare Fig.\thinspace\ref{fig:nz}).
Thus, we expect that our estimate provides a reasonable assessment of the approximate level of systematic uncertainty related to the redshift calibration.
However, given the limitations
explained above, we conservatively inflate our uncertainty estimate by a factor three, yielding 1.2\%.

Adding all the sources of systematic uncertainty listed above in quadrature, yields a combined 5\% systematic mass uncertainty from the weak lensing shear analysis,
dominated by the shear calibration uncertainty.
We remind the reader that this quoted uncertainty assumes a one-parameter spherical NFW density profile with fixed concentration \mbox{$c_\mathrm{200c}=4$}  and accurate centring.
While centring uncertainties are indeed negligible for our study given the tightly constrained
centre (see Sect.\thinspace\ref{sc:wl_shear_profile_fitting}),
the assumption of a  spherical NFW density profile with fixed concentration
can lead
to substantial additional scatter \citep[approximately 20\%, see e.g.][]{becker11,sommer22,giocoli24} when compared to other mass estimates.\footnote{The impact is likely smaller for our analysis given the good agreement of the employed concentration value with the results of earlier studies  (see Sect.\thinspace\ref{sc:comparison_earlier}).}
For cosmological cluster population studies that include adequate simulation-based weak lensing mass modelling corrections \citep[e.g.][]{bocquet24,bocquet24b,grandis24} this, however, only leads to scatter and not systematic uncertainty.
Therefore,
we choose to not
include this
scatter in the shear-related systematic error budget presented in this subsection
(for constraints that vary both mass and concentration see our companion paper, Diego et al., in prep.).

\subsubsection{Comparison of results from different catalogues and bins}

As can be seen in Table \ref{tab:results}, the WL
mass constraints derived from the \texttt{KSB+} and \texttt{SE++} catalogues are in excellent agreement,
and the \texttt{LensMC} best-fit mass is slightly higher.
However, when considering the systematic uncertainty due to shear calibration,  the \texttt{LensMC} and \texttt{KSB+} masses also agree at the  $1.2\thinspace\sigma$ consistency level.
We note that differences in the mass constraints are additionally caused by
differences in shear weights and selections, which is why the true level of consistency is even better.
In conclusion we find that the mass constraints obtained from the different shear catalogues are fully consistent.

As an important further check for the overall analysis
we investigate also the consistency of mass
constraints for different tomographic redshift bins and magnitude
bins.
Figure \ref{fig:mass_as_fct_of_bin} compares the overall mass constraints indicated by the dotted blue lines to \mbox{$M_\mathrm{200c}$} estimates computed from the different magnitude and photo-$z$ bin combinations separately (in this figure uncertainties only reflect the shape-noise component).
While individually much noisier, we find that the results obtained in the individual bins agree very well with the joint constraint.
This suggests that there are no strong magnitude-  or photometric redshift-dependent systematic trends in the data. Only the bright (\mbox{$22<\IE<24.5$}) galaxies in the highest tomographic redshift bin (\mbox{$1.5<z_\mathrm{ph}<2.8$}) yield a mass constraint that is high compared to the joint constraint by about $2\sigma$, but with a total of eight bin combinations such an outlier is not unexpected.

The shear profiles computed in the individual tomographic redshift and magnitude bin combinations are very noisy.
Therefore, we show combined profiles in Figs.\thinspace\ref{fig:combined_shear_profile_tomo} and \ref{fig:combined_shear_profile}.
Here we scale profiles to the same average $\langle\beta\rangle$ of the contributing bins and also rescale model predictions accordingly. We combine the two magnitude bins for each contributing photometric redshift bin in Fig.\thinspace\ref{fig:combined_shear_profile_tomo}.
For comparison, Fig.\thinspace\ref{fig:combined_shear_profile} shows the
combined profiles when including all photometric redshift and magnitude bins that contribute to the overall mass constraints, illustrating the overall constraining power of the data. In this figure we also show the 45$^\circ$-rotated ($B$-mode) $\langle g_\times \rangle$ profiles, which are broadly consistent with zero, as expected in case of an accurate removal of instrumental signatures.

\section{\label{sc:discussion}Discussion}

\subsection{\label{sc:comparison_earlier}Comparison to earlier studies}

An early detection of the A2390 WL signal was achieved by \citet{squires96}, who employed images taken with the Loral 3 CCD on CFHT. These early observations were limited to a small $7^\prime \times 7^\prime$ field-of-view and therefore only capable of probing the inner cluster regions.
This limitation was overcome in later years via the use of wide-field imaging data.
For example, the cluster was included in the
Weighing the Giants project  \citep{vonderlinden14,kelly14,applegate14}, which employed  imaging from Subaru/Suprime-Cam and CFHT/Megacam.
\citet{applegate14} report constraints for Abell 2390 assuming a spherical NFW density profile with fixed concentration \mbox{$c_\mathrm{200c}=4$}, which can directly be compared to our study.
Their constraint \mbox{$r_\mathrm{s}=(0.56\pm0.04)\,\mathrm{Mpc}$}  on the NFW scale radius (obtained for a colour-cut source selection) corresponds to a mass
$M_\mathrm{200c}=16.1^{+3.7}_{-3.2} \times 10^{14}M_\odot$, which is fully consistent with our derived mass constraints.

As part of the LoCuSS project, \citet{okabe16} also studied the WL shear profile of A2390 using Subaru/Suprime-Cam imaging.
They report a mass constraint
$M_\mathrm{200c}=15.1^{+2.7}_{-2.4} \times 10^{14}M_\odot$,
in excellent agreement with our measurements.\footnote{\citet{okabe16} estimated a concentration of \mbox{$c_\mathrm{200c}=4.1^{+1.1}_{-1.0}$}. When shifting to their best-fit concentration, our mass constraints decrease only marginally  by
$0.9\%$.}
Likewise, the cluster was part of the joint analysis of CFHT data from the CCCP \citep{hoekstra15} and MENeaCS programmes
by \citet{herbonnet20}, who
find
$M_\mathrm{200c}=(16.4\pm 3.0)\times 10^{14}M_\odot$, which is likewise
in good agreement with our constraints.\footnote{\citet{herbonnet20} employ the concentration--mass relation from
\citet{dutton14}, which yields a concentration \mbox{$c_\mathrm{200c}\simeq 3.8$} at the cluster mass and redshift. When using this concentration our derived masses increase  by only
$1.8\%$.}
We note that \citet{herbonnet20} scaled their mass estimates by a simulation-derived factor $0.93^{-1}$ to account for an average mass modelling bias for their cluster population study.
Such a  bias  not only depends on the adopted concentration--mass relation and fit range \citep{sommer22}, but  additionally  on factors such as  the dynamical state and the triaxiality \citep[e.g.,][]{giocoli24},
which are not fully known for  individual clusters. Accordingly, we do not apply a mass bias correction for our single-target study.
For a more direct comparison we therefore also consider the mass $M_{200c}=(15.3\pm 2.8)\times 10^{14}M_\odot$ that
 \citet{herbonnet20} would have found without mass  bias correction.
 This mass estimate is likewise fully consistent with our results.
In summary, our mass constraints agree very well with recent wide-field imaging analyses that also report  mass constraints  assuming NFW density profiles.
We note however that the \Euclid constraints are tighter by factors of approximately $0.6$--$0.8$ thanks to their higher WL source densities.

A recent publication studying the WL signature of A2390  was presented by \citet{dutta24}, who  combine shape estimates in 411 short exposures taken with the   WIYN/ODI imager.
Since this paper does not report
spherical overdensity mass estimates we cannot compare their and our results quantitatively.
Finally, first WL measurements based on the \Euclid ERO observations of A2390 were already presented in the  overview publication of this ERO data set \citep{EROLensData}. That analysis employed  the \texttt{KSB+} measurements discussed here to present an initial convergence reconstruction based on the ERO observations.
We have extended this analysis by adding two further shape catalogues, refined calibrations, a background selection using photometric redshifts, and an estimation and correction for cluster member contamination. The latter two items  increase the measured shear signal, which leads to an increased peak in the  convergence reconstruction compared to the results presented in \citet{EROLensData}.

\subsection{\label{sc:implications}Possible implications for future \Euclid cluster weak lensing studies}

Within the
Euclid Consortium substantial efforts are underway in order to
achieve highly accurate tomographic cosmic shear measurements.
With sources split into photometric redshift bins,
the required calibrations for shear measurements and true source redshift distributions are derived for the same split into tomographic bins.
The analysis strategy followed in our paper to derive accurate cluster WL mass constraints can directly be applied to such future \Euclid data sets, with WL sources split into  photometric redshift bins, avoiding the  need for custom source selections and calibrations.
A key requirement for this are accurate estimates for  residual cluster member contamination in the different photometric redshift bins.
Following \citet{kleinebreil25} we estimate this contamination from the source number density profile after applying corrections for the impact of foreground source obscuration (which reduces the detection probability) and magnification.
Here it is important to realise that the combination of cluster member contamination, source obscuration, and magnification can in  combination  lead to almost flat uncorrected number density profiles, which would naively suggest low contamination even if the true contamination is significant (compare Figs.\thinspace\ref{fig:nd_raw} and \ref{fig:nd_corr}).

Our study also demonstrates that \Euclid has the potential to deliver sensitive WL constraints significantly beyond the nominal \mbox{$\IE=24.5$} limit of the EWS.
While the analysed ERO observations have a 3 times longer coadded integration time compared to the EWS, this corresponds to
a depth difference of only $0.6$ mag.
Accordingly, since our shear catalogues
extended to  \mbox{$\IE
\simeq 26.5$},
we  expect that it will be possible to compute shear estimates (for incomplete source samples) in the EWS to \mbox{$\IE
\simeq 25.9$}, with some dependence on zodiacal background.
For early EWS WL studies the main limitation for exploiting these additional faint WL sources will likely be the depth of the ground-based observations included in the photometric redshift computation.
However, this limitation may be overcome in the future if \Euclid shear measurements are combined with deeper photometry from the  Vera C.~Rubin Observatory's Legacy Survey of Space and Time (LSST), as considered in \citet{guy22}.
As also discussed in  \citet{rhodes17},
such a faint extension of the EWS WL catalogue has the potential to extend sensitive WL mass measurements to higher redshifts thanks to the increased fraction of high-redshift source galaxies at fainter magnitudes.
Such fainter galaxies typically have more noisy photometric redshifts and may also have more uncertain calibrations of their true redshift distribution.
As a simple solution for this we propose to split sources into magnitude bins, as done in our study.
In this way, brighter galaxies have more precise photometric redshifts and more accurate calibrations than fainter ones.
This approach allows science analyses to choose which bins to include, depending on their individual systematic error requirements.

We note that future \Euclid cluster WL studies may need to revisit the question of a cluster-regime shear calibration. For clusters, shears are often non-weak, while blending is increased compared to the field.
Such effects may affect shear calibrations at the per-cent level \citep[see e.g.,][]{hernandez20}.
Considering a wider shear range  we present a first test for such nonlinear shear calibration corrections for \texttt{LensMC} in Appendix \ref{appendix:lensmc}, finding them to be small, but non-zero (e.g., at \mbox{$|g|=0.1$} they correspond to a multiplicative bias shift of 0.5\%).
In the future, this work should be extended to also capture the impact of increased blending.
For very massive low-redshift clusters
the shear is detected with high significance, which opens the possibility to apply corrections directly as a function of the radius- and redshift-dependent measured shear.
However, for clusters at lower mass or higher redshift, shear estimates are more noisy. In this case a more useful approach might  be to derive effective corrections for the WL mass bias that depend on cluster mass and redshift, in order to include them in the population modelling and mass calibration \citep[e.g.,][]{grandis24}.

\section{\label{sc:conclusions}Conclusions}

In this work we have presented the first detailed WL analysis using \Euclid data, analysing ERO observations of the massive galaxy cluster Abell 2390.
Thanks to the high spatial resolution and sensitivity of the \Euclid VIS observations,
as well as the depth of complementary photometric measurements derived from the \Euclid NISP images and
ground-based data, we were able to include a high density of sources  into our analysis down to magnitudes \mbox{$\IE\simeq 26.5$}.
As a result, we obtained constraints on the cluster mass that are significantly tighter compared to earlier studies using ground-based data.
Based on our analysis we conclude that \Euclid has the potential to provide sensitive WL measurements in its EWS
well beyond the nominal \mbox{$\IE=24.5$} limit, especially if it is combined with deep photometric data, as will be obtained within LSST.

In our analysis, which focuses on the WL measurements, we assumed a simple spherical NFW density profile with fixed concentration for simplicity. A more detailed analysis of the cluster, including constraints on its concentration, will be presented in a companion paper by
Diego et al.~(in prep.),
who also incorporate strong lensing measurements.

We conducted our analysis using tomographic redshift bins, closely mimicking the data structure expected for future \Euclid WL data sets. As an important validation for our tomographic WL measurements, we compared the cluster mass constraints that are derived jointly from all bins to those derived from the individual ones, finding excellent agreement. Likewise, we find very good agreement between the results derived from the three independent shape catalogues (within \mbox{$1.2\thinspace\sigma$} when including shear calibration uncertainties).

For our single-target study the error budget is fully dominated by statistical uncertainties from large-scale structure projections and intrinsic galaxy shapes.
Future {\Euclid} cluster WL studies will investigate larger samples, especially with the goal to constrain cosmological parameters using clusters.
These studies will have more stringent requirements regarding their systematic accuracy.
Fortunately, large efforts are underway to obtain highly accurate WL measurements for cosmic shear analyses, which require highly accurate PSF models, as well as accurate shear and redshift calibrations (for which we have demonstrated strategies in Appendix \ref{appendix:lensmc} and Sect.\thinspace\ref{sec:nofz}, respectively).
Cluster WL studies will be able to
make use of these calibrations
if their analysis is conducted in the same photometric redshift bins and complemented with an accurate estimation of cluster member contamination, as demonstrated in Sect.\thinspace\ref{sc:cluster_member_contamination},
although slightly more conservative scale cuts may be needed to ensure sub-percent accuracy given the impact of intra-cluster light \citep{gruen19}.
Cluster cosmology studies additionally require cluster samples with well-modelled selection functions \citep[e.g.,][]{bleem15,bleem20,bleem24,hilton21,bulbul24}, as well as accurate estimates for WL mass modelling biases from simulations \citep[e.g.,][]{grandis21,sommer22,sommer25}.
With all these ingredients in hand they are expected to become a key component in multi-probe cosmological analyses, as already demonstrated by \citet{bocquet25}, who combined data from the Dark Energy Survey and the South Pole Telescope.

\begin{acknowledgements}
The Innsbruck authors acknowledge support provided by the Austrian Research Promotion Agency (FFG) and the Federal Ministry of the Republic of Austria for Climate Action, Environment, Mobility, Innovation and Technology (BMK) via the Austrian Space Applications Programme with grant numbers 899537, 900565, and
911971.
Part of this work was made possible by utilising the CANDIDE cluster at the Institut d’Astrophysique de Paris. The cluster was funded through grants from the PNCG, CNES, DIM-ACAV, the Euclid Consortium, and the Danish National Research Foundation Cosmic Dawn Center (DNRF140); it is maintained by Stephane Rouberol.
JMD acknowledges support from project PID2022-138896NB-C51 (MCIU/AEI/MINECO/FEDER, UE) Ministerio de Ciencia, Investigación y Universidades.
LL acknowledges  support from  the Austrian Science Fund (FWF)
[ESP 357-N].
  \AckERO
  \AckEC
This research is based in part on archival data collected at the Subaru Telescope, which is operated by the National Astronomical Observatory of Japan.
This research is also based in part on archival observations obtained with MegaPrime/MegaCam, a joint project of CFHT and CEA/DAPNIA, at the Canada-France-Hawaii Telescope (CFHT) which is operated by the National Research Council (NRC) of Canada, the Institut National des Sciences de l'Univers of the Centre National de la Recherche Scientifique (CNRS) of France, and the University of Hawaii.
We are honoured and grateful for the opportunity of observing the Universe from Maunakea, which has the cultural, historical, and natural significance in Hawaii.

\end{acknowledgements}

\bibliography{Euclid,eroWL}

\begin{appendix}

\section{Further details on the \texttt{SE++} analysis}
\label{appendix:SE++-details}

\subsection{Star-galaxy separation}

The top panel of Fig.~\ref{fig:sepp_photom} shows the distribution of \texttt{SE++} sources in the plane spanned by effective radius and the $\IE$
single-S\'ersic magnitude.
Non-saturated
stars and galaxies, located in the
primary region of interest and selected according to size and magnitude,
are highlighted along with the broken line marking the boundary $r_{\rm eff}\simeq 0\farcs05$ between point sources and galaxies.
One can notice some bending of the stellar locus,
which is caused by
a log-size prior,  progressively kicking in at \mbox{$\IE<22$}.
For magnitudes \mbox{$\IE\gtrsim 27$} stars and galaxies merge in the diagram and cannot be distinguished any more in these ERO data.
In the lower panel, the morphological selection based on the best-fit effective radius is cast into the \mbox{$B-\IE$} versus \mbox{$\IE -\HE$} plane, mimicking the commonly used $BzK$ colour-colour diagram.
At the location of the stellar locus, we do not observe any noticeable excess of objects classified as galaxies underneath, which would hint at misclassified stars if detected.

\begin{figure}[htbp!]
\centering
\includegraphics[angle=0,width=0.95\hsize]{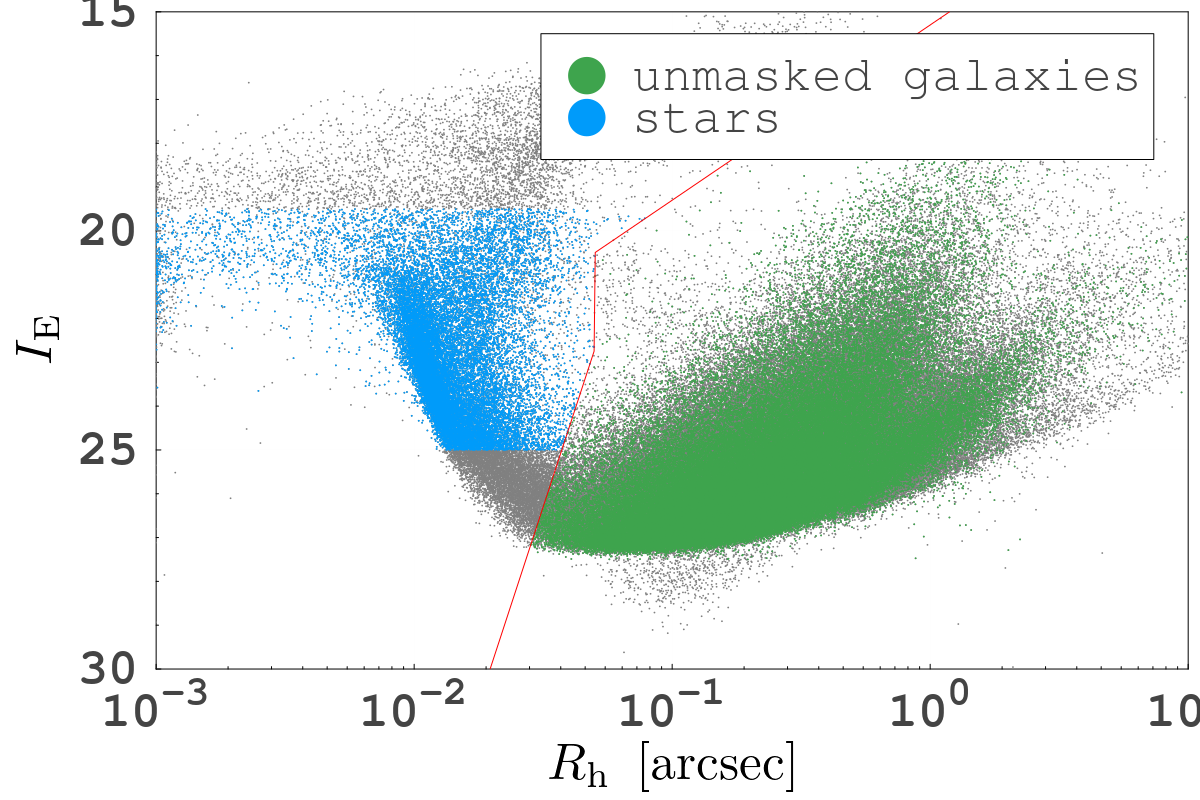}
\includegraphics[angle=0,width=0.95\hsize]{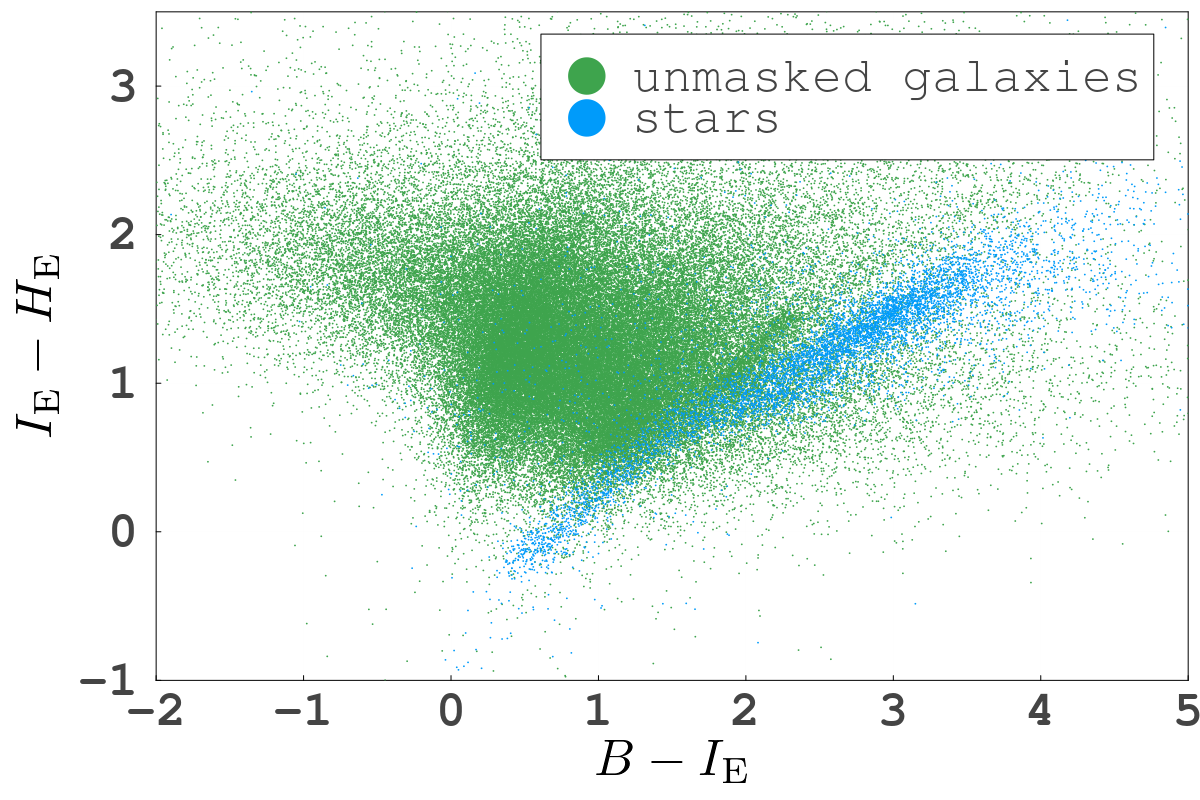}
\caption{Top:
Best-fit {\tt \texttt{SE++}} single-S\'ersic effective radius versus \IE magnitude  diagram
  with highlighted unsaturated stars (blue) and galaxies (green). The morphological distinction is based on the size boundary overlaid in red.
  Bottom:
  Galaxies (green) and stars (blue) distributed in the $B-\IE$ versus $\IE-\HE$ plane, showing the stellar locus and the red sequence of passive cluster member galaxies.}
\label{fig:sepp_photom}
\end{figure}

Figure \ref{fig:sepp_pm} further illustrates the quality of the star/galaxy discrimination by presenting the recovered best-fit proper motion of sources in the field of view, as measured with \texttt{SE++}.
Making use of the long time span between the Suprime-Cam observations (early 2000's) and the \Euclid~observations,
we fit band-to-band offsets in the multi-band bulge+disc modelling run. Proper motions smaller than $1\,{\rm mas\,yr}^{-1}$ can be measured down to
faint \mbox{$\IE \simeq 25$} magnitudes.
This analysis reveals a significantly larger spread and mean centroid shift for stars, suggestive of an apparent bulk motion of a large fraction of field stars, whereas galaxies are consistent with no apparent motion. This is verified down to faint magnitudes
and illustrates the effectiveness of the star/galaxy separation with limited contamination\footnote{One should nevertheless bear in mind that very faint stars are on average more distant,
progressively exhibiting smaller apparent motions.
Moreover,  at \mbox{$\IE> 25$}
a loose prior centred on null motion (with dispersion $10\,{\rm mas\,yr}^{-1}$)
starts to pull source proper motions towards that of galaxies, hence weakening the power of this diagnostic.}.

\begin{figure}[htbp!]
\centering
\includegraphics[angle=0,width=0.95\hsize]{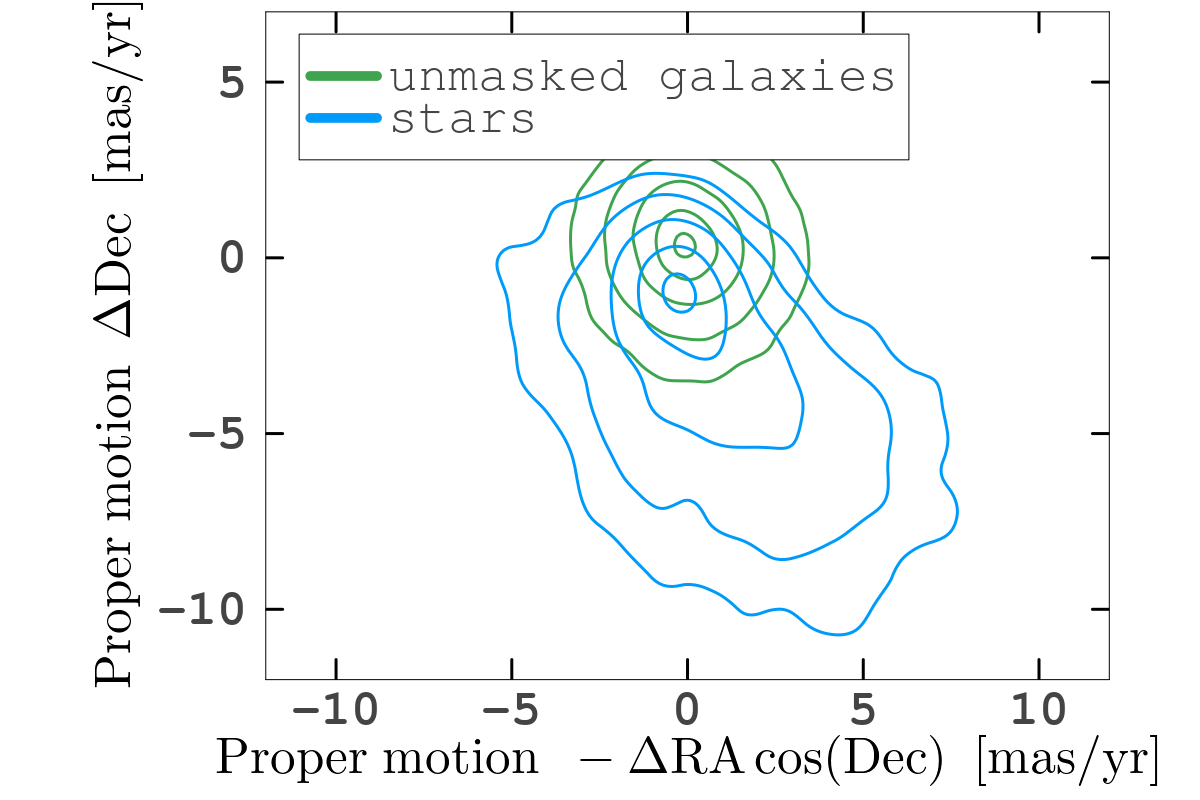}
\includegraphics[angle=0,width=0.95\hsize]{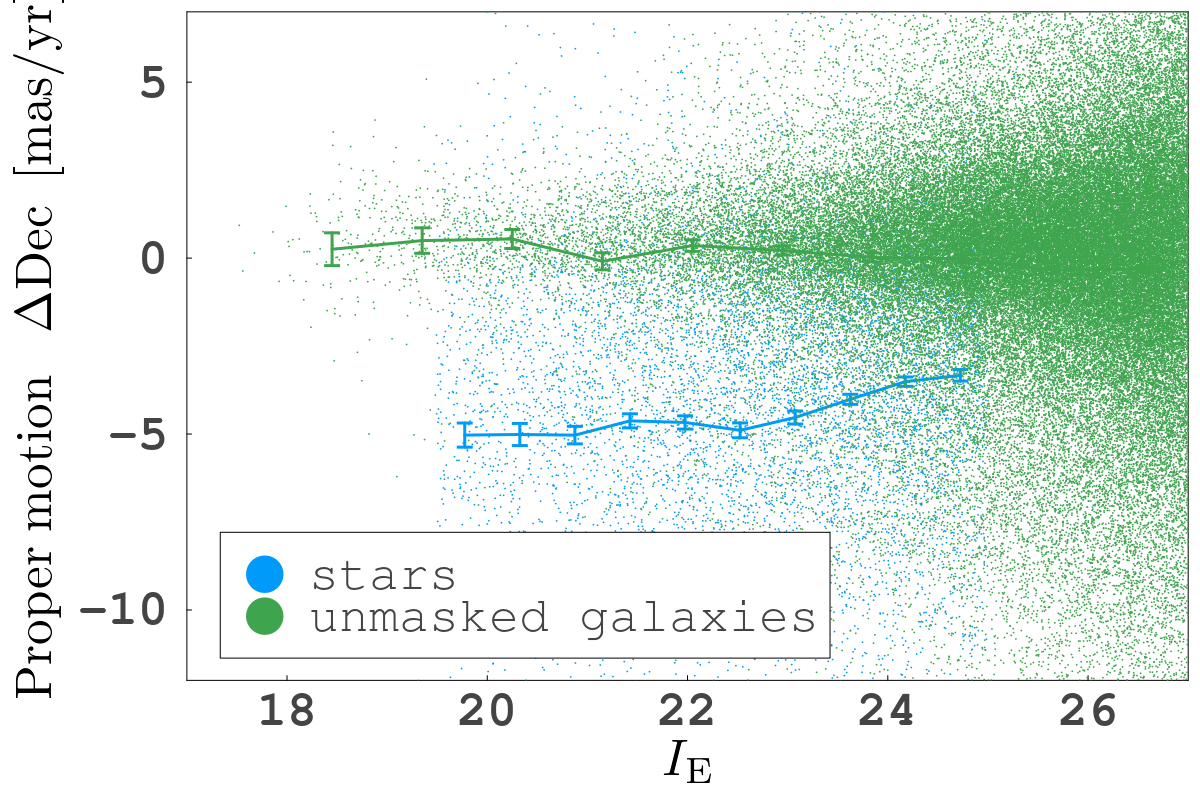}
\caption{Top:
Proper motion of stars (blue) and galaxies (green), showing clear evidence for more spread and shifted motion of the former population.
Bottom:
Proper motions of stars (blue) and galaxies (green) along the Declination direction as a function of $\IE$ magnitude, illustrative of a stable systematic trend with magnitude, and thus indicative of mild cross contamination.
}
\label{fig:sepp_pm}
\end{figure}

\subsection{\texttt{SE++} configuration for single-S\'ersic fitting}
The following code snippet details the important lines of the {\tt python} configuration file of {\tt SE++}, optimised for single-S\'ersic fitting. Note several internal changes of variables: use of log(radius) with size prior; use of Cartesian
$\epsilon_1,\epsilon_2$
ellipticity components with a simple Gaussian prior of standard deviation $0.25$ instead of the (axis ratio, angle) pair of variables; and finally an internal \verb!X_ser! variable casting the domain of S\'ersic indices onto a symmetric unbound support.
\begin{verbatim}
x,y=get_pos_parameters()
ra,dec=get_world_position_parameters(x, y)

r_range=Range(lambda v, o: (0.01, 10*v),
               RangeType.EXPONENTIAL )
rd = FreeParameter(lambda o: o.radius, r_range )
lrd=DependentParameter(lambda re:np.log10(re),rd)
add_prior( lrd, 0.16, 0.30 )

e_range=Range((-0.999, 0.999), RangeType.LINEAR)
e1 = FreeParameter( 0.0, e_range)
e2 = FreeParameter( 0.0, e_range)
add_prior( e1, 0.0, 0.25 )
add_prior( e2, 0.0, 0.25 )
ang = DependentParameter( lambda x,y:
                0.5*np.arctan2( y, x ), e1, e2 )
emod = DependentParameter( lambda x,y:
                np.sqrt( x*x + y*y ), e1, e2 )
ar = DependentParameter( lambda e:
                np.abs(1-e)/(1+e), emod )
X_ser=FreeParameter( -2.3,
         Range((-20, 20), RangeType.LINEAR) )
add_prior( X_ser, -2.3, 1.1 )
n_ser=DependentParameter( lambda x:
     (10*np.exp(x)+0.4)/(1+np.exp(x)), X_ser )
flux = get_flux_parameter()
ssercomp=SersicModel( x, y, flux, rd,
                      ar, ang, n_ser )
\end{verbatim}

\section{Testing \texttt{LensMC} in the non-weak shear regime}
\label{appendix:lensmc}

 \begin{figure}[htbp!]
        \centering
        \includegraphics[width=0.95\hsize]{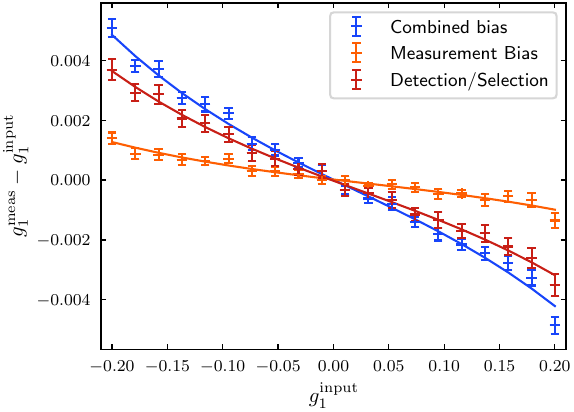}
        \caption{Multiplicative bias estimate for \texttt{LensMC} at non-weak shears. For this measurement we used a $3$-times oversampled PSF. We distinguish here between measurement bias and detection bias and fit a third-order polynomial to all data points.}
        \label{fig:lensmc_biases}
    \end{figure}

    In order to validate the shear measurement with \texttt{LensMC} in the regime of non-weak shears (i.e., beyond the cosmic shear regime with \mbox{$|g|\lesssim 0.05$}), we rendered images from the Flagship mock galaxy catalogue
    \citep{EuclidSkyFlagship},
    which was obtained from Cosmohub \citep{CosmoHub1, CosmoHub2}.
    The simulations follow the approach from \cite{Jansen24}, using an analytic {\Euclid}-like PSF model \citep[similarly to the model employed by][]{tewes19}, but their depth has been adjusted to match
    this ERO observation. We include shape and pixel noise cancellation to guarantee
    efficient bias estimation \citep[see][]{Jansen24}, where we
applied $20$ different constant values for the $g_1$ shear component, uniformly spaced between $-0.2$ and $0.2$, to
the Flagship inputs.
    We then took the weighted mean of the recovered ellipticities from \texttt{LensMC} as an estimate for the shear, taking the \texttt{LensMC} shear weights into account.
    We also generated a second smaller set of simulations with shear applied only to the  $g_2$ component
    for the determination of refined linear bias estimates (see below).

    The measurement with \texttt{LensMC} was conducted in two different ways on the same simulations.
    In the first run we provided \texttt{LensMC} with a $3$-times oversampled representation of the analytic input PSF model, following the standard procedure used in \citet{congedo24}  to account for the undersampled nature of the \Euclid PSF.
In the second run we provided  \texttt{LensMC} with a PSF model sampled at the native VIS pixel scale,
which matches the \texttt{LensMC} runs on the ERO data and
reflects the fact that an accurate super-resolution PSF model was not available for the ERO analysis.\footnote{ While the \texttt{PSFEx} model for VIS was generated with $2$-times oversampling (see Sect.\thinspace\ref{sc:PSFEXmodel}), the super-resolution information within the model is limited given that it was generated from the image stack. Since  the current version of \texttt{LensMC}
can only employ odd-number oversampling factors we therefore had to employ PSF models sampled at the native pixel scale (rather than a poorly constrained model with $3$-times oversampling).}
We only considered galaxies in the magnitude range between $20.5$ and $26.5$ for our simulation analysis, which resembles the magnitude range of the galaxies used in our weak lensing analysis (plus a small fraction of brighter galaxies).

    \begin{figure}[htbp!]
        \centering
        \includegraphics[width=0.95\hsize]{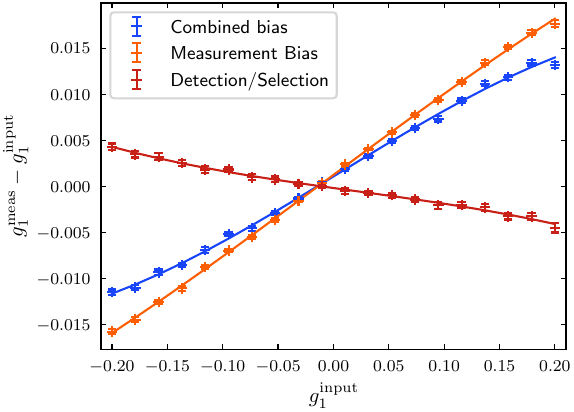}
        \caption{Multiplicative bias estimate for \texttt{LensMC} at non-weak shears. For this measurement we use the same simulations as for Fig.~\ref{fig:lensmc_biases}, but employed a PSF model sampled at the native VIS pixel scale for the  \texttt{LensMC} measurement. We distinguish here between measurement bias and detection bias and fit a third-order polynomial to all data points.}
        \label{fig:lensmc_biases_1x}
    \end{figure}

    \begin{table*}[htbp!]
        \caption{Fit parameters for the fits in Figs.~\ref{fig:lensmc_biases} and \ref{fig:lensmc_biases_1x} according to Eq.\thinspace(\ref{eq:cubic_shear_bias}).}
        \label{tab:lensmc_fit_parameters}
        \centering
      \begin{tabular}{l  c  c  c  c  c  c  c  c}
        \hline \hline
        \multicolumn{9}{c}{Measurement using a three-times oversampled PSF model} \\[1pt]
        \hline
         \rule{0pt}{2ex}
        Bias & $\alpha$ & $\sigma_\alpha$ & $\beta$ & $\sigma_\beta$ & $\mu$ & $\sigma_\mu$ & $c$ & $\sigma_c$\\
        & $[10^{-2}]$ & $[10^{-2}]$ & $[10^{-3}]$ & $[10^{-3}]$ & $[10^{-3}]$ & $[10^{-3}]$
        & $[10^{-4}]$ & $[10^{-4}]$ \\[1pt]
        \hline
         \rule{0pt}{2ex}
        Combined & $-$11.18 & 3.22 & 8.28 & 3.35 & $-$18.68 & 0.85 & 0.55 & 0.53 \\
        Measurement & $-$4.27 & 2.45 & 2.36 & 2.57 & $-$4.48 & 0.64 & 0.55 & 0.39 \\
        Detection / Selection & $-$6.79 & 4.07 & 7.15 & 4.26 & $-$14.27 & 1.07 & $-$0.20 & 0.66 \\[1pt]
        \hline
        \multicolumn{9}{c}{ \rule{0pt}{2ex} Measurement using a PSF model sampled at the native pixel scale} \\[1pt]
        \hline
         \rule{0pt}{2ex}
        Combined & $-$23.56 & 3.76 & 3.51 & 3.87 & 72.82 & 1.01 & 11.12 & 0.60 \\
        Measurement & $-$11.98 & 3.07 & $-$3.10 & 3.30 & 89.31 & 0.78 & 12.83 & 0.48 \\
        Detection / Selection & $-$10.34 & 4.88 & 7.25 & 5.11 & $-$16.93 & 1.28 & $-$1.75 & 0.78 \\[1pt]
        \hline
        \end{tabular}
    \end{table*}

    In Figs.~\ref{fig:lensmc_biases} and \ref{fig:lensmc_biases_1x} we show the resulting bias fits with and without PSF model oversampling, respectively. For these figures we simulated $\SI{40}{\deg\squared}$ of images, where $\SI{10}{\deg\squared}$ are unique and the remaining area corresponds to shape and pixel noise cancelled versions of the same $\SI{10}{\deg\squared}$ \citep[see][]{Jansen24}.
    Since we simulated large shears and make use of noise cancellations, this small area is already sufficient to estimate biases with high precision. We separate between measurement and detection bias by considering only complete noise cancellations to determine the measurement bias. This is done on the unweighted shears, since the different versions of a galaxy have different weights, which is a selection effect. All other biases are then captured by the difference between the total bias and the measurement bias. The fitted function is of the form
    \begin{equation}\label{eq:cubic_shear_bias}
        g_{1, \mathrm{meas}} - g_{1, \mathrm{input}} = \alpha\, g_{1, \mathrm{input}}^3 + \beta\, g_{1, \mathrm{input}}^2 + \mu\, g_{1, \mathrm{input}} + c \,.
    \end{equation}
    This is chosen to study the next higher-order symmetric and antisymmetric terms in addition to the linear bias model. We list the best-fit parameters in Table~\ref{tab:lensmc_fit_parameters}. The slight difference in the detection and selection bias originates from the shear weights, which are also impacted by the chosen oversampling.

    We want to highlight that the values of the multiplicative measurement bias are well in line with the results from \citet{congedo24} if we oversample the PSF. The combined detection and selection bias, on the other hand, deviates as expected, since we used a different simulation setup and a different magnitude selection. We find nonlinear terms, which are $1$--$3\,\sigma$ inconsistent with zero, depending on the type of bias and the oversampling. Overall these terms are small and can be neglected
    at the accuracy requirements of our current study.
In future cluster WL studies using larger samples it may become necessary to account for these nonlinear terms.
However, for this a more detailed analysis will we required to also captures the increased blending occurring in cluster environments \citep[see e.g.,][]{hernandez20}, which we ignored in our current investigation.

Since the nonlinear terms are small we employ a refined linear bias correction for our current analysis, which captures the shift in bias values occurring when using native PSF model sampling instead of a
$3$-times
oversampling.
Omitting the cubic and quadratic terms in Eq.~\eqref{eq:cubic_shear_bias} and fitting to the data in Fig.~\ref{fig:lensmc_biases_1x}, we find $\mu_1=(65.73\pm0.77)\times 10^{-3}$  and $c_1=(12.41\pm0.82)\times 10^{-4}$, i.e.~a shift by $\Delta\mu_1=(-7\pm 1)\times 10^{-3}$ in multiplicative bias compared to the cubic-fit results.
For the second shear component we find $\mu_2=(33.91\pm1.14)\times 10^{-3}$ and $c_2=(7.54\pm1.23)\times 10^{-4}$.
The difference in the multiplicative bias values for the two shear components is likely due to the limited sampling, which effectively differs for the two components.
In our cluster WL analysis the mass constraints are derived from the azimuthally averaged tangential reduced shear, which has equal contributions from both components.
We therefore correct the \texttt{LensMC} shear estimates using
the multiplicative bias estimate
derived when combining  both components,
$\mu=(49.82\pm0.69)\times 10^{-3}$.

\section{Description of the Wiener-filtered convergence reconstruction algorithm}
\label{appendix:wiener_filtered}

For the convergence reconstruction described in Sect.\thinspace\ref{sc:wl_results_reconstruction} we employ a
Wiener filter, which
yields the minimum variance estimate
\citep{seljak98,mcinnes09,simon09}
\begin{equation}
  \vec{\kappa}_{\rm mv}=
  \left(\mat{1}+
    \mat{S}\,\mat{P}_{\gamma\kappa}^\dagger\,\mat{N}^{-1}\,\mat{P}_{\gamma\kappa}\right)^{-1}\,
    \mat{S}\,\mat{P}_{\gamma\kappa}^\dagger\,\mat{N}^{-1}\,\vec{\epsilon}=:\mat{W}\,\vec{\epsilon}
    \;,
\end{equation}
of the lensing convergence on a grid, $\vec{\kappa}$, assuming: the
convolution $\mat{P}_{\gamma\kappa}$ in
$\vec{\epsilon}=\mat{P}_{\gamma\kappa}\,\vec{\kappa}+\vec{n}$ for
random shape noise, $\vec{n}$, and for the source ellipticity binned
within grid cells, $\vec{\epsilon}$; the covariance
$N_{ij}=\ave{n(\vec{\theta}_i)n^\ast(\vec{\theta}_j)}$ of shape noise
between grid cells at $\vec{\theta}_i$ and $\vec{\theta}_j$; and the
signal covariance
$S_{ij}:=\ave{\kappa(\vec{\theta_i})\kappa^\ast(\vec{\theta_j})}=\xi_+(|\vec{\theta}_i-\vec{\theta}_j|)$. The
result is a smoothed $\vec{\kappa}$ subject to a smoothing kernel
defined by
$\ave{\vec{\kappa}_{\rm
    mv}}=\mat{W}\,\mat{P}_{\gamma\kappa}\,\vec{\kappa}=([\mat{S}\,\mat{P}_{\gamma\kappa}^\dagger\,\mat{N}^{-1}\,\mat{P}_{\gamma\kappa}]^{-1}+\mat{1})^{-1}\,\vec{\kappa}$. Specifically,
we assume uncorrelated noise and infinite noise for grid cells without
sources, $n_{\rm gal}(\vec{\theta}_i)=0$, this means
$N^{-1}_{ii}=n_{\rm gal}(\vec{\theta}_i)\,\sigma_\epsilon^{-2}$ and
$N_{ij}=0$ for $i\ne j$. For $S_{ii}$, we average $\xi_+(\theta)$,
obtained from the measured shear-shear correlations in the cluster
field, over the solid angle of a grid cell. A best-fit of the generic
profile $\xi_+(\theta)=[a_0+a_1(\theta/^\prime)]\,[1+a_2\,(\theta/^\prime)]^{-1}$ to the data,  with fit
parameters
  $(a_0, a_1, a_2)=(0.00198397,
    -0.00029534,
    0.209271)
    $,
provides a smoother kernel for the Wiener
filter, which we truncate at $\theta=7^\prime$.
Given the large sky area of the ERO WL data, we employ a $512\times 512$ pixel grid for both the shear field binning and the convergence reconstruction.
Furthermore, to account for the reduced shear,
$\ave{\epsilon}=g=\gamma\,(1-\kappa)^{-1}$, the algorithm, similarly to
\citet{seljak98},
is run iteratively by approximately converting the
initial ellipticity grid at $\vec{\theta}_i$ into an estimator of
shear, $\epsilon^n_i=\epsilon_i\,(1-\kappa^{n-1}_{{\rm mv},i})$, for
the next iteration $n$, where we set $\kappa^0_{{\rm mv},i}\equiv0$
initially.

\section{\label{app:overlays}Additional convergence reconstructions}

 Figs.\thinspace\ref{fig:image_mass_overlay_KSB} and
\ref{fig:image_mass_overlay_SEpp}
show overlays of the \Euclid optical+NIR colour image of the cluster
with convergence reconstructions for the \texttt{KSB+} and \texttt{SE++} shear catalogues.

\begin{figure*}
    \centering
 \includegraphics[width=0.69\linewidth]{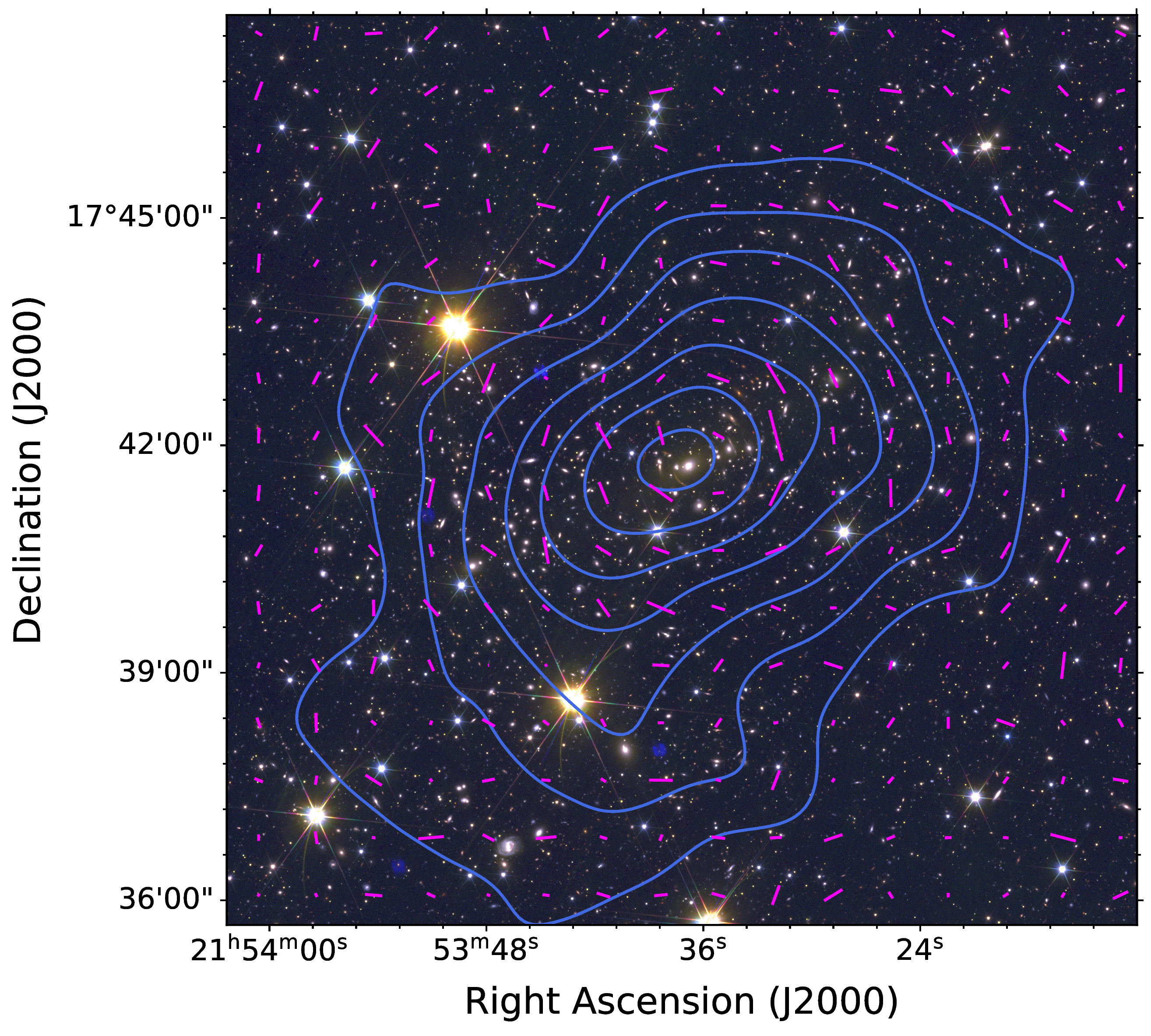}
    \caption{As Fig.\thinspace\ref{fig:image_mass_overlay_lensMC}, but employing the \texttt{KSB+} shear catalogue. }
    \label{fig:image_mass_overlay_KSB}
\end{figure*}

\begin{figure*}
    \centering
        \includegraphics[width=0.69\linewidth]{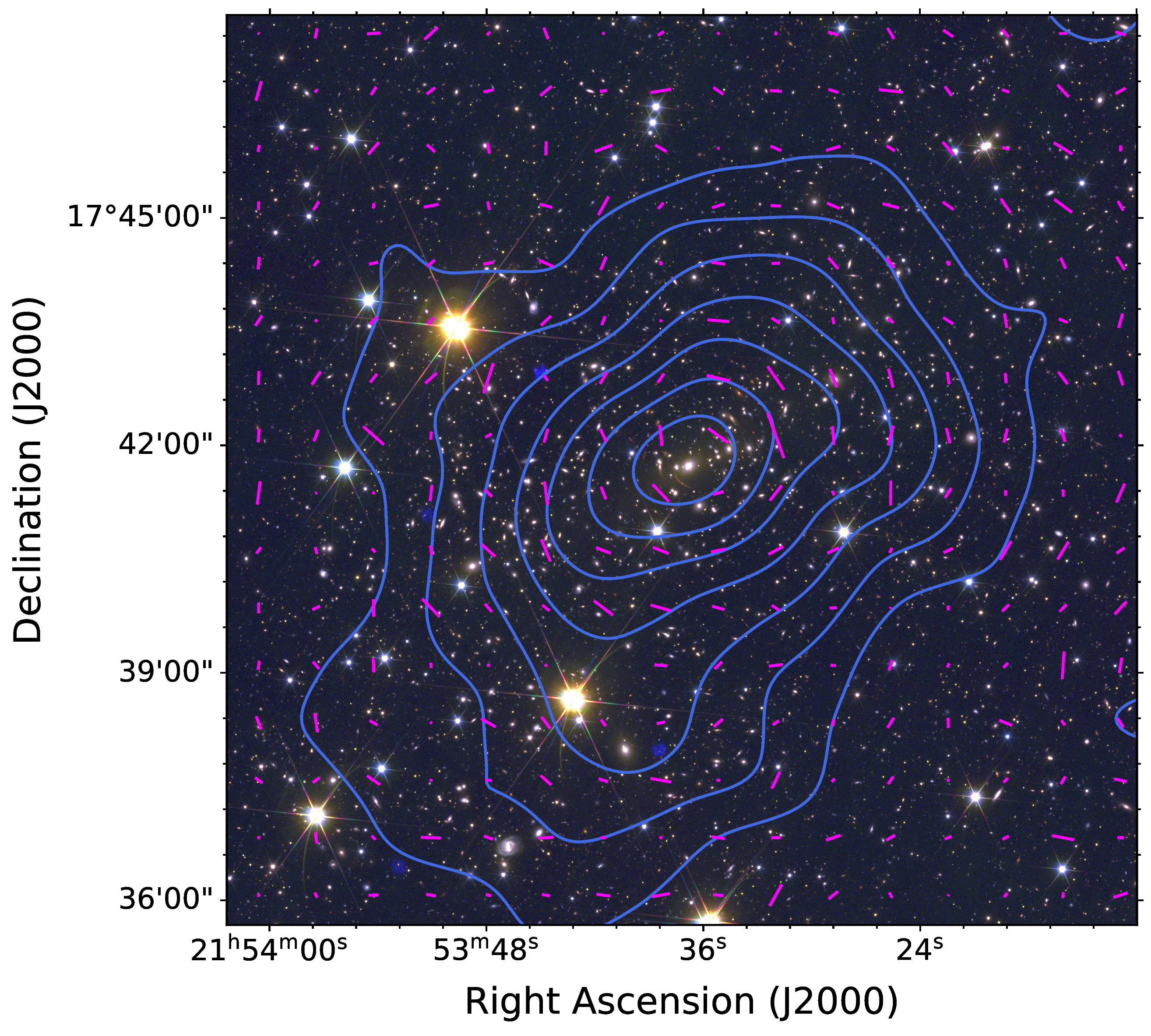}
    \caption{As Fig.\thinspace\ref{fig:image_mass_overlay_KSB}, but employing the \texttt{SE++} shear catalogue. }
    \label{fig:image_mass_overlay_SEpp}
\end{figure*}

\end{appendix}

\end{document}